\PassOptionsToPackage{unicode}{hyperref}
\PassOptionsToPackage{hyphens}{url}
\documentclass[
  10pt,
  a4paper,
]{article}
\usepackage{amsmath,amssymb}
\usepackage[]{tgpagella}
\usepackage{iftex}
\ifPDFTeX
  \usepackage[T1]{fontenc}
  \usepackage[utf8]{inputenc}
  \usepackage{textcomp} 
\else 
  \usepackage{unicode-math}
  \defaultfontfeatures{Scale=MatchLowercase}
  \defaultfontfeatures[\rmfamily]{Ligatures=TeX,Scale=1}
\fi
\IfFileExists{upquote.sty}{\usepackage{upquote}}{}
\IfFileExists{microtype.sty}{
  \usepackage[]{microtype}
  \UseMicrotypeSet[protrusion]{basicmath} 
}{}
\makeatletter
\@ifundefined{KOMAClassName}{
  \IfFileExists{parskip.sty}{%
    \usepackage{parskip}
  }{
    \setlength{\parindent}{0pt}
    \setlength{\parskip}{6pt plus 2pt minus 1pt}}
}{
  \KOMAoptions{parskip=half}}
\makeatother
\usepackage{xcolor}
\IfFileExists{xurl.sty}{\usepackage{xurl}}{} 
\IfFileExists{bookmark.sty}{\usepackage{bookmark}}{\usepackage{hyperref}}
\hypersetup{
  pdftitle={Combining chains of Bayesian models with Markov melding},
  pdfauthor={Andrew A. Manderson; Robert J. B. Goudie},
  pdfkeywords={Combining models, Markov melding, Bayesian graphical
models, Multi-stage estimation, Model/data integration, Integrated
population model},
  hidelinks,
  pdfcreator={LaTeX via pandoc}}
\usepackage[margin=2.25cm]{geometry}
\usepackage{graphicx}
\makeatletter
\def\maxwidth{\ifdim\Gin@nat@width>\linewidth\linewidth\else\Gin@nat@width\fi}
\def\maxheight{\ifdim\Gin@nat@height>\textheight\textheight\else\Gin@nat@height\fi}
\makeatother
\setkeys{Gin}{width=\maxwidth,height=\maxheight,keepaspectratio}
\makeatletter
\def\fps@figure{htbp}
\makeatother
\providecommand{\tightlist}{%
  \setlength{\itemsep}{0pt}\setlength{\parskip}{0pt}}
\usepackage{amsmath,mathtools}
\usepackage{tikz}
\usetikzlibrary{positioning, shapes, intersections, through, backgrounds, fit, decorations.pathmorphing, angles, quotes}
\usepackage{lineno}

\makeatletter
\@ifclassloaded{imsart}{}{
\usepackage{setspace}
\onehalfspacing
}
\makeatother

\usepackage{bbm}

\usepackage{enumitem}
\usepackage{relsize}
\usepackage{placeins}

\usepackage{pdflscape}

\usepackage{longtable}
\usepackage{booktabs}
\usepackage{caption}

\usepackage{color}
\definecolor{myredhighlight}{RGB}{180, 15, 32}
\definecolor{mydarkblue}{RGB}{0, 33, 79}
\definecolor{mymidblue}{RGB}{44, 127, 184}
\definecolor{mylightblue}{RGB}{166, 233, 255}
\definecolor{mygrey}{RGB}{68, 68, 68}

\usepackage{colortbl}

\let\Oldcap\cap
\renewcommand{\cap}{\mathrel{\mathsmaller{\Oldcap}}}

\newcommand{\pd}{\text{p}}
\newcommand{\q}{\text{q}}

\newcommand{\Nm}{M}

\newcommand{\modelindex}{m}

\newcommand{\indep}{\perp\!\!\!\perp}

\DeclareMathOperator*{\argmin}{arg\,min}

\newcommand{\paoii}{PaO\textsubscript{2}}
\newcommand{\fioii}{FiO\textsubscript{2}}

\newcommand{\pfratio}{\paoii/\fioii}

\ifLuaTeX
  \usepackage{selnolig}  
\fi
\newlength{\cslhangindent}
\newlength{\csllabelwidth}
\newenvironment{CSLReferences}[2] 
 {
  \setlength{\parindent}{0pt}
  \ifodd #1 \everypar{\setlength{\hangindent}{\cslhangindent}}\ignorespaces\fi
  \ifnum #2 > 0
  \setlength{\parskip}{#2\baselineskip}
  \fi
 }%
 {}
\usepackage{calc}

\title{Combining chains of Bayesian models with Markov melding}
\author{Andrew A. Manderson\footnote{MRC Biostatistics Unit, University
  of Cambridge, United Kingdom, and The Alan Turing Institute, The
  British Library, London.
  \href{mailto:andrew.manderson@mrc-bsu.cam.ac.uk}{\nolinkurl{andrew.manderson@mrc-bsu.cam.ac.uk}}} \and Robert
J. B. Goudie\footnote{MRC Biostatistics Unit, University of Cambridge,
  United Kingdom.
  \href{mailto:robert.goudie@mrc-bsu.cam.ac.uk}{\nolinkurl{robert.goudie@mrc-bsu.cam.ac.uk}}}}
\date{May, 2022}

\begin{document}
\maketitle
\begin{abstract}
A challenge for practitioners of Bayesian inference is specifying a
model that incorporates multiple relevant, heterogeneous data sets. It
may be easier to instead specify distinct submodels for each source of
data, then join the submodels together. We consider chains of submodels,
where submodels directly relate to their neighbours via common
quantities which may be parameters or deterministic functions thereof.
We propose \emph{chained Markov melding}, an extension of Markov
melding, a generic method to combine chains of submodels into a joint
model. One challenge we address is appropriately capturing the prior
dependence between common quantities within a submodel, whilst also
reconciling differences in priors for the same common quantity between
two adjacent submodels. Estimating the posterior of the resulting
overall joint model is also challenging, so we describe a sampler that
uses the chain structure to incorporate information contained in the
submodels in multiple stages, possibly in parallel. We demonstrate our
methodology using two examples. The first example considers an
ecological integrated population model, where multiple data sets are
required to accurately estimate population immigration and reproduction
rates. We also consider a joint longitudinal and time-to-event model
with uncertain, submodel-derived event times. Chained Markov melding is
a conceptually appealing approach to integrating submodels in these
settings.
\end{abstract}

\hypertarget{introduction}{%
\section{Introduction}\label{introduction}}

The Bayesian philosophy is appealing in part because the posterior
distribution quantifies all sources of uncertainty. However, a joint
model for all data and parameters is a prerequisite to posterior
inference, and in situations where multiple, heterogeneous sources of
data are available, specifying such a joint model is a formidable task.
Models that consider such data are necessary to describe complex
phenomena at a useful precision. One possible approach begins by
specifying individual submodels for each source of data. These submodels
could guide the statistician when directly specifying the joint model,
but to use the submodels only informally seems wasteful. Instead, it may
be preferable to construct a joint model by formally joining the
individual submodels together.

Some specific forms of combining data are well established.
Meta-analyses and evidence synthesis methods are widely used to
summarise data, often using hierarchical models (Ades and Sutton, 2006;
Presanis \emph{et al.}, 2014). Outside of the statistical literature, a
common name for combining multiple data is \emph{data fusion} (Kedem
\emph{et al.}, 2017; Lahat \emph{et al.}, 2015), though there are many
distinct methods that fall under this general name. Interest in
integrating data is not just methodological; applied researchers often
collect multiple disparate data sets, or data of different modalities,
and wish to combine them. For example, to estimate SARS-CoV-2 positivity
Donnat \emph{et al.} (2020) build an intricate hierarchical model that
integrates both testing data and self-reported questionnaire data, and
Parsons \emph{et al.} (2021) specify a hierarchical model of similar
complexity to estimate the number of injecting drug users in Ukraine.
Both applications specify Bayesian models with data-specific components,
which are united in a hierarchical manner. In conservation ecology,
\emph{integrated population models} (IPMs) (Besbeas \emph{et al.}, 2002;
Brooks \emph{et al.}, 2004; Maunder and Punt, 2013; Schaub and Abadi,
2011; Zipkin and Saunders, 2018) are used to estimate population
dynamics, e.g.~reproduction and immigration rates, using multiple data
on the same population. Such data have standard models associated with
them, such as the Cormack-Jolly-Seber model (Lebreton \emph{et al.},
1992) for capture-recapture data, and the IPM serves as the framework in
which the standard models are combined. More generally, the applications
we list illustrate the importance of generic, flexible methods for
combining data to applied researchers.

\emph{Markov melding} (Goudie \emph{et al.}, 2019) is a general
statistical methodology for combining submodels. Specifically, it
considers \(\Nm\) submodels that share some common quantity \(\phi\),
with each of the \(\modelindex = 1, \ldots, \Nm\) submodels possessing
distinct parameters \(\psi_{\modelindex}\), data \(Y_{\modelindex}\),
and form
\(\pd_{\modelindex}(\phi, \psi_{\modelindex}, Y_{\modelindex})\). Goudie
\emph{et al.} (2019) then propose to combine the submodels into a joint
model, denoted
\(\pd_{\text{meld}}(\phi, \psi_{1}, \ldots, \psi_{\Nm}, Y_{1}, \ldots, Y_{\Nm})\).
However, it is unclear how to integrate models where there is no single
quantity \(\phi\) common to all submodels, such as for submodels that
are linked in a chain structure.

We propose an extension to Markov melding, which we call \emph{chained
Markov melding}\footnote{\emph{``Chained graphs''} were considered by
  Lauritzen and Richardson (2002), however they are unrelated to our
  proposed model. We use ``chained'' to emphasise the nature of the
  relationships between submodels.}, which facilitates the combination
of \(\Nm\) submodels that are in a chain structure. For example, when
\(\Nm = 3\) we address the case in which submodel 1 and 2 share a common
quantity \(\phi_{1 \cap 2}\), and submodel 2 and 3 share a different
quantity \(\phi_{2 \cap 3}\). Our extension addresses previously
unconsidered complications including the distinct domains (and possibly
supports) of the common quantities, and the desire to capture possible
prior correlation between them. Two examples serve to illustrate our
methodology, which we introduce in the following section. The
computational effort required to fit a complex, multi-response model is
a burden to the model development process. We propose a multi-stage
posterior estimation method that exploits the properties of our chained
melded model to reduce this burden. We can parallelise aspects of the
computation across the submodels, using less computationally expensive
techniques for some submodels. Reusing existing software implementations
of submodels, and subposterior samples where available, is also
possible. Multi-stage samplers can aid in understanding the contribution
of each submodel to the final posterior, and are used in many applied
settings, including hierarchical modelling (Lunn \emph{et al.}, 2013)
and joint models (Mauff \emph{et al.}, 2020).

One contribution of our work is to clarify the informal process commonly
used in applied analyses of summarising and/or approximating submodels
for use in subsequent analyses. The two most common approximation
strategies seem to be (i) approximating the subposterior of the common
quantity with a normal distribution for use in subsequent models (see,
e.g. Jackson and White, 2018; Nicholson \emph{et al.}, 2021) and (ii)
taking only a point estimate of the subposterior, and treating it as a
known value in further models. These strategies may, but not always,
produce acceptable approximations to the chained melded model. Both the
chained melded model and these approximation strategies are examples of
`multi-phase' and `multi-source' inference (Meng, 2014), with the
melding approach most comprehensively accounting for uncertainty.

\hypertarget{example-introduction}{%
\subsection{Example introduction}\label{example-introduction}}

In this section we provide a high-level overview of two applications
that require integrating a chain of submodels, with more details in
Sections \ref{an-integrated-population-model-for-little-owls-1} and
\ref{survival-analysis-with-time-varying-covariates-and-uncertain-event-times-1}.
Our first example decomposes a joint model into its constituent
submodels and rejoins them. This simple situation allows us to compare
the output from the chained melding process to the complete joint model,
and is meant to illustrate both the `chain-of-submodels' notion and the
mechanics of chained melding. The second example is a realistic and
complex setting in which the combining of submodels without chained
Markov melding is nonobvious. Our comparator is the common technique of
summarising previously considered submodels with point estimates, and
demonstrates the importance of fully accounting for uncertainty.

\hypertarget{an-integrated-population-model-for-little-owls}{%
\subsubsection{An integrated population model for little
owls}\label{an-integrated-population-model-for-little-owls}}

\begin{figure}[tb]
  \centering
  \begin{tikzpicture}[> = stealth, node distance = 1cm, thick, state/.style={draw = none, minimum width = 1cm}]
  \node [state] (y-2) {$Y_{2}$};
  \node [state] [above = of y-2] (psi-2) {$\psi_{2}$};
  \node [state] [right = of psi-2] (phi-23) {$\phi_{2 \cap 3}$};
  \node [state] [left = of psi-2] (phi-12) {$\phi_{1 \cap 2}$};

  \node [state] [right = of phi-23] (phi-23-p3) {$\phi_{2 \cap 3}$};
  \node [state] [left = of phi-12] (phi-12-p1) {$\phi_{1 \cap 2}$};

  \node [state] [right = of phi-23-p3] (psi-3) {$\psi_{3}$};
  \node [state] [below = of psi-3] (y-3) {$Y_{3}$};

  \node [state] [left = of phi-12-p1] (psi-1) {$\psi_{1}$};
  \node [state] [below = of psi-1] (y-1) {$Y_{1}$};

  \path [->] (psi-1) edge (y-1);
  \path [->] (psi-2) edge (y-2);
  \path [->] (psi-3) edge (y-3);
  \path [->] (phi-12-p1) edge (y-1);
  \path [->] (phi-12) edge (y-2);
  \path [->] (phi-23) edge (y-2);
  \path [->] (phi-23-p3) edge (y-3);

  \node (model-1) [draw = mymidblue, fit = (psi-1) (y-1) (phi-12-p1), inner sep = 0.05cm, solid] {};
  \node (model-1-label) [yshift = 1.5ex, mymidblue] at (model-1.north west) {$\pd_{1}$};
  \node (model-2) [draw = black, fit = (phi-12) (psi-2) (phi-23) (y-2), inner sep = 0.10cm, solid] {};
  \node (model-2-label) [yshift = 1.5ex] at (model-2.north) {$\pd_{2}$};
  \node (model-3) [draw = myredhighlight, fit = (phi-23-p3) (psi-3) (y-3), inner sep = 0.05cm, solid] {};
  \node (model-3-label) [yshift = 1.5ex, myredhighlight] at (model-3.north east) {$\pd_{3}$};
  \end{tikzpicture}
  \caption{A simplified DAG of the integrated population model (IPM) for the little owls. The capture-recapture submodel ($\pd_{1}$) is surrounded by the blue line, the population count submodel ($\pd_{2}$) by the black line, and the fecundity submodel ($\pd_{3}$) by the red line. The capture-recapture and population count submodels share parameters affecting the juvenile and adult survival rate ($\phi_{1 \cap 2}$), whilst the parameter for fecundity is common to both the population count and fecundity submodels ($\phi_{2 \cap 3}$). The combination of all the submodels forms the IPM.}
  \label{fig:owls-simple-dag}
\end{figure}
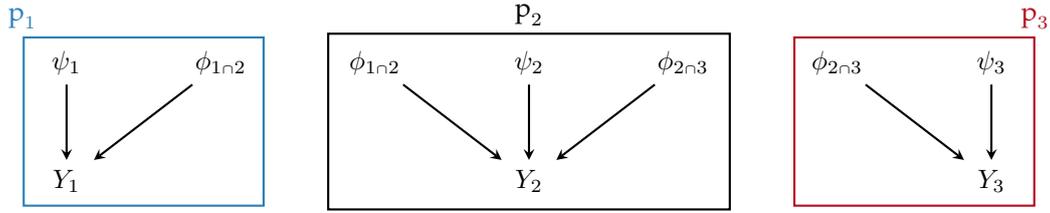

Integrated population models (IPMs) (Zipkin and Saunders, 2018) combine
multiple data to estimate key quantities governing the dynamics of a
specific population. Schaub \emph{et al.} (2006) and Abadi \emph{et al.}
(2010) used an IPM to estimate fecundity, immigration, and yearly
survival rates for a population of little owls. These authors collect
and model three types of data, illustrated in Figure
\ref{fig:owls-simple-dag}. Capture-recapture data \(Y_{1}\), and
associated capture-recapture submodel
\(\pd_{1}(\phi_{1 \cap 2}, \psi_{1}, Y_{1})\), are acquired by capturing
and tagging owls each year, and then counting the number of tagged
individuals recaptured in subsequent years. Population counts \(Y_{2}\)
are obtained by observing the number of occupied nesting sites, and are
modelled in
\(\pd_{2}(\phi_{1 \cap 2}, \phi_{2 \cap 3}, \psi_{2}, Y_{2})\). Finally,
nest-record data \(Y_{3}\) counts both the number of reproductive
successes and possible breading pairs, and is associated with a submodel
for fecundity \(\pd_{3}(\phi_{2 \cap 3}, \psi_{3}, Y_{3})\). The
population count model \(\pd_{2}\) shares the parameter
\(\phi_{1 \cap 2}\) with the capture-recapture model \(\pd_{1}\), and
the parameter \(\phi_{2 \cap 3}\) with the fecundity model \(\pd_{3}\);
each of the \(\modelindex = 1, 2, 3\) submodels has distinct,
submodel-specific parameters \(\psi_{\modelindex}\). No single source of
data is sufficient to estimate all quantities of interest, so it is
necessary to integrate the three submodels into a single joint model to
produce acceptably precise estimates of fecundity and immigration rates.
We will show that the chained Markov melding framework developed in
Section \ref{chained-model-specification} encapsulates the process of
integrating these submodels, producing results that are concordant with
the original joint IPM.

\hypertarget{survival-analysis-with-time-varying-covariates-and-uncertain-event-times}{%
\subsubsection{Survival analysis with time varying covariates and
uncertain event
times}\label{survival-analysis-with-time-varying-covariates-and-uncertain-event-times}}

Our second example considers the time to onset of respiratory failure
(RF) amongst patients in intensive care units, and factors that
influence the onset of RF. A patient can be said to be experiencing RF
if the ratio of the partial pressure of arterial blood oxygen (\paoii)
to the faction of inspired oxygen (\fioii) is less than 300mmHg (The
ARDS Definition Task Force, 2012), though this is not the only
definition of RF. Patients' \pfratio~(P/F) ratios are typically measured
only a few times a day. The relative infrequency of P/F ratio data, when
combined with the intrinsic variability in each individual's blood
oxygen level, results in significant uncertainty in about the time of
onset of RF.

Factors that influence the time to onset of RF are both longitudinal and
time invariant. Both types of data can be considered in \emph{joint
models} (Rizopoulos, 2012), which are composed of two distinct
submodels, one for each data type. However, existing joint models are
not able to incorporate the uncertainty surrounding the event time,
which may result in overconfident and/or biased estimates of the
parameters in the joint model.

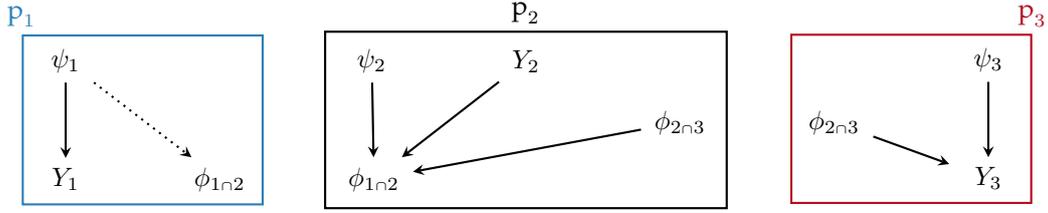
\begin{figure}[tb]
  \centering
  \begin{tikzpicture}[> = stealth, node distance = 1cm, thick, state/.style={draw = none, minimum width = 1cm}]
  \node [state] (y-1) {$Y_{1}$};
  \node [state] [above = of y-1] (psi-1) {$\psi_{1}$};
  \node [state] [right = of y-1] (phi-12) {$\phi_{1 \cap 2}$};

  \node [state] [right = of phi-12] (phi-12-p2) {$\phi_{1 \cap 2}$};
  
  \node [state] [right = 3cm of psi-1] (psi-2) {$\psi_{2}$};
  \node [state] [right = of psi-2] (y-2) {$Y_{2}$};
  \node [state] [below right = 0.25cm and 1cm of y-2] (phi-23-p2) {$\phi_{2 \cap 3}$};
  \node [state] [right = of phi-23-p2] (phi-23) {$\phi_{2 \cap 3}$};
  \node [state] [above right = 0.25cm and 1cm of phi-23] (psi-3) {$\psi_{3}$};
  \node [state] [below = of psi-3] (y-3) {$Y_{3}$};

  \path [->] (psi-1) edge (y-1);
  \path [->, dotted] (psi-1) edge (phi-12);
  \path [->] (psi-2) edge (phi-12-p2);
  \path [->] (psi-3) edge (y-3);
  \path [->] (phi-23-p2) edge (phi-12-p2);
  \path [->] (phi-23) edge (y-3);
  \path [->] (y-2) edge (phi-12-p2);

  \node (model-1) [draw = mymidblue, fit = (psi-1) (y-1) (phi-12), inner sep = 0.05cm, solid] {};
  \node (model-1-label) [yshift = 1.5ex, mymidblue] at (model-1.north west) {$\pd_{1}$};
  \node (model-2) [draw = black, fit = (phi-12-p2) (psi-2) (phi-23-p2) (y-2), inner sep = 0.1cm, solid] {};
  \node (model-2-label) [yshift = 1.5ex] at (model-2.north) {$\pd_{2}$};
  \node (model-3) [draw = myredhighlight, fit = (phi-23) (psi-3) (y-3), inner sep = 0.05cm, solid] {};
  \node (model-3-label) [yshift = 1.5ex, myredhighlight] at (model-3.north east) {$\pd_{3}$};
  \end{tikzpicture}
  \caption{A simplified DAG of the submodels considered in the survival analysis example. The event time submodel $\pd_{1}$ defines the event time $\phi_{1 \cap 2}$ as noninvertible function of the other model parameters (denoted by the dotted line), whilst the survival submodel $\pd_{2}$ considers $\phi_{1 \cap 2}$ as the response. The longitudinal submodel $\pd_{3}$ has parameters $\phi_{2 \cap 3}$ in common with the survival submodel.}
  \label{fig:surv-simple-dag}
\end{figure}

Chained Markov melding offers a conceptually straightforward, Bayesian
approach to incorporating uncertain event times into joint models.
Specifically, we consider the event time as a submodel-derived quantity
from a hierarchical regression model akin to Lu and Meeker (1993). We
call this submodel the \emph{uncertain event time} submodel and denote
it \(\pd_{1}(\phi_{1 \cap 2}, \psi_{1}, Y_{1})\), where
\(\phi_{1 \cap 2}\) incorporates the event time. The survival submodel
\(\pd_{2}(\phi_{1 \cap 2}, \phi_{2 \cap 3}, \psi_{2}, Y_{2})\) uses the
event time within \(\phi_{1 \cap 2}\), the common quantity, as the
response. We treat the longitudinal submodel,
\(\pd_{3}(\phi_{2 \cap 3}, \psi_{3}, Y_{3})\), separately from the
survival submodel, as is common in two-stage joint modelling (Mauff
\emph{et al.}, 2020), and denote the subject-specific parameters that
also appear in the survival model as \(\phi_{2 \cap 3}\). Each of the
\(\modelindex = 1, 2, 3\) has submodel-specific data \(Y_{\modelindex}\)
and parameters \(\psi_{\modelindex}\). The high level submodel
relationships are displayed as a DAG in Figure
\ref{fig:surv-simple-dag}.

It is in examples such as this one that we foresee the most use for
chained Markov melding; a fully Bayesian approach is desired and the
submodels are nontrivial in complexity, with no previously existing or
obvious joint model.

\hypertarget{markov-melding}{%
\subsection{Markov melding}\label{markov-melding}}

We now review Markov melding (Goudie \emph{et al.}, 2019) before
detailing our proposed extension. As noted in the introduction, Markov
melding is a method for combining \(\Nm\) submodels
\(\pd_{1}(\phi, \psi_{1}, Y_{1}), \ldots, \pd_{\Nm}(\phi, \psi_{\Nm}, Y_{\Nm})\)
which share the same \(\phi\). When the submodel prior marginals
\(\pd_{\modelindex}(\phi)\) are identical,
i.e.~\(\pd_{\modelindex}(\phi) = \pd(\phi)\) for all \(\modelindex\), it
is possible to combine the submodels using \emph{Markov combination}
(Dawid and Lauritzen, 1993; Massa and Lauritzen, 2010)
\begin{equation}
  \begin{aligned}
    \pd_{\text{comb}}\left(\phi, \psi_{1}, \ldots, \psi_{\Nm}, Y_{1}, \ldots, Y_{\Nm}\right)
    &= \pd(\phi) \prod_{\modelindex = 1}^{\Nm} \pd_{\modelindex}\left(\psi_{\modelindex}, Y_{\modelindex} \mid \phi\right), \\
    &= \frac{\prod_{\modelindex = 1}^{\Nm} \pd_{\modelindex}\left(\phi, \psi_{\modelindex}, Y_{\modelindex}\right)}{\pd(\phi)^{\Nm - 1}}.
  \end{aligned}
  \label{eqn:markov-combination}
\end{equation}\noindent
Markov combination is not immediately applicable when submodel prior
marginals are distinct, so Goudie \emph{et al.}~define a \emph{marginal
replacement} procedure, where individual submodel prior marginals are
replaced with a common marginal
\(\pd_{\text{pool}}(\phi) = h(\pd_{1}(\phi), \ldots, \pd_{\Nm}(\phi))\)
which is the result of a pooling function \(h\) that appropriately
summarises all prior marginals (the choice of which is described below).
The result of marginal replacement is
\begin{equation}
  \pd_{\text{repl}, \modelindex}(\phi, \psi_{\modelindex}, Y_{\modelindex}) =
  \pd_{\text{pool}}(\phi)
  \frac{
    \pd_{\modelindex}(\phi, \psi_{\modelindex}, Y_{\modelindex})
  } {
    \pd_{\modelindex}(\phi)
  }.
  \label{eqn:marginal-replacement}
\end{equation}\noindent
Goudie \emph{et al.}~show that
\(\pd_{\text{repl}, \modelindex}(\phi, \psi_{\modelindex}, Y_{\modelindex})\)
minimises the Kullback--Leibler (KL) divergence between a distribution
\(\q(\phi, \psi_{\modelindex}, Y_{\modelindex})\) and
\(\pd_{\modelindex}(\phi, \psi_{\modelindex}, Y_{\modelindex})\) under
the constraint that \(\q(\phi) = \pd_{\text{pool}}(\phi)\), and that
marginal replacement is valid when \(\phi\) is a deterministic function
of the other parameters in submodel \(\modelindex\). Markov melding
joins the submodels via the Markov combination of the marginally
replaced submodels
\begin{equation}
\begin{aligned}
  \pd_{\text{meld}}(\phi, \psi_{1}, \ldots, \psi_{\Nm}, Y_{1}, \ldots, Y_{\Nm}) &= 
  \pd_{\text{pool}}(\phi) \prod_{\modelindex = 1}^{\Nm} \pd_{\text{repl}, \modelindex}(\psi_{\modelindex}, Y_{\modelindex} \mid \phi), \\ &= 
  \pd_{\text{pool}}(\phi) \prod_{\modelindex = 1}^{\Nm} \frac{
    \pd_{\modelindex}(\phi, \psi_{\modelindex}, Y_{\modelindex})
  } {
    \pd_{\modelindex}(\phi)
  }.
\end{aligned}
\end{equation}

\hypertarget{pooled-prior}{%
\subsubsection{Pooled prior}\label{pooled-prior}}

Goudie \emph{et al.}~proposed forming \(\pd_{\text{pool}}(\phi)\) using
linear or logarithmic prior pooling (Genest \emph{et al.}, 1986; O'Hagan
\emph{et al.}, 2006)
\begin{align}
  \pd_{\text{pool, lin}}(\phi) &=
    \frac{1}{K_{\text{lin}}(\lambda)}
    \sum_{\modelindex = 1}^{\Nm} \lambda_{\modelindex} \pd_{\modelindex}(\phi), \quad
  K_{\text{lin}}(\lambda) = \int \sum_{\modelindex = 1}^{\Nm} \lambda_{\modelindex} \pd_{\modelindex}(\phi) \text{d}\phi \label{eqn:orig-pooling-linear}\\
  \pd_{\text{pool, log}}(\phi) &=  
    \frac{1}{K_{\text{log}}(\lambda)}
    \prod_{\modelindex = 1}^{\Nm} \pd_{\modelindex}(\phi)^{\lambda_{\modelindex}}, \quad
  K_{\text{log}}(\lambda) = \int \prod_{\modelindex = 1}^{\Nm} \pd_{\modelindex}(\phi)^{\lambda_{\modelindex}} \text{d}\phi
  \label{eqn:orig-pooling-logarithmic}
\end{align}\noindent
where \(\lambda = (\lambda_{1}, \ldots, \lambda_{\Nm})\) are nonnegative
weights, which are chosen subjectively to ensure
\(\pd_{\text{pool}}(\phi)\) appropriately represents prior knowledge
about the common quantity. Two special cases of pooling are of
particular interest. \emph{Product of experts (PoE) pooling} (Hinton,
2002) is a special case of logarithmic pooling that occurs when
\(\lambda_{\modelindex} = 1\) for all \(\modelindex\). \emph{Dictatorial
pooling} is a special case of either pooling method when
\(\lambda_{\modelindex'} = 1\) and, for all
\(\modelindex \neq \modelindex'\), \(\lambda_{\modelindex} = 0\).

\hypertarget{chained-model-specification}{%
\section{Chained model
specification}\label{chained-model-specification}}

Consider \(\modelindex = 1, \ldots, \Nm\) submodels each with data
\(Y_{\modelindex}\) and parameters \(\theta_{\modelindex}\) denoted
\(\pd_{\modelindex}(\theta_{\modelindex}, Y_{\modelindex})\), with
\(\Nm \geq 3\). We assume that the submodels are connected in a manner
akin to a chain and so can be ordered such that only `adjacent'
submodels in the chain have parameters in common. Specifically we assume
that submodels \(\modelindex\) and \(\modelindex + 1\) have some
parameter \(\phi_{\modelindex \cap \modelindex + 1}\) in common for
\(\modelindex = 1, \ldots, \Nm - 1\). For notational convenience define
\(\phi_{1} = \phi_{1 \cap 2}, \phi_{\Nm} = \phi_{\Nm-1 \cap \Nm}\) and
\(\phi_{\modelindex} = (\phi_{\modelindex - 1 \cap \modelindex}, \, \phi_{\modelindex \cap \modelindex + 1})\)
for \(\modelindex = 2, \ldots, \Nm - 1\), so that
\(\phi_{\modelindex} \subseteq \theta_{\modelindex}\) denotes the
parameters in model \(\modelindex\) shared with another submodel. The
submodel-specific parameters of submodel \(\modelindex\) are thus
\(\psi_\modelindex = \theta_\modelindex \setminus \phi_\modelindex\).
Define the vector of all common quantities
\(\boldsymbol{\phi} = (\phi_{1}, \ldots, \phi_{\Nm}) = (\phi_{1 \cap 2}, \ldots, \phi_{\Nm - 1 \cap \Nm})\),
and denote by \(\boldsymbol{\phi}_{-\modelindex}\) the subvector of
\(\boldsymbol{\phi}\) excluding the \(\modelindex\)\textsuperscript{th}
element. It will also be convenient to define
\(\boldsymbol{\psi} = (\psi_{1}, \ldots, \psi_{\Nm})\) and likewise
\(\boldsymbol{Y} = (Y_{1}, \ldots, Y_{\Nm})\). Note that all components
of \(\boldsymbol{\phi}, \boldsymbol{\psi}\) and \(\boldsymbol{Y}\) may
themselves be multivariate. Additionally, because
\(\phi_{\modelindex \cap \modelindex + 1}\) may be a deterministic
function of either \(\theta_{\modelindex}\) or
\(\theta_{\modelindex + 1}\) we refer to
\(\phi_{\modelindex \cap \modelindex + 1}\) as a common parameter or a
common quantity as appropriate.

All submodels, and marginal and conditional distributions thereof, have
density functions that are assumed to exist and integrate to one. When
considering conditional distributions we assume that the parameter being
conditioned on has support in the relevant region. We define the
\(\modelindex\)\textsuperscript{th} \emph{subposterior} as
\(\pd_{\modelindex}(\phi_{\modelindex}, \psi_{\modelindex} \mid Y_{\modelindex})\).

\hypertarget{extending-marginal-replacement}{%
\subsection{Extending marginal
replacement}\label{extending-marginal-replacement}}

We now define the chained melded model by extending the marginal
replacement procedure to submodels linked in a chain-like way. The
proposed chained marginal replacement operation modifies the submodels
to enforce a common prior for \(\boldsymbol{\phi}\). This consistency
allows us to employ Markov combination to unite the submodels.

Specifically, the \(\modelindex\)\textsuperscript{th} marginally
replaced submodel is
\begin{equation}
  \pd_{\text{repl}, \modelindex}(\boldsymbol{\phi}, \psi_{\modelindex}, Y_{\modelindex}) 
  = 
    \pd_{\text{pool}}(\boldsymbol{\phi})
    \pd_{\modelindex}(\psi_{\modelindex}, Y_{\modelindex} \mid \boldsymbol{\phi})
  = 
    \pd_{\text{pool}}(\boldsymbol{\phi})
    \frac{
      \pd_{\modelindex}(\phi_{\modelindex}, \psi_{\modelindex}, Y_{\modelindex})
    } {
      \pd_{\modelindex}(\phi_{\modelindex})
    },
    \label{eqn:chained-marginal-replacement}
\end{equation}\noindent
where
\(\pd_{\text{pool}}(\boldsymbol{\phi}) = g(\pd_{1}(\phi_{1}), \pd_{2}(\phi_{2}), \ldots, \pd_{\modelindex}(\phi_{\modelindex}))\)
is a pooling function that appropriately summarises all submodel prior
marginals. The second equality in Equation
\eqref{eqn:chained-marginal-replacement} is because of the conditional
independence
\((\psi_{\modelindex}, Y_{\modelindex} \indep \boldsymbol{\phi}_{-\modelindex}) \mid \phi_{\modelindex}\)
that exists due to the chained relationship between submodels. It is
important to note that
\(\pd_{\text{repl}, \modelindex}(\boldsymbol{\phi}, \psi_{\modelindex}, Y_{\modelindex})\)
is defined on a larger parameter space than
\(\pd_{\modelindex}(\phi_{\modelindex}, \psi_{\modelindex}, Y_{\modelindex})\),
as it includes \(\boldsymbol{\phi}_{-\modelindex}\).

Define
\(\pd_{\text{repl}, \modelindex}(\phi_{\modelindex}, \psi_{\modelindex}, Y_{\modelindex}) = \int \pd_{\text{repl}, \modelindex}(\boldsymbol{\phi}, \psi_{\modelindex}, Y_{\modelindex})\text{d}\boldsymbol{\phi}_{-\modelindex}\).
Each marginally replaced submodel, as defined in Equation
\eqref{eqn:chained-marginal-replacement}, minimises the following KL
divergence\footnote{This is shown in Appendix B of the online supplement
  to Goudie \emph{et al.} (2019).}
\begin{equation}
  \pd_{\text{repl}, \modelindex}(\phi_{\modelindex}, \psi_{\modelindex}, Y_{\modelindex}) =
  \argmin_{\q} \left\{
    D_{\mathrm{KL}}\left(\q \, \| \, \pd_{\modelindex}\right) \mid \q(\phi_{\modelindex}) = \pd_{\text{pool}}(\phi_{\modelindex}) \text { for all } \phi_{\modelindex}
  \right\},
  \label{eqn:chained-marginal-replacement-kl}
\end{equation}\noindent
where
\(\pd_{\text{pool}}(\phi_{\modelindex}) = \int \pd_{\text{pool}}(\boldsymbol{\phi})\text{d}\boldsymbol{\phi}_{-\modelindex}\).
We can thus interpret
\(\pd_{\text{repl}, \modelindex}(\phi_{\modelindex}, \psi_{\modelindex}, Y_{\modelindex})\)
as a minimally modified
\(\pd_{\modelindex}(\phi_{\modelindex}, \psi_{\modelindex}, Y_{\modelindex})\)
which admits \(\pd_{\text{pool}}(\phi_{\modelindex})\) as a marginal.
Note that it is the combination of
\(\pd_{\text{repl}, \modelindex}(\phi_{\modelindex}, \psi_{\modelindex}, Y_{\modelindex})\)
and
\(\pd_{\text{pool}}(\boldsymbol{\phi}_{-\modelindex} \mid \phi_{\modelindex})\)
that uniquely determine \eqref{eqn:chained-marginal-replacement}.

We form the chained melded model by taking the Markov combination of the
marginally replaced submodels
\begin{align}
  \pd_{\text{meld}}(\boldsymbol{\phi}, \boldsymbol{\psi}, \boldsymbol{Y}) &=
    \pd_{\text{pool}}(\boldsymbol{\phi})
    \prod_{\modelindex = 1}^{\Nm}
      \pd_{\text{repl}, \modelindex}(\psi_{\modelindex}, Y_{\modelindex} \mid \boldsymbol{\phi}) \\
    &=
    \pd_{\text{pool}}(\boldsymbol{\phi})
    \prod_{\modelindex =1}^{\Nm}
    \frac {
      \pd_{\modelindex}(\phi_{\modelindex}, \psi_{\modelindex}, Y_{\modelindex})
    } {
      \pd_{\modelindex}(\phi_{\modelindex})
    }.
  \label{eqn:melded-model-full}
\end{align} Rewriting
\eqref{eqn:melded-model-full} in terms of
\(\phi_{\modelindex \cap \modelindex + 1}\) for
\(\modelindex = 1, \ldots, \Nm - 1\) yields
\begin{equation}
  \begin{aligned}
  \pd_{\text{meld}} (\boldsymbol{\phi}, \boldsymbol{\psi}, \boldsymbol{Y}) =
    \pd_{\text{pool}}&(\boldsymbol{\phi})
    \frac {
      \pd_{1} (\phi_{1 \cap 2}, \psi_{1}, Y_{1})
    } {
      \pd_{1}(\phi_{1 \cap 2})
    }
    \frac {
      \pd_{\Nm} (\phi_{\Nm - 1 \cap \Nm}, \psi_{\Nm}, Y_{\Nm})
    } {
      \pd_{\Nm}(\phi_{\Nm - 1 \cap \Nm})
    } \\
    &\times \prod_{\modelindex = 2}^{\Nm - 1} \left(
      \frac {
        \pd_{\modelindex} (\phi_{\modelindex - 1 \cap \modelindex}, \phi_{\modelindex \cap \modelindex + 1}, \psi_{\modelindex}, Y_{\modelindex})
      } {
        \pd_{\modelindex}(\phi_{\modelindex - 1 \cap \modelindex}, \phi_{\modelindex \cap \modelindex + 1})
      }
    \right).
  \end{aligned}
  \label{eqn:melded-model-general}
\end{equation} Finally, we
use \emph{chained melded posterior}
\(\pd_{\text{meld}}(\boldsymbol{\phi}, \boldsymbol{\psi} \mid \boldsymbol{Y}) \propto \pd_{\text{meld}}(\boldsymbol{\phi}, \boldsymbol{\psi}, \boldsymbol{Y})\)
to refer to posterior of the chained melded model conditioned on all
data.

\hypertarget{pooled-prior-1}{%
\subsection{Pooled prior}\label{pooled-prior-1}}

Specifying \eqref{eqn:melded-model-full} requires a joint prior for
\(\boldsymbol{\phi}\). As in Markov melding we form the joint prior by
pooling the marginal priors, selecting a pooling function \(g\) that
appropriately represents prior knowledge about the common quantities. We
define \(\pd_{\text{pool}}(\boldsymbol{\phi})\) as a generic function of
all prior marginals
\begin{align}
  \pd_{\text{pool}}(\boldsymbol{\phi})
  &= g(\pd_{1}(\phi_{1}), \pd_{2}(\phi_{2}), \ldots, \pd_{\Nm}(\phi_{\Nm})) \\
  &= g(\pd_{1}(\phi_{1 \cap 2}), \pd_{2}(\phi_{1 \cap 2}, \phi_{2 \cap 3}), \ldots, \pd_{\Nm}(\phi_{\Nm - 1 \cap \Nm})),
  \label{eqn:pooled-prior-general-def-two}
\end{align}
because we do not always wish to assume independence between the
components of \(\boldsymbol{\phi}\).

Two special cases of Equation \eqref{eqn:pooled-prior-general-def-two}
are noteworthy. Firstly, if all components of \(\boldsymbol{\phi}\) are
independent, then we can form \(\pd_{\text{pool}}(\boldsymbol{\phi})\)
as the product of \(\Nm - 1\) standard pooling functions
\(h_{\modelindex}\) defined in Section \ref{pooled-prior}
\begin{gather}
  \pd_{\text{pool}}(\boldsymbol{\phi}) = 
    \prod_{\modelindex = 1}^{\Nm - 1}
    \pd_{\text{pool}, \modelindex}(\phi_{\modelindex \cap \modelindex + 1}),
  \\
  \pd_{\text{pool}, \modelindex}(\phi_{\modelindex \cap \modelindex + 1}) =
  h_{\modelindex}(\pd_{\modelindex}(\phi_{\modelindex \cap \modelindex + 1}), \pd_{\modelindex + 1}(\phi_{\modelindex \cap \modelindex + 1})).
  \label{eqn:bad-alternative-one}
\end{gather}\noindent
A second case, in between complete dependence
\eqref{eqn:pooled-prior-general-def-two} and independence
\eqref{eqn:bad-alternative-one}, is that if
\(\pd_{\modelindex}(\phi_{\modelindex - 1 \cap \modelindex}, \phi_{\modelindex \cap \modelindex + 1}) = \pd_{\modelindex}(\phi_{\modelindex - 1 \cap \modelindex}) \pd_{\modelindex}(\phi_{\modelindex \cap \modelindex + 1})\)
then we can define
\begin{equation}
  \pd_{\text{pool}}(\boldsymbol{\phi}) = 
  g_{1}(\pd_{1}(\phi_{1 \cap 2}), \ldots, \pd_{\modelindex}(\phi_{\modelindex - 1 \cap \modelindex}))
  g_{2}(\pd_{\modelindex}(\phi_{\modelindex \cap \modelindex + 1}), \ldots, \pd_{\Nm}(\phi_{\Nm}))
  \label{eqn:pooled-prior-split-def}
\end{equation}\noindent
without any additional assumptions. That is, if any two consecutive
components of \(\boldsymbol{\phi}\) are independent in the submodel
containing both of them, we can divide the pooled prior specification
problem into two pooling functions. The smaller number of arguments to
\(g_{1}\) and \(g_{2}\) make it easier to choose appropriate forms for
those functions.

Selecting a specific form of \(g\) is not trivial given the many choices
of functional form and pooling weights (the latter of which we discuss
momentarily). One complication is that standard linear and logarithmic
pooling, as defined in Equations \eqref{eqn:orig-pooling-linear} and
\eqref{eqn:orig-pooling-logarithmic}, are not immediately applicable
when the submodel marginal distributions consider different quantities.
We now propose extensions to logarithmic, linear, and dictatorial
pooling for use in the case of chained melding.

\hypertarget{chained-logarithmic-pooling}{%
\subsubsection{Chained logarithmic
pooling}\label{chained-logarithmic-pooling}}

Extending logarithmic pooling for chained Markov melding is
straightforward. We define the logarithmically pooled prior to be
\begin{equation}
\begin{gathered}
  \pd_{\text{pool, log}}(\boldsymbol{\phi}) =
  \frac{1}{K_{\text{log}}(\lambda)}
  \prod_{\modelindex = 1}^{\Nm}
    \pd_{\modelindex}(\phi_{\modelindex})^{\lambda_{\modelindex}}, 
  \label{eqn:pooled-prior-overall}
\end{gathered}
\end{equation}\noindent
with
\(K_{\text{log}}(\lambda) = \int \prod_{\modelindex = 1}^{\Nm} \pd_{\modelindex}(\phi_{\modelindex})^{\lambda_{\modelindex}} \text{d}\boldsymbol{\phi}\)
for nonnegative weight vector
\(\lambda = (\lambda_{1}, \ldots, \lambda_{\Nm})\) and
\(\sum_{\modelindex = 1}^{\Nm} \lambda_{\modelindex} \geq 1\). Note that
\eqref{eqn:pooled-prior-overall} does not imply independence between the
elements of \(\boldsymbol{\phi}\) because
\begin{equation}
  \prod_{\modelindex = 1}^{\Nm}
    \pd_{\modelindex}(\phi_{\modelindex})^{\lambda_{\modelindex}}
  = 
  \pd_{1}(\phi_{1 \cap 2})^{\lambda_{1}}
  \prod_{\modelindex = 2}^{\Nm - 1}
  \left(
    \pd_{\modelindex}(\phi_{\modelindex - 1 \cap \modelindex}, \phi_{\modelindex \cap \modelindex + 1})^{\lambda_{\modelindex}}
  \right)
  \pd_{\Nm}(\phi_{\Nm - 1 \cap \Nm})^{\lambda_{\Nm}}.
  \label{eqn:pooled-prior-log-alt-def}
\end{equation}\noindent
When \(\lambda_{1} = \lambda_{2} = \ldots = \lambda_{\Nm} = 1\) we
obtain a special case which we call product-of-experts (PoE) pooling
(Hinton, 2002).

\hypertarget{chained-linear-pooling}{%
\subsubsection{Chained linear pooling}\label{chained-linear-pooling}}

Our generalisation of linear pooling to handle marginals of different
quantities is a two step procedure. The first step forms intermediary
pooling densities via standard linear pooling, using appropriate
marginals of the relevant quantity
\begin{gather}
  \pd_{\text{pool}, \modelindex}(\phi_{\modelindex \cap \modelindex + 1}) \, \propto \,
  \lambda_{\modelindex, 1}\pd_{\modelindex}(\phi_{\modelindex \cap \modelindex + 1}) +
  \lambda_{\modelindex, 2}\pd_{\modelindex + 1}(\phi_{\modelindex \cap \modelindex + 1}),
  \label{eqn:M-model-linear-pooling-one}
\end{gather}\noindent
where
\(\lambda_{\modelindex} = (\lambda_{\modelindex, 1}, \lambda_{\modelindex, 2})\)
are nonnegative pooling weights, and for
\(\modelindex = 2, \ldots, \Nm - 1\)
\begin{gather}
  \pd_{\modelindex}(\phi_{\modelindex \cap \modelindex + 1}) =
    \int
    \pd_{\modelindex}(\phi_{\modelindex - 1 \cap \modelindex}, \phi_{\modelindex \cap \modelindex + 1})
    \text{d}\phi_{\modelindex - 1 \cap \modelindex}.
  \label{eqn:M-model-linear-pooling-component-def}
\end{gather}\noindent
For \(\modelindex = 1\) and \(\modelindex = \Nm\) the relevant marginals
are \(\pd_{1}(\phi_{1 \cap 2})\) and
\(\pd_{\Nm}(\phi_{\Nm - 1 \cap \Nm})\). In step two we form the pooled
prior as the product of the intermediaries
\begin{equation}
  \pd_{\text{pool, lin}} (\boldsymbol{\phi}) =
  \frac{1}{K_{\text{lin}}(\lambda)}
  \prod_{\modelindex = 1}^{\Nm - 1}
  \pd_{\text{pool}, \modelindex}(\phi_{\modelindex \cap \modelindex + 1}),
  \label{eqn:silly-linear-overall}
\end{equation}\noindent
with
\(K_{\text{lin}}(\lambda) = \int \prod_{\modelindex = 1}^{\Nm - 1} \pd_{\text{pool}, \modelindex}(\phi_{\modelindex \cap \modelindex + 1}) \text{d}\boldsymbol{\phi}\),
for \(\lambda = (\lambda_{1}, \ldots, \lambda_{\Nm})\). Clearly, this
assumes prior independence amongst all components of
\(\boldsymbol{\phi}\) which may be undesirable, particularly if this
independence was not present under one or more of the submodel priors.
We discuss extensions to linear pooling that enable prior dependence
between the components of \(\boldsymbol{\phi}\) in Section
\ref{conclusion}.

\hypertarget{dictatorial-pooling}{%
\subsubsection{Dictatorial pooling}\label{dictatorial-pooling}}

Chained Markov melding does not admit a direct analogue to dictatorial
pooling as defined in Section \ref{pooled-prior} because not all
submodel prior marginals contain all common quantities. For example,
consider the logarithmically pooled prior of Equation
\eqref{eqn:pooled-prior-overall} with, say, the
\(\modelindex\)\textsuperscript{th} entry in \(\lambda\) set to \(1\)
and all others set to \(0\). This choice of \(\lambda\) results in
\(\pd_{\text{pool}}(\boldsymbol{\phi}) = \pd(\phi_{\modelindex})\),
which is flat for \(\boldsymbol{\phi}_{-\modelindex}\). It seems
reasonable to require any generalisation of dictatorial pooling to
result in a reasonable prior for all components in
\(\boldsymbol{\phi}\). Such a generalisation should also retain the
original intention of dictatorial pooling, i.e.~`\emph{the authoritative
prior for} \(\phi_{\modelindex}\) \emph{is}
\(\pd_{\modelindex}(\phi_{\modelindex})\)'.

We propose two possible forms of dictatorial pooling that satisfy the
aforementioned criteria. \emph{Partial dictatorial pooling} enforces a
single submodel prior for the relevant components of
\(\boldsymbol{\phi}\), with no restrictions on the pooling of the
remaining components; and \emph{complete dictatorial pooling} which
requires selecting one of the two possible submodel priors for each
component of \(\boldsymbol{\phi}\).

Partial dictatorial pooling considers
\(\pd_{\modelindex}(\phi_{\modelindex})\) as the authoritative prior for
\(\phi_{\modelindex}= (\phi_{\modelindex - 1 \cap \modelindex}, \phi_{\modelindex \cap \modelindex + 1})\).
This results in,
\begin{equation}
  \begin{aligned}
    \pd_{\text{pool,dict}}(\boldsymbol{\phi}) = \,\,
      & g_{1} \! \left(\pd_{1}(\phi_{1 \cap 2}), \ldots, \pd_{\modelindex - 1}(\phi_{\modelindex - 2 \cap \modelindex - 1})\right)  \\
      \,\, & \times \pd_{\modelindex}(\phi_{\modelindex - 1 \cap \modelindex}, \phi_{\modelindex \cap \modelindex + 1}) \\
      \,\, & \times g_{2}\!\left(\pd_{\modelindex + 1}(\phi_{\modelindex + 1 \cap \modelindex + 2}), \ldots, \pd_{\Nm}(\phi_{\Nm - 1 \cap \Nm})\right),
  \end{aligned}
  \label{eqn:dictatorial-one-submodel}
\end{equation}\noindent
where \(g_{1}\) and \(g_{2}\) are linear or logarithmic pooling
functions as desired\footnote{Some care is required if the authoritative
  submodel is \(\pd_{\modelindex}\) for
  \(\modelindex \in \{1, 2, M - 1, M\}\). If it is taken to be
  \(\modelindex \in \{1, 2\}\), then \(g_{1}\) does not exist, and
  additionally in the \(\modelindex = 1\) case
  \(\pd_{1}(\phi_{0 \cap 1}, \phi_{1 \cap 2}) \coloneqq \pd_{1}(\phi_{1 \cap 2})\).
  The \(\modelindex \in \{\Nm - 1, \Nm\}\) cases have analogous
  definitions.}.

Complete dictatorial pooling requires the marginal pooled prior for each
component in \(\boldsymbol{\phi}\) to be chosen solely on the basis of
only one of the two priors specified for it under the submodels. For
\(\modelindex = 1, \ldots, \Nm - 1\), the
\(\modelindex\)\textsuperscript{th} marginal of the pooled prior is
either
\begin{equation}
  \pd_{\text{pool, dict}}(\phi_{\modelindex \cap \modelindex + 1}) \coloneqq
  \begin{cases}
    \pd_{\modelindex}(\phi_{\modelindex \cap \modelindex + 1}) & \text{or} \\
    \pd_{\modelindex + 1}(\phi_{\modelindex \cap \modelindex + 1}). &
  \end{cases}
  \label{eqn:dictatorial-def-three}
\end{equation}\noindent
If two consecutive marginals are chosen to have the same submodel prior
then we instead define
\begin{equation}
  \begin{aligned}
    \pd_{\text{pool, dict}}(\phi_{\modelindex - 1 \cap \modelindex})
    \pd_{\text{pool, dict}}(\phi_{\modelindex \cap \modelindex + 1})
      &= \pd_{\modelindex}(\phi_{\modelindex - 1 \cap \modelindex}) \pd_{\modelindex}(\phi_{\modelindex \cap \modelindex + 1}) \\
      &= \pd_{\modelindex}(\phi_{\modelindex - 1 \cap \modelindex}, \phi_{\modelindex \cap \modelindex + 1}),
  \end{aligned}
  \label{eqn:dictatorial-def-three-part-three}
\end{equation}\noindent
to preserve any dependence between
\(\phi_{\modelindex - 1 \cap \modelindex}\) and
\(\phi_{\modelindex \cap \modelindex + 1}\) that may be present under
\(\pd_{\modelindex}\). The complete dictatorially pooled prior is thus
\begin{equation}
  \pd_{\text{pool, dict}}(\boldsymbol{\phi}) =
    \prod_{\modelindex = 1}^{\Nm - 1} \pd_{\text{pool, dict}}(\phi_{\modelindex \cap \modelindex + 1}),
  \label{eqn:complete-dict-joint-def}
\end{equation}\noindent
where, subject to the potential modification in Equation
\eqref{eqn:dictatorial-def-three-part-three}, the terms in the product
are as defined in Equation \eqref{eqn:dictatorial-def-three}. For
example, if \(\Nm = 5\) and we wish to ignore \(\pd_{2}\) and
\(\pd_{4}\) when constructing the pooled prior and instead associate
\(\phi_{1 \cap 2}\) with \(\pd_{1}\), both \(\phi_{2 \cap 3}\) and
\(\phi_{3 \cap 4}\) with \(\pd_{3}\), and \(\phi_{4 \cap 5}\) with
\(\pd_{5}\), then
\begin{equation}
  \begin{aligned}
    \pd_{\text{pool,dict}}(\boldsymbol{\phi})
    &= \pd_{1, \text{dict}}(\phi_{1 \cap 2})
      \pd_{3, \text{dict}}(\phi_{2 \cap 3})
      \pd_{3, \text{dict}}(\phi_{3 \cap 4})
      \pd_{5, \text{dict}}(\phi_{4 \cap 5}) \\
    &= \pd_{1}(\phi_{1 \cap 2})
      \pd_{3}(\phi_{2 \cap 3}, \phi_{3 \cap 4})
      \pd_{5}(\phi_{4 \cap 5}).
  \end{aligned}
  \label{eqn:dictatorial-def-two-example-one}
\end{equation}

\hypertarget{pooling-weights}{%
\subsubsection{Pooling weights}\label{pooling-weights}}

\begin{figure}

{\centering \includegraphics[width=0.95\linewidth]{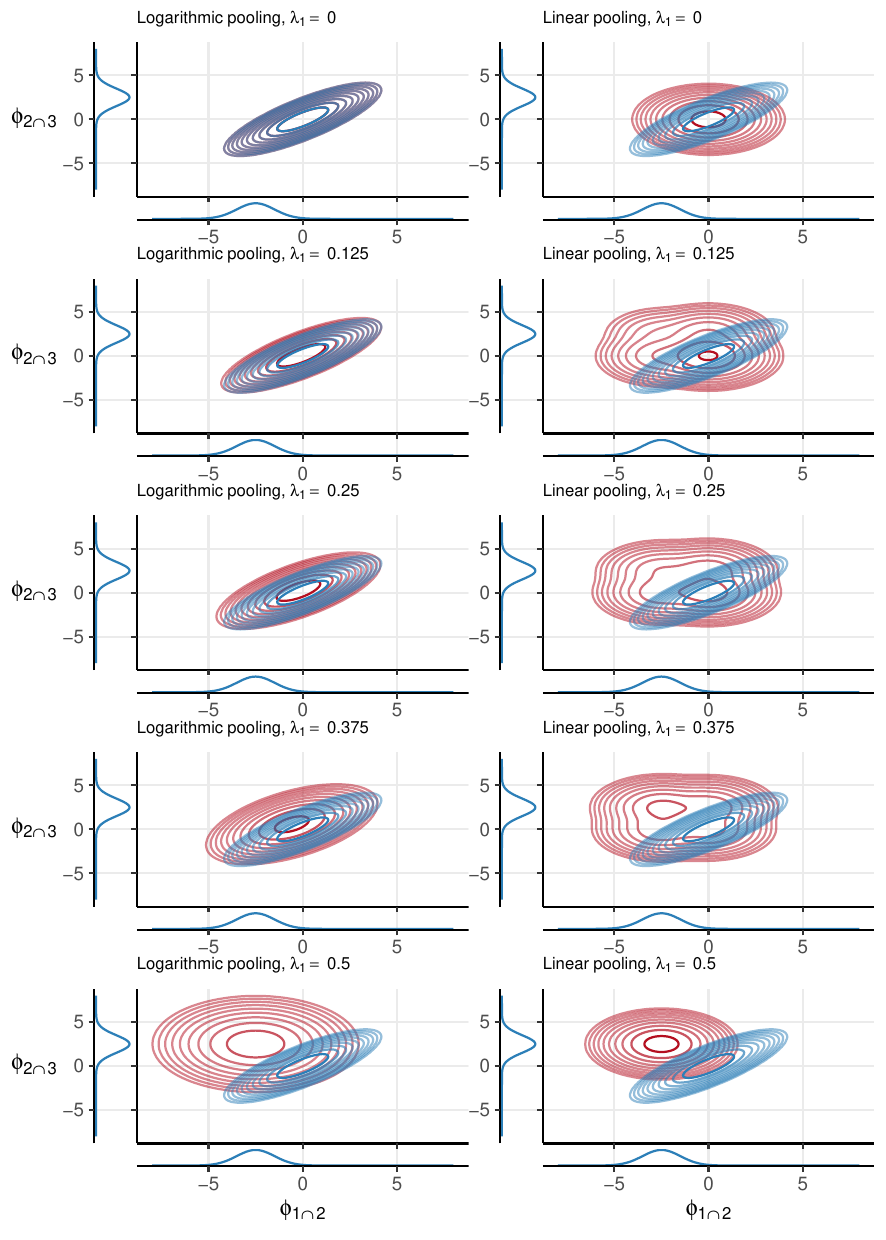} 

}

\caption{Contour plots of $\pd_{\text{pool}}(\boldsymbol{\phi})$ (red) under logarithmic and linear pooling (left and right column respectively). The three original densities $\pd_{1}(\phi_{1 \cap 2})$, $\pd_{3}(\phi_{2 \cap 3})$ and $\pd_{2}(\phi_{1 \cap 2}, \phi_{2 \cap 3})$ are shown in blue, with the univariate densities shown on the appropriate axis. The pooling weight parameter $\lambda_{1}$ is indicated in the plot titles.}\label{fig:pooled_densities_plot}
\end{figure}

Choosing values for the pooling weights is an important step in
specifying the pooled prior (Abbas, 2009; Carvalho \emph{et al.}, 2022;
María Jesús Rufo \emph{et al.}, 2012; MJ Rufo \emph{et al.}, 2012).
Because appropriate values for the weights depend on the submodels being
pooled and the information available \emph{a priori}, universal
recommendations are impossible, so we illustrate the impact of different
choices in a straightforward example. It is important that prior
predictive visualisations of the pooled prior are produced (Gabry
\emph{et al.}, 2019; Gelman \emph{et al.}, 2020) to guide the choice of
pooling weights and ensure that the result suitably represents the
available information. Figure \ref{fig:pooled_densities_plot}
illustrates how \(\lambda\) and the choice of pooling method impacts
\(\pd_{\text{pool}}(\boldsymbol{\phi})\) when pooling normal
distributions. Specifically, we consider \(\Nm = 3\) submodels and pool
\begin{equation}
  \begin{gathered}
  \pd_{1}(\phi_{1 \cap 2}) = \text{N}(\phi_{1 \cap 2}; \mu_{1}, \sigma_{1}^{2}), \quad
  \pd_{3}(\phi_{2 \cap 3}) = \text{N}(\phi_{2 \cap 3}; \mu_{3}, \sigma_{3}^{2}), \\[0.3em]
  \pd_{2}(\phi_{1 \cap 2}, \phi_{2 \cap 3}) = \text{N}\left(
    \begin{bmatrix} \phi_{1 \cap 2} \\ \phi_{2 \cap 3} \end{bmatrix}\!\!;
    \begin{bmatrix} \mu_{2, 1} \\ \mu_{2, 2} \end{bmatrix}\!\!,
    \begin{bmatrix} \sigma_{2}^{2} & \rho \sigma_{2}^{2} \\ \rho \sigma_{2}^{2} & \sigma_{2}^{2} \end{bmatrix}
  \right),
  \label{eqn:marginal-gaussian-example}
  \end{gathered}
\end{equation}\noindent
where \(\text{N}(\phi; \mu, \sigma^{2})\) is the normal density function
with mean \(\mu\) and variance \(\sigma^{2}\) (or covariance matrix
where appropriate). The two dimensional density function \(\pd_{2}\) has
an additional parameter \(\rho\), which controls the intra-submodel
marginal correlation. We set
\(\mu_{1} = -2.5, \mu_{2} = \left[\mu_{2, 1} \,\, \mu_{2, 2}\right]' = \left[0 \,\, 0\right]', \mu_{3} = 2.5, \sigma_{1}^{2} = \sigma_{2}^{2} = \sigma_{3}^{2} = 1\)
and \(\rho = 0.8\). In the logarithmic case we set
\(\lambda_{1} = \lambda_{3}\) and parameterise
\(\lambda_{2} = 1 - 2\lambda_{1}\), so that
\(\lambda_{1} + \lambda_{2} + \lambda_{3} = 1\) whilst limiting
ourselves to varying only \(\lambda_{1}\). Similarly, in the linear case
we set \(\lambda_{1, 1} = \lambda_{2, 2} = \lambda_{1}\) and
\(\lambda_{1, 2} = \lambda_{2, 1} = 1 - 2 \lambda_{1}\). We consider 5
evenly spaced values of \(\lambda_{1} \in [0, 0.5]\).

For both pooling methods, as the weight \(\lambda_{1}\) associated with
models \(\pd_{1}\) and \(\pd_{3}\) increases, the relative contributions
of \(\pd_{1}(\phi_{1 \cap 2})\) and \(\pd_{3}(\phi_{2 \cap 3})\)
increase. Note the lack of correlation in \(\pd_{\text{pool}}\) under
linear pooling (right column of Figure \ref{fig:pooled_densities_plot})
due to Equation \eqref{eqn:silly-linear-overall}. A large, near-flat
plateau is visible in the \(\lambda_{1} = 0.25\) and
\(\lambda_{1} = 0.375\) cases, which is a result of the mixture of four,
2-D normal distributions that linear pooling produces in this example.
The logarithmic pooling process produces a more concentrated prior for
small values of \(\lambda_{1}\), and does not result in \emph{a priori}
independence between \(\phi_{1 \cap 2}\) and \(\phi_{2 \cap 3}\).
Appendix~\ref{log-pooling-gaussian-densities} shows analytically that
\(\lambda_{2}\) controls the quantity of correlation present in
\(\pd_{\text{pool}}\) in this setting.

\hypertarget{posterior-estimation}{%
\section{Posterior estimation}\label{posterior-estimation}}

We now present a multi-stage MCMC method for generating samples from the
melded posterior. Whilst the melded posterior is a standard Bayesian
posterior and so can, in principle, be targetted using any suitable
Monte Carlo method, in practice this may be cumbersome or infeasible.
More specifically, it may be feasible to fit each submodel separately
using standard methods, but when the submodels are combined -- either
through Markov melding, or by expanding the definition of one submodel
to include another -- the computation required to estimate the posterior
in a single step poses an insurmountable barrier. In such settings we
can employ multi-stage posterior estimation methods including Tom
\emph{et al.} (2010), Lunn \emph{et al.} (2013), Hooten \emph{et al.}
(2019), and Mauff \emph{et al.} (2020). We propose a multi-stage
strategy that uses the chain-like relationship to both avoid evaluating
all submodels simultaneously, and parallelise the computation required
in the first stage to produce posterior samples in less time than an
equivalent sequential method\footnote{For completeness, Appendix
  \ref{sequential-sampler} describes such a sequential MCMC sampler. We
  do not use the sequential sampler in this paper.}. Avoiding
concurrently evaluating all submodels also enables the reuse of existing
software, minimising the need for custom submodel and/or sampler
implementations.

We also describe an approximate method, where stage one submodels are
summarised by normal distributions for use in stage two.

We consider the \(\Nm = 3\) case, as this setting includes both of our
examples. Our approach can be extended to \(\Nm > 3\) settings, although
we anticipate that it is unlikely to be suitable for large \(\Nm\)
settings. We discuss some of difficulties associated with generic,
parallel methodology for efficient posterior sampling in Section
\ref{conclusion}.

\hypertarget{parallel-sampler}{%
\subsection{Parallel sampler}\label{parallel-sampler}}

\begin{figure}[htb]
  \centering
  \begin{tikzpicture}[> = stealth, node distance = 1.5cm, thick, state/.style={draw = none, minimum width = 1.25cm}]
  \node [state] (model-1) {$\pd_{1}$};
  \node [state] [right = of model-1](model-2) {$\pd_{2}$};
  \node [state] [right = of model-2](model-3) {$\pd_{3}$};

  \path[-] (model-1) edge node[above] (phi-12) {$\phi_{1 \cap 2}$} (model-2);
  \path[-] (model-2) edge node[above] (phi-23) {$\phi_{2 \cap 3}$} (model-3);

  \node (stage-1) [draw = mymidblue, fit = (model-1) (phi-12), inner sep = 0.05cm, dashed] {};
  \node (stage-1-label) [yshift = 1.5ex, mymidblue] at (stage-1.north) {$s_{1}$};

  \node (stage-1-prime) [draw = mymidblue, fit = (model-3) (phi-23), inner sep = 0.05cm, dashed] {};
  \node (stage-1-prime-label) [yshift = 1.5ex, mymidblue] at (stage-1-prime.north) {$s_{1}$};

  \node (stage-2) [draw = myredhighlight, fit = (model-1) (model-2) (model-3) (phi-12) (phi-23) (stage-1-prime-label) (stage-1-prime-label), inner sep = 0.15cm, dashed] {};
  \node (stage-2-label) [yshift = 1.5ex, myredhighlight] at (stage-2.north) {$s_{2}$};
  \end{tikzpicture}
  \caption{A graphical depiction of the submodels and their shared quantities, with the parallel sampling strategy overlaid. The stage one ($s_{1}$) targets are surrounded by blue dashed lines, with the stage two ($s_{2}$) target in red.}
  \label{fig:parallel-dag}
\end{figure}
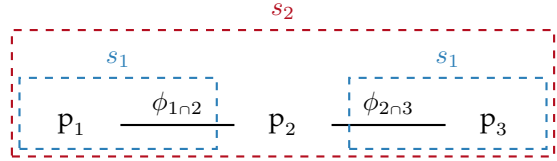

Our proposed strategy involves obtaining in stage one samples from
submodels 1 and 3 in parallel. Stage two reuses these samples in a
Metropolis-within-Gibbs sampler, which targets the full melded
posterior. The stage specific targets are displayed in Figure
\ref{fig:parallel-dag}.

The parallel sampler assumes that the pooled prior decomposes such that
\begin{equation}
  \pd_{\text{pool}}(\boldsymbol{\phi}) = 
    \pd_{\text{pool}, 1}(\phi_{1 \cap 2})
    \pd_{\text{pool}, 2}(\phi_{1 \cap 2}, \phi_{2 \cap 3})
    \pd_{\text{pool}, 3}(\phi_{2 \cap 3}).
  \label{eqn:parallel-decomposition}
\end{equation}\noindent
All pooled priors trivially satisfy \eqref{eqn:parallel-decomposition}
by assuming \(\pd_{\text{pool}, 1}(\phi_{1 \cap 2})\) and
\(\pd_{\text{pool}, 3}(\phi_{2 \cap 3})\) are improper and/or flat
distributions. Alternatively we may choose
\(\pd_{\text{pool}, 1}(\phi_{1 \cap 2}) = \pd_{1}(\phi_{1 \cap 2})\) and
\(\pd_{\text{pool}, 3}(\phi_{2 \cap 3}) = \pd_{3}(\phi_{2 \cap 3})\),
with appropriate adjustments to
\(\pd_{\text{pool}, 2}(\phi_{1 \cap 2}, \phi_{2 \cap 3})\). This choice
targets, in stage one, the subposteriors of \(\pd_{1}\) and \(\pd_{3}\)
under their original prior for \(\phi_{1 \cap 2}\) and
\(\phi_{2 \cap 3}\) respectively.

\hypertarget{stage-one}{%
\paragraph{Stage one}\label{stage-one}}

Two independent, parallel sampling processes occur in stage one. Terms
from the melded model that pertain to \(\pd_{1}\) and \(\pd_{3}\) are
isolated
\begin{gather}
  \pd_{\text{meld}, 1} (\phi_{1 \cap 2}, \psi_{1} \mid Y_{1}) \propto
  \pd_{\text{pool}, 1} (\phi_{1 \cap 2})
  \frac {
    \pd_{1}(\phi_{1 \cap 2}, \psi_{1}, Y_{1})
  } {
    \pd_{1}(\phi_{1 \cap 2})
  }
  \label{eqn:stage-one-targets-one} \\
  \pd_{\text{meld}, 3} (\phi_{2 \cap 3}, \psi_{3} \mid Y_{3}) \propto
  \pd_{\text{pool}, 3} (\phi_{2 \cap 3})
  \frac {
    \pd_{3}(\phi_{2 \cap 3}, \psi_{3}, Y_{3})
  } {
    \pd_{3}(\phi_{2 \cap 3})
  },
  \label{eqn:stage-one-targets-two}
\end{gather}\noindent and
targeted using standard MCMC methodology. This produces \(N_{1}\)
samples \(\{(\phi_{1 \cap 2}, \psi_{1})_{n}\}_{n = 1}^{N_{1}}\) from
\(\pd_{\text{meld}, 1}(\phi_{1 \cap 2}, \psi_{2} \mid Y_{1})\) and
\(N_{3}\) samples
\(\{(\phi_{2 \cap 3}, \psi_{3})_{n}\}_{n = 1}^{N_{3}}\) from
\(\pd_{\text{meld}, 3}(\phi_{2 \cap 3}, \psi_{3} \mid Y_{3})\).

\hypertarget{stage-two}{%
\paragraph{Stage two}\label{stage-two}}

Stage two targets the melded posterior of Equation
\eqref{eqn:melded-model-full} using a Metropolis-within-Gibbs sampler,
where the proposal distributions are
\begin{align}
  \phi_{1 \cap 2}^{*}, \psi_{1}^{*} \mid \phi_{2 \cap 3}, \psi_{2}, \psi_{3} &\sim \pd_{\text{meld}, 1}(\phi_{1 \cap 2}^{*}, \psi_{1}^{*} \mid Y_{1}) \label{eqn:stage-two-proposals-one} \\
  \phi_{2 \cap 3}^{*}, \psi_{3}^{*} \mid \phi_{1 \cap 2}, \psi_{1}, \psi_{2} &\sim \pd_{\text{meld}, 3}(\phi_{2 \cap 3}^{*}, \psi_{3}^{*} \mid Y_{3}) \label{eqn:stage-two-proposals-two} \\
  \psi_{2}^{*} \mid \phi_{1 \cap 2}, \phi_{2 \cap 3}, \psi_{1}, \psi_{3} &\sim \q(\psi_{2}^{*} \mid \psi_{2}),
  \label{eqn:stage-two-proposals-three}
\end{align}\noindent where
\(\q(\psi_{2}^{*} \mid \psi_{2})\) is a generic proposal distribution
for \(\psi_{2}\). We draw an index \(n_{1}^{*}\) uniformly from
\(\{1, \ldots, N_{1}\}\) and use the corresponding value
\((\phi_{1 \cap 2}^{*}, \psi_{1}^{*})_{n_{1}^{*}}\) as the proposal,
doing likewise for \(n_{3}^{*}\) and
\((\phi_{2 \cap 3}^{*}, \psi_{3}^{*})_{n_{3}^{*}}\). The acceptance
probabilities for these updates are
\begin{align}
  \alpha((\phi_{1 \cap 2}^{*}, \psi_{1}^{*})_{n_{1}^{*}}, (\phi_{1 \cap 2}, \psi_{1})_{n_{1}}) &=
    \frac {
      \pd_{\text{pool}, 2}(\phi_{1 \cap 2}^{*}, \phi_{2 \cap 3})
    } {
      \pd_{\text{pool}, 2}(\phi_{1 \cap 2}, \phi_{2 \cap 3})
    }
    \frac {
      \pd_{2}(\phi_{1 \cap 2}^{*}, \phi_{2 \cap 3}, \psi_{2}, Y_{2})
    } {
      \pd_{2}(\phi_{1 \cap 2}, \phi_{2 \cap 3}, \psi_{2}, Y_{2})
    }
    \frac {
      \pd_{2} (\phi_{1 \cap 2}, \phi_{2 \cap 3})
    } {
      \pd_{2} (\phi_{1 \cap 2}^{*}, \phi_{2 \cap 3})
    }
  \\
  \alpha((\phi_{2 \cap 3}^{*}, \psi_{3}^{*})_{n_{3}^{*}}, (\phi_{2 \cap 3}, \psi_{3})_{n_{3}}) &=
    \frac {
      \pd_{\text{pool}, 2}(\phi_{1 \cap 2}, \phi_{2 \cap 3}^{*})
    } {
      \pd_{\text{pool}, 2}(\phi_{1 \cap 2}, \phi_{2 \cap 3})
    }
    \frac {
      \pd_{2}(\phi_{1 \cap 2}, \phi_{2 \cap 3}^{*}, \psi_{2}, Y_{2})
    } {
      \pd_{2}(\phi_{1 \cap 2}, \phi_{2 \cap 3}, \psi_{2}, Y_{2})
    }
    \frac {
      \pd_{2} (\phi_{1 \cap 2}, \phi_{2 \cap 3})
    } {
      \pd_{2} (\phi_{1 \cap 2}, \phi_{2 \cap 3}^{*})
    }
  \\
  \alpha(\psi_{2}^{*}, \psi_{2}) &=
    \frac {
      \pd_{2}(\phi_{1 \cap 2}, \phi_{2 \cap 3}, \psi_{2}^{*}, Y_{2})
    } {
      \pd_{2}(\phi_{1 \cap 2}, \phi_{2 \cap 3}, \psi_{2}, Y_{2})
    }
    \frac {
      \q(\psi_{2} \mid \psi_{2}^{*})
    } {
      \q(\psi_{2}^{*} \mid \psi_{2})
    },
  \label{eqn:stage-two-acceptance}
\end{align}\noindent
where \(\alpha(x, z)\) denotes the probability associated with a move
from \(z\) to \(x\). Note that all stage two acceptance probabilities
only contain terms from the second submodel and the pooled prior, and
thus do not depend on \(\psi_{1}\) or \(\psi_{3}\). If a move is
accepted then we also store the index, i.e.~\(n_{1}^{*}\) or
\(n_{3}^{*}\), associated with the move, otherwise we store the current
value of the index. The stored indices are used to appropriately
resample \(\psi_{1}\) and \(\psi_{3}\) from the stage one samples.

\hypertarget{normal-approximations-to-submodel-components}{%
\subsection{Normal approximations to submodel
components}\label{normal-approximations-to-submodel-components}}

Normal approximations are commonly employed to summarise submodels for
subsequent use in more complex models. For example, two-stage
meta-analyses often use a normal distribution centred on each studies'
effect estimate (Burke \emph{et al.}, 2017). Suppose we employ such an
approximation to summarise the prior and posterior of
\(\phi_{1 \cap 2}\) and \(\phi_{2 \cap 3}\) under \(\pd_{1}\) and
\(\pd_{3}\) respectively. In addition, assume that (a) such
approximations are appropriate for
\(\pd_{1}(\phi_{1 \cap 2}), \pd_{1}(\phi_{1 \cap 2} \mid Y_{1}), \pd_{3}(\phi_{2 \cap 3})\),
and \(\pd_{3}(\phi_{2 \cap 3} \mid Y_{3})\), (b) we are not interested
in \(\psi_{1}\) and \(\psi_{3}\), and can integrate them out of all
relevant densities, and (c) we employ our second form of dictatorial
pooling and choose \(\pd_{2}(\phi_{1 \cap 2}, \phi_{2 \cap 3})\) as the
authoritative prior. The latter two assumptions imply that the melded
posterior of interest is proportional to
\begin{equation}
  \pd_{\text{meld}} (\phi_{1 \cap 2}, \phi_{2 \cap 3}, \psi_{2} \mid \boldsymbol{Y})
  \, \propto \, 
  \frac {
    \pd_{1}(\phi_{1 \cap 2} \mid Y_{1})
  } {
    \pd_{1}(\phi_{1 \cap 2})
  }
  \pd_{2}(\phi_{1 \cap 2}, \phi_{2 \cap 3}, \psi_{2} \mid Y_{2})
  \frac {
    \pd_{3}(\phi_{2 \cap 3} \mid Y_{3})
  } {
    \pd_{3}(\phi_{2 \cap 3})
  }.
  \label{eqn:normal-approx-melded-posterior-target}
\end{equation}

Denote the normal approximation of
\(\pd_{1}(\phi_{1 \cap 2} \mid Y_{1})\) as
\(\widehat{\pd}_{1}(\phi_{1 \cap 2} \mid \widehat{\mu}_{1}, \widehat{\Sigma}_{1})\),
which is a normal distribution with mean \(\widehat{\mu}_{1}\) and
covariance matrix \(\widehat{\Sigma}_{1}\). The corresponding normal
approximation of the prior \(\pd_{1}(\phi_{1 \cap 2})\) is
\(\widehat{\pd}_{1}(\phi_{1 \cap 2} \mid \widehat{\mu}_{1, 0}, \widehat{\Sigma}_{1, 0})\).
The equivalent approximations for the subposterior and prior of
\(\pd_{3}\) are
\(\widehat{\pd}_{3}(\phi_{2 \cap 3} \mid \widehat{\mu}_{3}, \widehat{\Sigma}_{3})\)
and
\(\widehat{\pd}_{3}(\phi_{2 \cap 3} \mid \widehat{\mu}_{3, 0}, \widehat{\Sigma}_{3, 0})\)
respectively. Substituting in the approximations and using standard
results for Gaussian density functions (see Bromiley (2003) and Appendix
\ref{normal-approximation-calculations}) results in
\begin{equation}
\begin{gathered}
  \widehat{\pd}_{\text{meld}} (\phi_{1 \cap 2}, \phi_{2 \cap 3}, \psi_{2} \mid \boldsymbol{Y})
  \, \propto \,
  \widehat{\pd}\left((\phi_{1 \cap 2}, \phi_{2 \cap 3}) \mid \widehat{\mu}, \, \widehat{\Sigma}\right)
  \pd_{2}(\phi_{1 \cap 2}, \phi_{2 \cap 3}, \psi_{2} \mid Y_{2}),
\end{gathered}
\label{eqn:final-normal-approx}
\end{equation}\noindent
where
\begin{equation}
  \begin{gathered}
    \widehat{\mu}_{\text{nu}} = \begin{bmatrix}
      \widehat{\mu}_{1} \\
      \widehat{\mu}_{3}
    \end{bmatrix}\!\!, \quad
    \widehat{\Sigma}_{\text{nu}} = \begin{bmatrix}
      \widehat{\Sigma}_{1} & 0 \\
      0 & \widehat{\Sigma}_{3}
    \end{bmatrix}\!\!, \quad
    \widehat{\mu}_{\text{de}} = \begin{bmatrix}
      \widehat{\mu}_{1, 0} \\
      \widehat{\mu}_{3, 0}
    \end{bmatrix}\!\!, \quad
    \widehat{\Sigma}_{\text{de}} = \begin{bmatrix}
      \widehat{\Sigma}_{1, 0} & 0 \\
      0 & \widehat{\Sigma}_{3, 0}
    \end{bmatrix}\!\!, \\[1.5ex]
    \widehat{\Sigma} = \left(
      \widehat{\Sigma}_{\text{nu}}^{-1} - \widehat{\Sigma}_{\text{de}}^{-1}
    \right)^{-1}, \quad
    \widehat{\mu} = \widehat{\Sigma} \left(
      \widehat{\Sigma}_{\text{nu}}^{-1}\widehat{\mu}_{\text{nu}} - \widehat{\Sigma}_{\text{de}}^{-1}\widehat{\mu}_{\text{de}}
    \right).
  \end{gathered}
  \label{eqn:in-text-normal-approx}
\end{equation}\noindent
Standard MCMC methods can be used to sample from the approximate melded
posterior. If instead we opt for product-of-experts pooling, all
\(\widehat{\mu}_{\text{de}}\) and \(\widehat{\Sigma}_{\text{de}}\) terms
disappear from the parameter definitions in Equation
\eqref{eqn:in-text-normal-approx}.

\hypertarget{an-integrated-population-model-for-little-owls-1}{%
\section{An integrated population model for little
owls}\label{an-integrated-population-model-for-little-owls-1}}

We now return to the integrated population model (IPM) for the little
owls introduced in Section
\ref{an-integrated-population-model-for-little-owls}. Finke \emph{et
al.} (2019) consider a number of variations on the original model of
Schaub \emph{et al.} (2006) and Abadi \emph{et al.} (2010): here we
consider only the variant from Finke \emph{et al.} (2019) with the
highest marginal likelihood (Model 4 of their online supplement). This
example is particularly interesting to us as, for a certain choice of
pooling function and pooling weights, the chained Markov melded model
and the IPM are identical. This coincidence allows us to use the
posterior from the IPM as a benchmark for our multi-stage sampler.

Before we detail the specifics of each submodel, we must introduce some
notation. Data and parameters are stratified into two age-groups
\(a \in \{J, A\}\) where \(J\) denotes juvenile owls (less than one year
old) and \(A\) adults, two sexes \(s \in \{M, F\}\), and observations
occur annually at times \(t \in \{1, \ldots, T\}\), with \(T = 25\). The
sex- and age-specific probability of an owl surviving from time \(t\) to
\(t + 1\) is \(\delta_{a, s, t}\), and the sex-specific probability of a
previously captured owl being recaptured at time \(t + 1\) is
\(\pi_{s, t + 1}\) so long as the owl is alive at time \(t + 1\).

\hypertarget{capture-recapture-pd_1}{%
\subsection{\texorpdfstring{Capture recapture:
\(\pd_{1}\)}{Capture recapture: \textbackslash pd\_\{1\}}}\label{capture-recapture-pd_1}}

Capture-recapture data pertain to owls that are released at time \(t\)
(having been captured and tagged), and then recaptured at time
\(u = t + 1, \dots, T\), or not recaptured before the conclusion of the
study, in which case \(u = T + 1\). Define \(M_{a, s, t, u}\) as the
number of owls of age-group \(a\) and sex \(s\) released at time \(t\)
and recaptured at time \(u\). We aggregate these observations into age-
and sex-specific matrices \(\boldsymbol{M}_{a, s}\), with \(T\) rows,
corresponding to release times, and \(T + 1\) columns, corresponding to
recapture times. Let
\(R_{a, s, t} = \sum_{u = 1}^{T + 1} M_{a, s, t, u}\) be the number of
owls released at time \(t\), i.e.~a vector containing the row-wise sum
of the entries in \(\boldsymbol{M}_{a, s}\). The recapture times for
owls released at time \(t\) follow an age- and sex-specific multinomial
likelihood
\begin{equation}
  (M_{a, s, t, 1}, \ldots, M_{a, s, t, T + 1})
  \sim
  \text{Multinomial}(R_{a, s, t}, \boldsymbol{Q}_{a, s, t}),
  \label{eqn:capture-recapture-submodel}
\end{equation}\noindent
with probabilities
\(\boldsymbol{Q}_{a, s, t} = (Q_{a, s, t, 1}, \ldots, Q_{a, s, t, T + 1})\)
such that
\begin{equation}
  Q_{a, s, t, u} =
    \left\{
    \begin{array}{ll}
      0, 
      & 
      \text{for} \,\, u = 1, \, \ldots \, , t \\
      \delta_{a, s, t} 
      \pi_{s, u} 
      \prod_{r = t + 1}^{u - 1} 
        \delta_{a, s, r} \left(1 - \pi_{s, r} \right),
      & 
      \text{for} \,\, u = t + 1, \, \ldots \, , T \\
      1 - \sum_{r = 1}^{T} Q_{a, s, t, r},
      & 
      \text{if} \,\, u = T + 1.
    \end{array}\right.
  \label{eqn:multinomial-probabilities}
\end{equation}

\hypertarget{count-data-model-pd_2}{%
\subsection{\texorpdfstring{Count data model:
\(\pd_{2}\)}{Count data model: \textbackslash pd\_\{2\}}}\label{count-data-model-pd_2}}

To estimate population abundance, a two level model is used: the first
level models the observed (counted) number of females at each point in
time denoted \(y_{t}\), with a second, latent process modelling the
total number of females in population. The observation model is
\begin{equation}
  y_{t} \mid x_{t}
  \sim 
  \text{Poisson}\left(
    x_{t}
  \right),
  \label{eqn:observation-process}
\end{equation}\noindent
where we denote the number of juvenile and adult females in the
population at time \(t\) as \(x_{J, t}\) and \(x_{A, t}\) respectively,
with \(x_{t} = x_{J, t} + x_{A, t}\). If \(\text{sur}_{t}\) adult
females survive from time \(t - 1\) to time \(t\), and
\(\text{imm}_{t}\) adult females immigrate over the same time period,
then the latent, population level model is
\begin{equation}
  \begin{aligned}
    x_{J, t} \mid x_{t - 1}, \rho, \delta_{J, F, t - 1}
    \,\, &\sim \,\,
    \text{Poisson}\left(
      x_{t - 1}
      \, \frac{\rho}{2} \,
      \delta_{J, F, t - 1}
    \right), \\
    \text{sur}_{t} \mid x_{t - 1}, \delta_{A, F, t - 1}
    \,\, &\sim \,\,
    \text{Binomial}\left(
      x_{t - 1}, \,
      \delta_{A, F, t - 1}
    \right), \\
    \text{imm}_{t} \mid x_{t - 1}, \eta_{t}
    \,\, &\sim\,\,
    \text{Poisson}\left(
      x_{t - 1} \eta_{t}\right
    ), \\
    x_{A, t} &= \text{sur}_{t} + \text{imm}_{t},
    \end{aligned}
  \label{eqn:count-data-submodel}
\end{equation}\noindent
where \(\eta_{t}\) is the immigration rate. The initial population sizes
\(x_{J, 1}\) and \(x_{A, 1}\) have independent discrete uniform priors
on \(\{0, 1, \ldots, 50\}\). If \(x_{t - 1} = 0\) then we assume that
the Poisson and binomial distributions become point masses at zero.

\hypertarget{fecundity-pd_3}{%
\subsection{\texorpdfstring{Fecundity:
\(\pd_{3}\)}{Fecundity: \textbackslash pd\_\{3\}}}\label{fecundity-pd_3}}

The fecundity submodel considers the number of breeding females at time
\(t\) denoted \(N_{t}\), and the number of chicks produced that survive
and leave the nest denoted \(n_{t}\). A Poisson model is employed to
estimate fecundity (reproductive rate) \(\rho\)
\begin{equation}
  n_{t} \sim \text{Poisson}(N_{t} \rho).
  \label{eqn:fecundity-submodel}
\end{equation}

\hypertarget{parameterisation-and-melding-quantities}{%
\subsection{Parameterisation and melding
quantities}\label{parameterisation-and-melding-quantities}}

Abadi \emph{et al.} (2010) parameterise the time dependent quantities
via linear predictors to minimise the number of parameters in the
submodels. The specific parameterisation of Finke \emph{et al.} (2019)
we employ is
\begin{equation}
  \begin{gathered}
    \text{logit}(\delta_{a, s, t}) = \alpha_{0} + \alpha_{1}\mathbb{I}(s = M) + \alpha_{2}\mathbb{I}(a = A), \quad
    \text{log}(\eta_{t}) = \alpha_{6}, \\
    \text{logit}(\pi_{s, u}) = \alpha_{4} \mathbb{I}(s = M) + \alpha_{5, u}, \,\, \text{for} \,\, u = 2, \ldots T,
  \end{gathered}
  \label{eqn:parameterisation-info}
\end{equation}\noindent
thus the quantities in common between the submodels are
\(\phi_{1 \cap 2} = (\alpha_{0}, \alpha_{2})\) and
\(\phi_{2 \cap 3} = \rho\). To align the notation of this example with
our chained melding notation we define, for all permitted values of
\(a, s\) and \(t\), \(Y_{1} = (\boldsymbol{M}_{a, s})\),
\(\psi_{1} = \left(\alpha_{1}, \alpha_{4}, (\alpha_{5, u})_{u = 2}^{T}\right)\);
\(Y_{2} = (y_{t})\),
\(\psi_{2} = (x_{J, t}, \alpha_{6}, \text{sur}_{t}, \text{imm}_{t})\);
and \(Y_{3} = (N_{t}, n_{t})\), \(\psi_{3} = \varnothing\). Note that
the definition of \(\phi_{1 \cap 2}\) does not include \(\alpha_{1}\) as
it is male specific and does not exist in \(\pd_{2}\). The model variant
of Finke \emph{et al.} (2019) we consider does not include
\(\alpha_{3}\), and for comparability we keep the other parameter
indices the same.

\hypertarget{priors}{%
\subsection{Priors}\label{priors}}

We use the priors of Finke \emph{et al.} (2019) for the parameters in
each submodel. Denote
\(\boldsymbol{\alpha} = (\alpha_{0}, \alpha_{1}, \alpha_{2}, \alpha_{4}, \alpha_{6})\).
In both \(\pd_{1}\) and \(\pd_{2}\) the elements of
\(\boldsymbol{\alpha}\) are assigned independent
\(\text{Normal}(0, 2^2)\) priors truncated to \([-10, 10]\). The time
varying recapture probabilities \(\alpha_{5, u}\) also have
\(\text{Normal}(0, 2^2)\) priors truncated to \([-10, 10]\). A
\(\text{Uniform}(0, 10)\) prior is assigned to \(\rho\) in \(\pd_{2}\)
and \(\pd_{3}\).

To completely specify \(\pd_{\text{meld}}\) we must choose how to form
\(\pd_\text{pool}(\phi_{1 \cap 2}, \phi_{2 \cap 3})\). We form
\(\pd_\text{pool}(\phi_{1 \cap 2}, \phi_{2 \cap 3})\) using three
different pooling methods and estimate the melded posterior in each
case. The first pooling method is product-of-experts (PoE) pooling,
which is logarithmic pooling with \(\lambda = (1, 1, 1)\), and we denote
the melded posterior as \(\pd_{\text{meld, PoE}}\). We also use
logarithmic pooling with
\(\lambda = (\frac{1}{2}, \frac{1}{2}, \frac{1}{2})\), which is denoted
\(\pd_{\text{meld, log}}\) and results in the chained melded model being
identical to the IPM. The final pooling method is linear pooling with
\(\lambda = (\frac{1}{2}, \frac{1}{2}, \frac{1}{2}, \frac{1}{2})\),
denoted \(\pd_{\text{meld, lin}}\).

\hypertarget{posterior-estimation-1}{%
\subsection{Posterior estimation}\label{posterior-estimation-1}}

We estimate the melded posterior --
\(\pd_{\text{meld}}(\boldsymbol{\phi}, \boldsymbol{\psi} \mid \boldsymbol{Y})\),
proportional to Equation \eqref{eqn:melded-model-full} -- using both the
parallel sampler (Section \ref{parallel-sampler}) and normal
approximation (Section
\ref{normal-approximations-to-submodel-components}). This allows us to
use pre-existing implementations of the submodels. Specifically, the
capture-recapture submodel is written in BUGS (Lunn \emph{et al.}, 2009)
and sampled via \texttt{rjags} (Plummer, 2019). The fecundity submodel
is written in Stan (Carpenter \emph{et al.}, 2017) and sampled via
\texttt{rstan} (Stan Development Team, 2021). The count data submodel is
also written in \texttt{BUGS}, and we reuse this implementation in stage
two of the multi-stage sampler via \texttt{NIMBLE} (de Valpine \emph{et
al.}, 2017) and its \texttt{R} interface (NIMBLE Development Team,
2019). The approximate melded posterior obtained by Section
\ref{normal-approximations-to-submodel-components} is sampled using
\texttt{rjags}. Code and data for this example, as well as trace plots
and numerical convergence measures (Vehtari \emph{et al.}, 2020) for
both stages of the parallel sampling process, are available in the
accompanying online repository\footnote{\url{https://doi.org/10.5281/zenodo.6552714}}.

\hypertarget{results}{%
\subsection{Results}\label{results}}

\begin{figure}

{\centering \includegraphics{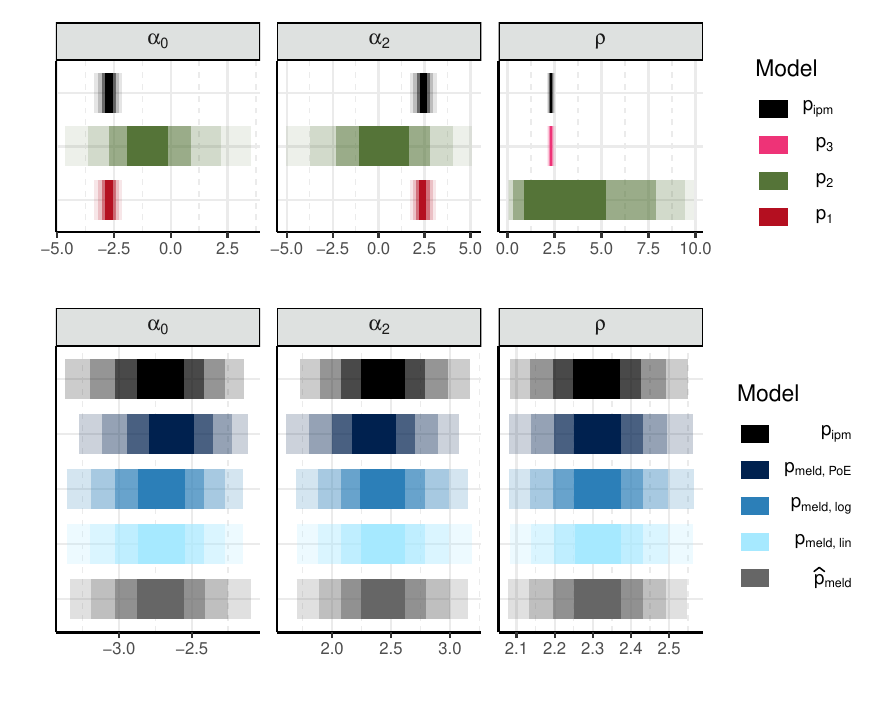} 

}

\caption{Top row: credible intervals for $\phi_{1 \cap 2} = (\alpha_{0}, \alpha_{2})$ and $\phi_{2 \cap 3} = \rho$ from the posterior of the original integrated population model $\pd_{\text{ipm}}$, and the individual subposteriors from submodels $\pd_{1}, \pd_{2}$, and $\pd_{3}$. Bottom row: credible intervals for the same quantities, but with a different x-axis scale, from the original IPM (repeated from top row); the chained melded posteriors using product-of-experts pooling, logarithmic pooling, and linear pooling denoted $\pd_{\text{meld}}$, $\pd_{\text{meld, log}}$ and $\pd_{\text{meld, lin}}$; and the melded posterior using the normal approximation $\widehat{\pd}_{\text{meld}}$. Intervals are 50\%, 80\%, 95\%, and 99\% wide.}\label{fig:phi_subpost}
\end{figure}

We empirically validate our methodology and sampler by comparing the
melded posterior samples to a large sample -- 6 chains, each containing
\(1 \times 10^5\) post-warmup iterations -- from the original IPM
posterior. Similarity in the posteriors is expected as the IPM is
effectively the joint model we wish to approximate with the chained
melded model. It is simply fortunate, from a modelling standpoint, that
this example's joint model is easy to construct and computationally
feasible with standard tools. Note that under logarithmic pooling with
\(\lambda = (\frac{1}{2}, \frac{1}{2}, \frac{1}{2})\) the melded
posterior is identical to the original IPM, so any differences between
the two posteriors are attributable to the multi-stage sampler. Figure
\ref{fig:phi_subpost} depicts the posterior credible intervals (Gabry
\emph{et al.}, 2021; Kay, 2020) for the common quantities from the
individual submodels, the melded models, and the original IPM. The top
row in Figure \ref{fig:phi_subpost} indicates that the count data alone
(\(\pd_{2}\)) contain minimal information about
\(\alpha_{0}, \alpha_{2}\) and \(\rho\); incorporating the data from the
other submodels is essential for precise estimates.

The multi-stage sampler works well by producing melded posterior
estimates generally similar to the original IPM estimate, and are near
identical for logarithmic pooling. PoE pooling produces the posterior
most different from the original IPM, as it yields a prior for
\((\alpha_{0}, \alpha_{2})\) that is more concentrated around zero than
the other pooling methods. The lack of large differences between the
melded posteriors that use different pooled priors indicates that the
prior has almost no effect on the posterior. The similarity of the
approximate approach (\(\widehat{\pd}_{\text{meld}}\) - bottom row of
Figure \ref{fig:phi_subpost}) to the melding approaches suggests that
the normal approximations are good summaries of the subposteriors, and
that the approximate melding procedure of Section
\ref{normal-approximations-to-submodel-components} is suitable for this
example.

\hypertarget{survival-analysis-with-time-varying-covariates-and-uncertain-event-times-1}{%
\section{Survival analysis with time varying covariates and uncertain
event
times}\label{survival-analysis-with-time-varying-covariates-and-uncertain-event-times-1}}

We return now to the respiratory failure example introduced in Section
\ref{survival-analysis-with-time-varying-covariates-and-uncertain-event-times}.
Our intention is to illustrate the application of chained Markov melding
to an example of realistic complexity, and explore empirically the
importance of accounting for all sources of uncertainty by comparing
chained Markov melding to equivalent analyses which use only a point
estimate summary of the uncertainty. Specifically, event times and
indicators are a noninvertible function of other parameters in the first
submodel, and are an uncertain response in the survival submodel.
Chained Markov melding enables us to specify a suitable joint model
despite these complications.

There are \(i = 1, \ldots, N\) individuals in the data set. Each
individual is admitted to the ICU at time \(0\), and is discharged or
dies at time \(C_{i}\). See Appendix \ref{cohort-selection-criteria} for
information on how the \(N = 37\) individuals were selected from
MIMIC-III (Johnson \emph{et al.}, 2016).

\hypertarget{pf-ratio-submodel-b-spline-pd_1}{%
\subsection{\texorpdfstring{P/F ratio submodel (B-spline):
\(\pd_{1}\)}{P/F ratio submodel (B-spline): \textbackslash pd\_\{1\}}}\label{pf-ratio-submodel-b-spline-pd_1}}

The first submodel fits a B-spline to the \pfratio~data to calculate if
and when an individual experiences respiratory failure. Each individual
has \pfratio~ratio observations \(z_{i, j}\) (in units of mmHg) at times
\(t_{i, j}\), with \(j = 1, \ldots, J_{i}\). For each individual denote
the vector of observations
\(\boldsymbol{z}_{i} = (z_{i, 1}, \ldots, z_{i, J_{i}})\) and
observation times
\(\boldsymbol{t}_{i} = (t_{i, 1}, \ldots, t_{i, J_{i}})\). To improve
computational performance, we standardise the P/F ratio data for each
individual such that
\(z_{i, j} = \frac{\tilde{z}_{i, j} - \overline{z}_{i}}{\hat{s}_{i}}\),
where \(\tilde{z}_{i, j}\) is the underlying unstandardised observation
with mean \(\overline{z}_{i}\) and standard deviation \(\hat{s}_{i}\).
Similarly we rescale the threshold for respiratory failure:
\(\tau_{i} = \frac{300 - \overline{z}_{i}}{\hat{s}_{i}}\).

We choose to model the P/F ratio using cubic B-splines and 7 internal
knots, and do not include an intercept column in the spline basis (for
background on B-splines see: Chapter 2 in Hastie and Tibshirani, 1999;
and the supplementary material of Wang and Yan, 2021). The internal
knots are evenly spaced between two additional boundary knots at
\(\min(\boldsymbol{t}_{i})\) and \(\max(\boldsymbol{t}_{i})\). These
choices result in \(k = 1, \ldots, 10\) spline basis terms per
individual, with coefficients \(\zeta_{i, k}\) where
\(\boldsymbol{\zeta}_{i} = (\zeta_{i, 1}, \ldots, \zeta_{i, 10})\). We
denote the individual specific B-spline basis evaluated at time
\(t_{i, j}\) as \(B_{i}(t_{i, j}) \in [0, \infty)^{10}\) so that the
submodel can be written as
\begin{equation}
  z_{i, j} = \beta_{0, i} + B_{i}(t_{i, j})^{\top}\boldsymbol{\zeta}_{i} + \varepsilon_{i, j}.
  \label{eqn:pf-model-def}
\end{equation}
\noindent We employ
a weakly informative prior for the intercept
\(\beta_{0, i} \sim \text{N}(0, 1^2)\), a heavy tailed distribution for
the error term\footnote{P/F data contain many outliers for, amongst many
  possible reasons, arterial/venous blood sample mislabelling;
  incorrectly recorded oxygenation support information; and differences
  between sample collection time, lab result time, and the observation
  time as recorded in the EHR.}
\(\varepsilon_{i, j} \sim t_{5}(0, \omega_{i})\), and a weakly
informative half-normal prior for the unknown scale parameter
\(\omega_{i} \sim \text{N}_{+}(0, 1^2)\). For the spline basis
coefficients we set \(\zeta_{i, 1} \sim \text{N}(0, 0.1^2)\), and for
\(k = 2, \ldots, 10\) we employ the random-walk prior
\(\zeta_{i, k} \sim \text{N}(\zeta_{i, k - 1}, 0.1^2)\) from
Kharratzadeh (2017).

\begin{figure}
  \begin{minipage}{.499\textwidth}
\begin{tikzpicture}[x=1pt,y=1pt,scale=0.9]
\definecolor{fillColor}{RGB}{255,255,255}
\path[use as bounding box,fill=fillColor,fill opacity=0.00] (0,0) rectangle (199.17,142.26);
\begin{scope}
\path[clip] (  0.00,  0.00) rectangle (199.17,142.26);
\definecolor{drawColor}{RGB}{255,255,255}
\definecolor{fillColor}{RGB}{255,255,255}

\path[draw=drawColor,line width= 0.6pt,line join=round,line cap=round,fill=fillColor] (  0.00,  0.00) rectangle (199.17,142.26);
\end{scope}
\begin{scope}
\path[clip] ( 36.11, 30.69) rectangle (193.67,136.76);
\definecolor{fillColor}{RGB}{255,255,255}

\path[fill=fillColor] ( 36.11, 30.69) rectangle (193.67,136.76);
\definecolor{drawColor}{RGB}{0,0,0}
\definecolor{fillColor}{RGB}{0,0,0}

\path[draw=mymidblue,line width= 0.4pt,line join=round,line cap=round,fill=mymidblue] ( 43.27,131.94) circle (  1.96);

\path[draw=drawColor,line width= 0.4pt,line join=round,line cap=round,fill=fillColor] (164.77, 76.78) circle (  1.96);

\path[draw=drawColor,line width= 0.6pt,line join=round] ( 43.27,131.94) --
	( 44.72,127.42) --
	( 46.17,123.73) --
	( 47.61,120.78) --
	( 49.06,118.49) --
	( 50.51,116.78) --
	( 51.95,115.56) --
	( 53.40,114.75) --
	( 54.85,114.26) --
	( 56.29,114.01) --
	( 57.74,113.92) --
	( 59.19,113.90) --
	( 60.63,113.87) --
	( 62.08,113.74) --
	( 63.53,113.49) --
	( 64.98,113.14) --
	( 66.42,112.70) --
	( 67.87,112.18) --
	( 69.32,111.60) --
	( 70.76,110.98) --
	( 72.21,110.33) --
	( 73.66,109.66) --
	( 75.10,108.99) --
	( 76.55,108.33) --
	( 78.00,107.70) --
	( 79.44,107.11) --
	( 80.89,106.58) --
	( 82.34,106.11) --
	( 83.78,105.69) --
	( 85.23,105.32) --
	( 86.68,105.02) --
	( 88.12,104.78) --
	( 89.57,104.59) --
	( 91.02,104.47) --
	( 92.46,104.41) --
	( 93.91,104.41) --
	( 95.36,104.47) --
	( 96.81,104.61) --
	( 98.25,104.80) --
	( 99.70,105.04) --
	(101.15,105.30) --
	(102.59,105.57) --
	(104.04,105.83) --
	(105.49,106.04) --
	(106.93,106.19) --
	(108.38,106.26) --
	(109.83,106.22) --
	(111.27,106.06) --
	(112.72,105.75) --
	(114.17,105.28) --
	(115.61,104.61) --
	(117.06,103.77) --
	(118.51,102.78) --
	(119.95,101.70) --
	(121.40,100.57) --
	(122.85, 99.42) --
	(124.29, 98.30) --
	(125.74, 97.25) --
	(127.19, 96.31) --
	(128.63, 95.52) --
	(130.08, 94.93) --
	(131.53, 94.58) --
	(132.98, 94.50) --
	(134.42, 94.72) --
	(135.87, 95.19) --
	(137.32, 95.85) --
	(138.76, 96.64) --
	(140.21, 97.49) --
	(141.66, 98.36) --
	(143.10, 99.18) --
	(144.55, 99.89) --
	(146.00,100.43) --
	(147.44,100.75) --
	(148.89,100.78) --
	(150.34,100.46) --
	(151.78, 99.75) --
	(153.23, 98.62) --
	(154.68, 97.11) --
	(156.12, 95.21) --
	(157.57, 92.95) --
	(159.02, 90.33) --
	(160.46, 87.39) --
	(161.91, 84.12) --
	(163.36, 80.54) --
	(164.81, 76.67) --
	(166.25, 72.52) --
	(167.70, 68.12) --
	(169.15, 63.46) --
	(170.59, 58.64) --
	(172.04, 53.84) --
	(173.49, 49.22) --
	(174.93, 44.99) --
	(176.38, 41.32) --
	(177.83, 38.39) --
	(179.27, 36.40) --
	(180.72, 35.51) --
	(182.17, 35.91) --
	(183.61, 37.80) --
	(185.06, 41.34) --
	(186.51, 46.72);

\path[draw=drawColor,line width= 0.6pt,dash pattern=on 4pt off 4pt ,line join=round] ( 36.11, 76.78) -- (193.67, 76.78);
\end{scope}
\begin{scope}
\path[clip] (  0.00,  0.00) rectangle (199.17,142.26);
\definecolor{drawColor}{RGB}{0,0,0}

\path[draw=drawColor,line width= 0.6pt,line join=round] ( 36.11, 30.69) --
	( 36.11,136.76);
\end{scope}
\begin{scope}
\path[clip] (  0.00,  0.00) rectangle (199.17,142.26);
\definecolor{drawColor}{gray}{0.30}

\node[text=drawColor,anchor=base east,inner sep=0pt, outer sep=0pt, scale=  0.88] at ( 31.16, 49.24) {250};

\node[text=drawColor,anchor=base east,inner sep=0pt, outer sep=0pt, scale=  0.88] at ( 31.16, 73.75) {300};

\node[text=drawColor,anchor=base east,inner sep=0pt, outer sep=0pt, scale=  0.88] at ( 31.16, 98.26) {350};

\node[text=drawColor,anchor=base east,inner sep=0pt, outer sep=0pt, scale=  0.88] at ( 31.16,122.77) {400};
\end{scope}
\begin{scope}
\path[clip] (  0.00,  0.00) rectangle (199.17,142.26);
\definecolor{drawColor}{gray}{0.20}

\path[draw=drawColor,line width= 0.6pt,line join=round] ( 33.36, 52.27) --
	( 36.11, 52.27);

\path[draw=drawColor,line width= 0.6pt,line join=round] ( 33.36, 76.78) --
	( 36.11, 76.78);

\path[draw=drawColor,line width= 0.6pt,line join=round] ( 33.36,101.29) --
	( 36.11,101.29);

\path[draw=drawColor,line width= 0.6pt,line join=round] ( 33.36,125.80) --
	( 36.11,125.80);
\end{scope}
\begin{scope}
\path[clip] (  0.00,  0.00) rectangle (199.17,142.26);
\definecolor{drawColor}{RGB}{0,0,0}

\path[draw=drawColor,line width= 0.6pt,line join=round] ( 36.11, 30.69) --
	(193.67, 30.69);
\end{scope}
\begin{scope}
\path[clip] (  0.00,  0.00) rectangle (199.17,142.26);
\definecolor{drawColor}{gray}{0.20}

\path[draw=drawColor,line width= 0.6pt,line join=round] ( 44.64, 27.94) --
	( 44.64, 30.69);

\path[draw=drawColor,line width= 0.6pt,line join=round] ( 74.08, 27.94) --
	( 74.08, 30.69);

\path[draw=drawColor,line width= 0.6pt,line join=round] (103.51, 27.94) --
	(103.51, 30.69);

\path[draw=drawColor,line width= 0.6pt,line join=round] (132.94, 27.94) --
	(132.94, 30.69);

\path[draw=drawColor,line width= 0.6pt,line join=round] (162.37, 27.94) --
	(162.37, 30.69);

\path[draw=drawColor,line width= 0.6pt,line join=round] (191.81, 27.94) --
	(191.81, 30.69);
\end{scope}
\begin{scope}
\path[clip] (  0.00,  0.00) rectangle (199.17,142.26);
\definecolor{drawColor}{gray}{0.30}

\node[text=drawColor,anchor=base,inner sep=0pt, outer sep=0pt, scale=  0.88] at ( 44.64, 19.68) {0};

\node[text=drawColor,anchor=base,inner sep=0pt, outer sep=0pt, scale=  0.88] at ( 74.08, 19.68) {5};

\node[text=drawColor,anchor=base,inner sep=0pt, outer sep=0pt, scale=  0.88] at (103.51, 19.68) {10};

\node[text=drawColor,anchor=base,inner sep=0pt, outer sep=0pt, scale=  0.88] at (132.94, 19.68) {15};

\node[text=drawColor,anchor=base,inner sep=0pt, outer sep=0pt, scale=  0.88] at (162.37, 19.68) {20};

\node[text=drawColor,anchor=base,inner sep=0pt, outer sep=0pt, scale=  0.88] at (191.81, 19.68) {25};
\end{scope}
\begin{scope}
\path[clip] (  0.00,  0.00) rectangle (199.17,142.26);
\definecolor{drawColor}{RGB}{0,0,0}

\node[text=drawColor,anchor=base,inner sep=0pt, outer sep=0pt, scale=  1.10] at (114.89,  7.64) {$t$};
\end{scope}
\begin{scope}
\path[clip] (  0.00,  0.00) rectangle (199.17,142.26);
\definecolor{drawColor}{RGB}{0,0,0}

\node[text=drawColor,rotate= 90.00,anchor=base,inner sep=0pt, outer sep=0pt, scale=  1.10] at ( 13.08, 83.72) {P/F ratio};
\end{scope}
\begin{scope}
\definecolor{drawColor}{RGB}{0,0,0}

\node[text=mymidblue,anchor=base,inner sep=0pt, outer sep=0pt, scale=  1.10] at ( 59, 131) {$\beta_{0, i}$};
\node[text=myredhighlight,anchor=base,inner sep=0pt, outer sep=0pt, scale=  1.10] at (145, 45) {$\genfrac{}{}{0pt}{}{T_{i}}{(d_{i} = 1)}$};
\node[text=drawColor,anchor=base,inner sep=0pt, outer sep=0pt, scale=  1.10] at (132,110) {${B_{i}(t)}^{\top}\boldsymbol{\zeta}_{i}$};

\draw[dotted,draw=myredhighlight,line width= 0.6pt] (164.77, 76.78) -- (164.77, 30.69);

\end{scope}
\end{tikzpicture}
  \end{minipage}%
  \begin{minipage}{.499\textwidth}
    \input{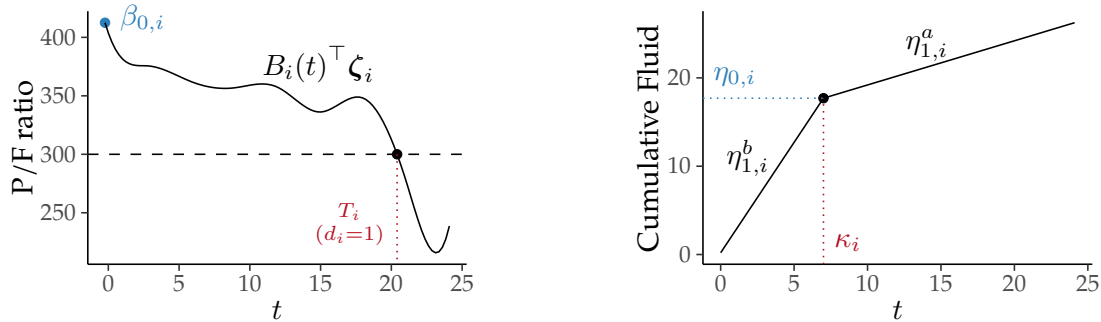}
  \end{minipage}
  \caption{Parameters and form for the P/F ratio submodel ($\pd_{1}$, left) and cumulative fluid submodel ($\pd_{3}$, right).}
  \label{fig:submodel-schematics}
\end{figure}

We identify that a respiratory failure event occurred (which we denote
by \(d_{i} = 1\)) at event time \(T_{i}\) if a solution to the following
optimisation problem exists
\begin{equation}
  T_{i} = \min_{t} \left\{
    \tau_{i} = \beta_{0, i} + B_{i}(t)\boldsymbol{\zeta}_{i}
    \mid
    t \in [\max(0, \min(\boldsymbol{t}_{i})),\, \max(\boldsymbol{t}_{i})]
  \right\},
  \label{eqn:event-time-model-def}
\end{equation}\noindent
We attempt to solve Equation \ref{eqn:event-time-model-def} using a
standard multiple root finder (Soetaert \emph{et al.}, 2020). If there
are no roots then the individual died or was discharged before
respiratory failure occurred so we set \(T_{i} = C_{i}\) and
\(d_{i} = 0\). The relationship between \(T_{i}\) and other model
coefficients is displayed in the left hand panel of Figure
\ref{fig:submodel-schematics}.

\hypertarget{cumulative-fluid-submodel-piecewise-linear-pd_3}{%
\subsection{\texorpdfstring{Cumulative fluid submodel (piecewise linear)
\(\pd_{3}\)}{Cumulative fluid submodel (piecewise linear) \textbackslash pd\_\{3\}}}\label{cumulative-fluid-submodel-piecewise-linear-pd_3}}

The rate of fluid administration reflects the clinical management of
patients by ICU staff, and hence changes to the rate reflect decisions
to change treatment strategy. We employ a breakpoint regression model to
capture the effect of such decisions, and consider only one breakpoint
as this appears sufficient to fit the observed data. Specifically, we
model the 8-hourly cumulative fluid balance data \(x_{i, l}\) (in
litres) at times \(u_{i, l}\), \(l = 1, \ldots, L_{i}\). The cumulative
data are derived from the raw fluid input/output observations, which we
detail in
Appendix~\ref{calculating-the-cumulative-fluid-balance-from-the-raw-fluid-data}.
We denote the complete vector of observations by
\(\boldsymbol{x}_{i} = (x_{i, 1}, \ldots, x_{i, L_{i}})\) and times by
\(\boldsymbol{u}_{i} = (u_{i, 1}, \ldots, u_{i, L_{i}})\).

We assume a piecewise linear model with \(\eta_{0, i}\) as the value at
the breakpoint at time \(\kappa_{i}\), slope \(\eta_{1, i}^{b}\) before
the breakpoint, and slope \(\eta_{1, i}^{a}\) after the breakpoint. We
write this submodel as
\begin{equation}
  \begin{gathered}
    x_{i, l} = m_{i}(u_{i, l}) + \epsilon_{i, l}, \\
    m_{i}(u_{i, l}) =
      \eta_{0, i} +
      \eta^{b}_{1, i}(u_{i, l} - \kappa_{i}) \mathbbm{1}_{\{u_{i, l} < \kappa_{i}\}} +
      \eta^{a}_{1, i}(u_{i, l} - \kappa_{i}) \mathbbm{1}_{\{u_{i, l} \geq \kappa_{i}\}}.
  \end{gathered}
  \label{eqn:piecewise-fluid-model}
\end{equation}\noindent
It will be useful to refer to the fitted value of this submodel at
arbitrary time as \(m_{i}(t)\). We assume a weakly informative prior for
the error term \(\epsilon_{i, l} \sim \text{N}(0, \sigma^{2}_{x, i})\),
with individual-specific error variances
\(\sigma_{x, i} \sim \text{N}_{+}(0, 5^2)\), and specific, informative
priors for the slope before the breakpoint
\(\eta^{b}_{1, i} \sim \text{Gamma}(1.53, 0.24)\) and after
\(\eta^{a}_{1, i} \sim \text{Gamma}(1.53, 0.24)\). An appropriate prior
for \(\kappa_{i}\) and \(\eta_{0, i}\) is challenging to specify due to
the relationship between the two parameters and the individual-specific
support for \(\kappa_{i}\). We address both challenges by
reparameterisation, resulting in a prior for \(\kappa_{i}\) that, in the
absence of other information, places the breakpoint in the middle of an
individual's ICU stay, and a prior for \(\eta_{0, i}\) that captures the
diverse pathways into ICU that an individual can experience. Details and
justifications for all the informative priors are available in
Appendix~\ref{priors-and-justification-for-the-cumulative-fluid-submodel}.
Figure \ref{fig:submodel-schematics} displays the parameters and their
relationship to the fitted regression line.

\hypertarget{survival-submodel-pd_2}{%
\subsection{\texorpdfstring{Survival submodel
\(\pd_{2}\)}{Survival submodel \textbackslash pd\_\{2\}}}\label{survival-submodel-pd_2}}

The rate at which fluid is administered is thought to influence the time
to respiratory failure (Seethala \emph{et al.}, 2017), so we explore
this relationship using a survival model. Individuals experience
respiratory failure (\(d_{i} = 1\)) at time \(0 < t < C_{i}\), or are
censored \((d_{i} = 0, t = C_{i})\). We assume a Weibull hazard with
shape parameter \(\gamma\) for the event times. All individuals have
baseline (time invariant) covariates \(w_{i, a}\), \(a = 1, \ldots, A\),
with \(\boldsymbol{w}_{i} = (1, w_{i, 1}, \ldots, w_{i, A})\)
(i.e.~including an intercept term), and common coefficients
\(\boldsymbol{\theta} = (\theta_{0}, \ldots, \theta_{A})\). The hazard
is assumed to be influenced by these covariates and the rate of increase
\(\frac{\partial}{\partial t} m_{i}(t)\) in the cumulative fluid
balance. The strength of the latter relationship is captured by
\(\alpha\). Hence, the hazard is
\begin{gather}
  h_{i}(t) =
    \gamma t^{\gamma - 1} \exp\left\{
      \boldsymbol{w}_{i}^{\top}\boldsymbol{\theta} + \alpha \frac{\partial}{\partial t} m_{i}(t)
    \right\} \label{eqn:surv-submodel-def-one}\\
  \frac{\partial}{\partial t} m_{i}(t) =
    \eta^{b}_{1, i} \mathbbm{1}_{\{t < \kappa_{i}\}} +
    \eta^{a}_{1, i} \mathbbm{1}_{\{t \geq \kappa_{i}\}}, \label{eqn:surv-submodel-def-two}
\end{gather}\noindent The
survival function at an individual's observed event time and status,
\((T_{i}, d_{i})\), denoted
\(S_{i}(T_{i}) = \exp\{-\int_{0}^{T_{i}}h_{i}(u)\text{d}u\}\), has an
analytic form which we derive in
Appendix~\ref{analytic-form-for-the-survival-function}. Hence the
likelihood for individual \(i\) is \begin{equation}
  \pd(T_{i}, d_{i} \mid \gamma, \boldsymbol{\theta}, \alpha, \kappa_{i}, \eta_{1, i}^{b}, \eta_{1, i}^{a}, \boldsymbol{w}_{i}) = h_{i}(T_{i})^{d_{i}} S_{i}(T_{i}), \\
\end{equation} where we suppress the dependence on the parameters on the
right hand side for brevity.

Our priors, which we justify in Appendix~\ref{p2-prior-justification},
for the submodel specific parameters are
\(\gamma \sim \text{Gamma}(9.05, 8.72)\),
\(\alpha \sim \text{SkewNormal}(0, 0.5, -2)\),
\(\theta_{a} \sim \text{SkewNormal}(0, 0.5, -1)\), and
\(\theta_{0} \sim \text{N}(\hat{E}, 0.5^2)\) where \(\hat{E}\) is the
log of the crude event rate (Brilleman \emph{et al.}, 2020). We adopt
the same priors as the cumulative fluid balance submodel for
\(\kappa_{i}, \eta_{1, i}^{b}\), and \(\eta_{1, i}^{a}\).

\hypertarget{chained-markov-melding-details}{%
\subsection{Chained Markov melding
details}\label{chained-markov-melding-details}}

\begin{figure}

{\centering \includegraphics{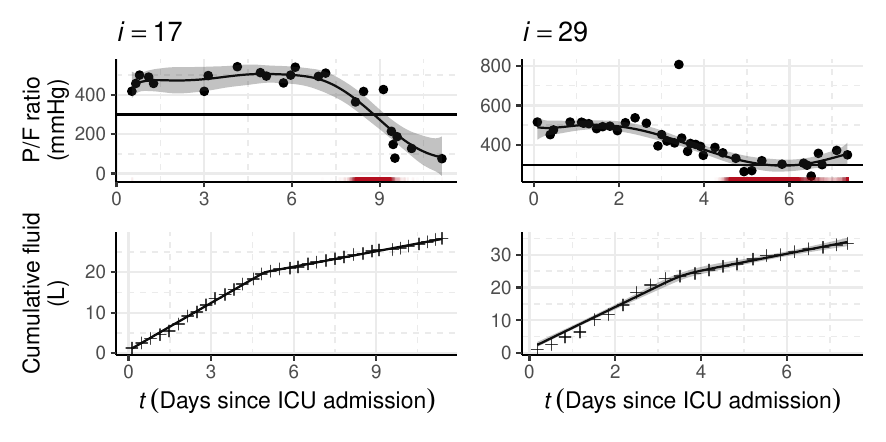} 

}

\caption{The P/F ratio data ($Y_{1}$, top row); cumulative fluid data ($Y_{3}$, bottom row); subposterior means and 95\% credible intervals for each of the submodels (black solid lines and grey intervals); and stage one event times ($T_{i}$, red rug in the top row) for individuals $i = 17$ and $29$.}\label{fig:pf_fit_and_fluid_fit_plot}
\end{figure}

To combine the submodels with chained Markov melding we must define the
common quantities \(\phi_{1 \cap 2}\) and \(\phi_{2 \cap 3}\). We meld
\(\pd_{1}\) and \(\pd_{2}\) by treating the derived event times and
indicators \(\{(T_{i}, d_{i})\}_{i = 1}^{N}\) under \(\pd_{1}\) as the
``response'', i.e.~event times, in \(\pd_{2}\). Care is required when
defining \(\phi_{1 \cap 2}\) under \(\pd_{1}\) as it is a deterministic
function of \(\beta_{0, i}\) and \(\boldsymbol{\zeta}_{i}\). Define
\(\chi_{1, i} = (\beta_{0, i}, \boldsymbol{\zeta}_{i})\) and
\(\phi_{1 \cap 2, i} = f(\chi_{1, i}) = (T_{i}, d_{i})\), where \(f\) is
the output from attempting to solve Equation
\eqref{eqn:event-time-model-def}, so that
\(\phi_{1 \cap 2} = (f(\chi_{1, i}), \ldots, f(\chi_{1, N}))\). The
parameters shared by Equations \eqref{eqn:piecewise-fluid-model} and
\eqref{eqn:surv-submodel-def-two} constitute
\(\phi_{2 \cap 3} = (\eta^{b}_{1, i}, \eta^{a}_{1, i}, \kappa_{i})_{i = 1}^{N}\).

To completely align with our chained melding notation we also define,
for the P/F submodel,
\(Y_{1} = (\boldsymbol{z}_{i}, \boldsymbol{t}_{i})_{i = 1}^{N}\) and
\(\psi_{1} = (\omega_{i})_{i = 1}^{N}\), noting that \(\psi_{1}\) and
\((\chi_{1, i}, \ldots, \chi_{1, N})\) have no components in common. For
the cumulative fluid submodel we define
\(Y_{3} = (\boldsymbol{x}_{i}, \boldsymbol{u}_{i})_{i = 1}^{N}\), and
\(\psi_{3} = (\eta_{0, i}, \sigma^{2}_{x, i})_{i = 1}^{N}\). Finally,
for the survival submodel we define
\(Y_{2} = (\boldsymbol{w}_{i})_{i = 1}^{N}\) and
\(\psi_{2} = (\gamma, \boldsymbol{\theta}, \alpha)\).

\hypertarget{pooling-and-estimation}{%
\subsection{Pooling and estimation}\label{pooling-and-estimation}}

We consider logarithmic pooling with
\(\lambda = (\frac{4}{5}, \frac{4}{5}, \frac{4}{5})\) (any smaller value
of \(\lambda\) results in a prior that is so uninformative that it
causes computational problems) and with \(\lambda = (1, 1, 1)\)
(Product-of-Experts). Because the correlation between
\(\phi_{1 \cap 2}\) and \(\phi_{2 \cap 3}\) in
\(\pd_{2}(\phi_{1 \cap 2}, \phi_{2 \cap 3})\) is important, we do not
consider linear pooling in this example. Logarithmic pooling requires
estimates of \(\pd_{1}(\phi_{1 \cap 2})\) and
\(\pd_{2}(\phi_{1 \cap 2}, \phi_{2 \cap 3})\). Because these are mixed
distributions, with both discrete and continuous components, standard
kernel density estimation, as suggested by Goudie \emph{et al.} (2019),
is inappropriate. Instead, we fit appropriate parametric mixture
distributions to the unknown prior marginal distributions. Details for
all the parametric mixture distribution estimates are contained in
Appendix~\ref{estimating-submodel-prior-marginal-distributions}.

We use the parallel multi-stage sampler with
\(\pd_{\text{pool}, 1}(\phi_{1 \cap 2}) = \pd_{1}(\phi_{1 \cap 2}), \,\, \pd_{\text{pool}, 3}(\phi_{2 \cap 3}) = \pd_{3}(\phi_{2 \cap 3})\)
and
\(\pd_{\text{pool}, 2}(\phi_{1 \cap 2}, \phi_{2 \cap 3}) = \pd_{\text{pool}}(\boldsymbol{\phi}) \mathop{/} \left(\pd_{1}(\phi_{1 \cap 2}) \pd_{3}(\phi_{2 \cap 3}) \right)\).
That is, in stage one we target the subposteriors
\(\pd_{1}(\phi_{1 \cap 2}, \psi_{1} \mid Y_{1})\) and
\(\pd_{3}(\phi_{2 \cap 3}, \psi_{3} \mid Y_{3})\); in stage two we
target the full melded model. Targeting
\(\pd_{1}(\phi_{1 \cap 2}, \psi_{1} \mid Y_{1})\) in stage one
alleviates the need to solve Equation \eqref{eqn:event-time-model-def}
within an MCMC iteration, instead turning the production of
\(\phi_{1 \cap 2}\) into an embarrassingly parallel, post-stage-one
processing step. Attempting to sample the melded posterior directly
would involve solving \eqref{eqn:event-time-model-def} many times within
each iteration, presenting a sizeable computational hurdle which we
avoid. It is crucial for the convergence of our multi-stage sampler that
the components of \(\phi_{1 \cap 2}\) and \(\phi_{2 \cap 3}\) are
updated \emph{individual-at-a-time} in stage two. This is possible due
to the conditional independence between individuals in the stage one
posterior, and Appendix~\ref{one-at-a-time} contains the details of this
scheme. The stage one subposteriors are sampled using \texttt{Stan},
using 5 chains with \(10^{3}\) warm-up iterations and \(10^{4}\) post
warm-up iterations. We use \texttt{Stan} to sample \(\psi_{2}\) where,
in every MH-within-Gibbs step, we run \texttt{Stan} for 9 warm-up
iterations and 1 post warm-up iteration\footnote{We also initialise Stan
  at the previous value of \(\psi_{2}\), and disable all adaptive
  procedures as the default (identity) mass matrix and step size are
  suitable for this example.}. We run 5 chains of \(10^{4}\) iterations
for all stage two targets. Visual and numerical diagnostics (Vehtari
\emph{et al.}, 2020) are assessed and are available in the repository
accompanying this paper\footnote{\url{https://doi.org/10.5281/zenodo.6552714}}.

\hypertarget{results-1}{%
\subsection{Results}\label{results-1}}

We first inspect the subposterior fitted values for \(\pd_{1}\) and
\(\pd_{3}\). The top row of Figure \ref{fig:pf_fit_and_fluid_fit_plot}
displays the P/F data, the fitted submodel, and derived event times for
individuals \(i = 17\) and \(29\). The spline appears to fit the raw P/F
data well, with the heavy tailed error term accounting for the larger
deviations away from the fitted value. It is interesting to see the
relatively wide, multimodal distribution for \((T_{29}, d_{29})\) (there
is a second mode at \((T_{29} = C_{29}, d_{29} = 0)\) and for other
individuals not shown here). The bottom row of Figure
\ref{fig:pf_fit_and_fluid_fit_plot} displays the cumulative fluid data
and the fitted submodel, with the little noise in the data resulting in
minimal uncertainty about the fitted value and a concentrated
subposterior distribution.

To assess the importance of fully accounting for the uncertainty in
\(\phi_{1 \cap 2}\) and \(\phi_{2 \cap 3}\), we compare the posterior
for \(\psi_{2}\) obtained using the chained melding approach with the
posterior obtained by fixing \(\phi_{1 \cap 2}\) and
\(\phi_{2 \cap 3}\). Plugging in a point estimate reflects common
applied statistical practice when combining submodels, particularly when
a distributional approximation is difficult to obtain (as it is for
\(\pd_{1}(\phi_{1 \cap 2} \mid Y_{1})\)). Additionally, standard
survival models and software typically do not permit uncertainty in
event times and indicators, rendering such a plug-in approach necessary.

\begin{figure}

{\centering \includegraphics{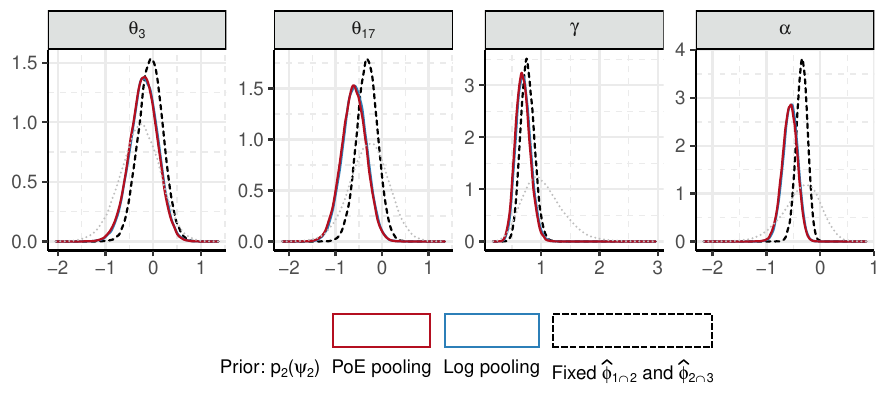} 

}

\caption{Density estimates for a subset of $\psi_{2}$. The submodel marginal prior $\pd_{2}(\psi_{2})$ is shown as the grey dotted line (note that this is not the marginal prior under the melded model). The figure also contains the subposteriors obtained from chained melding using PoE pooling (red, solid line) and logarithmic pooling (blue, solid line), as well as the posterior using the fixed values $\pd(\psi_{2} \mid \widehat{\phi}_{1 \cap 2}, \widehat{\phi}_{2 \cap 3}, Y_{2})$ (black, dashed line).}\label{fig:psi_2_comparison_plot}
\end{figure}

Specifically, we fix \(\phi_{1 \cap 2}\) to the median value\footnote{For
  each individual the samples of \((T_{i}, d_{i})_{i = 1}^{N}\)-pairs
  are sorted by \(T_{i}\), and the
  \(\lfloor \frac{N}{2}\rfloor\)\textsuperscript{th} tuple
  \((\widehat{T}_{i}, \widehat{d}_{i})\) is chosen as the median.} for
each individual under \(\pd_{1}(\phi_{1 \cap 2} \mid Y_{1})\) and denote
it \(\widehat{\phi}_{1 \cap 2}\), and use the subposterior mean of
\(\pd_{3}(\phi_{2 \cap 3} \mid Y_{3})\) denoted
\(\widehat{\phi}_{2 \cap 3}\). With these fixed values we sample
\(\pd(\psi_{2} \mid \widehat{\phi}_{1 \cap 2}, \widehat{\phi}_{2 \cap 3}, Y_{2})\).
We also compare the melded posterior to the submodel marginal prior
\(\pd_{2}(\psi_{2})\), but we note that this comparison is difficult to
interpret, as the melding process alters the prior for \(\psi_{2}\).
Figure \ref{fig:psi_2_comparison_plot} displays the aforementioned
densities for
\((\theta_{3}, \theta_{17}, \gamma, \alpha) \subset \psi_{2}\), with
\((\theta_{3}, \theta_{17})\) chosen as they exhibit the greatest
sensitivity to the fixing of \(\phi_{1 \cap 2}\) and
\(\phi_{2 \cap 3}\). For the baseline coefficients
(\(\theta_{3}, \theta_{17}\)) the chained melding posterior differs
slightly in location from
\(\pd(\psi_{2} \mid \widehat{\phi}_{1 \cap 2}, \widehat{\phi}_{2 \cap 3}, Y_{2})\),
with a small increase in uncertainty. A more pronounced change is
visible for \(\alpha\), where the melding process has added a notable
degree of uncertainty and shifted the posterior leftwards.

\begin{figure}

{\centering \includegraphics{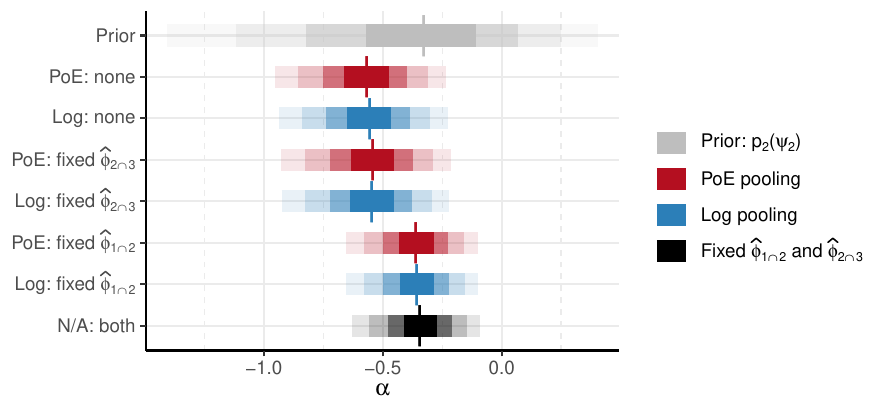} 

}

\caption{Median (vertical line), 50\%, 80\%, 95\%, and 99\% credible intervals (least transparent to most transparent) for $\alpha$. The marginal prior (grey, top row) and posterior using fixed $\widehat{\phi}_{1 \cap 2}$ and $\widehat{\phi}_{2 \cap 3}$ (black, bottom row) are as in Figure \ref{fig:psi_2_comparison_plot}. For the chained melded posteriors (red and blue, rows 2 and 3) and the melded posteriors (red and blue, rows 4 -- 7), the tick label on the y-axis denotes the type of pooling used, and which of $\phi_{1 \cap 2}$ and/or $\phi_{2 \cap 3}$ are fixed.}\label{fig:alpha_only_comparision_plot}
\end{figure}

To investigate which part of the melding process causes this change in
the posterior of \(\alpha\), we consider fixing either one of
\(\phi_{1 \cap 2}\) and \(\phi_{2 \cap 3}\) to their respective point
estimates. That is, we employ Markov melding as described in Section
\ref{markov-melding}, using either logarithmic or PoE pooling, to obtain
\(\pd_{\text{meld}}(\alpha \mid \widehat{\phi}_{1 \cap 2}, Y_{2}, Y_{3})\)
and
\(\pd_{\text{meld}}(\alpha \mid \widehat{\phi}_{2 \cap 3}, Y_{1}, Y_{2})\).
Figure \ref{fig:alpha_only_comparision_plot} displays the same
distributions for \(\alpha\) as Figure \ref{fig:psi_2_comparison_plot},
and adds the posteriors obtained using one fixed value
(\(\widehat{\phi}_{1 \cap 2}\) or \(\widehat{\phi}_{2 \cap 3}\)) whilst
melding the other non-fixed parameter.

Evident for both choices of pooling is the importance of incorporating
the uncertainty in \(\phi_{1 \cap 2}\). This is expected given the large
uncertainty and multimodal nature of \(\phi_{1 \cap 2}\) compared to
\(\phi_{2 \cap 3}\) (see Figure \ref{fig:pf_fit_and_fluid_fit_plot}). We
suspect that it is the multimodality in
\(\pd_{1}(\phi_{1 \cap 2} \mid Y_{1})\) that produces the shift in
posterior mode of \(\phi_{1 \cap 2}\), with the width of
\(\pd_{1}(\phi_{1 \cap 2} \mid Y_{1})\) affecting the increase in
uncertainty. Because we prefer the chained melded posterior, under
either pooling method, for its full accounting of uncertainty we
conclude that
\(\pd(\alpha \mid \widehat{\phi}_{1 \cap 2}, \widehat{\phi}_{2 \cap 3}, Y_{2})\)
is both overconfident and biased.

\begin{figure}

{\centering \includegraphics{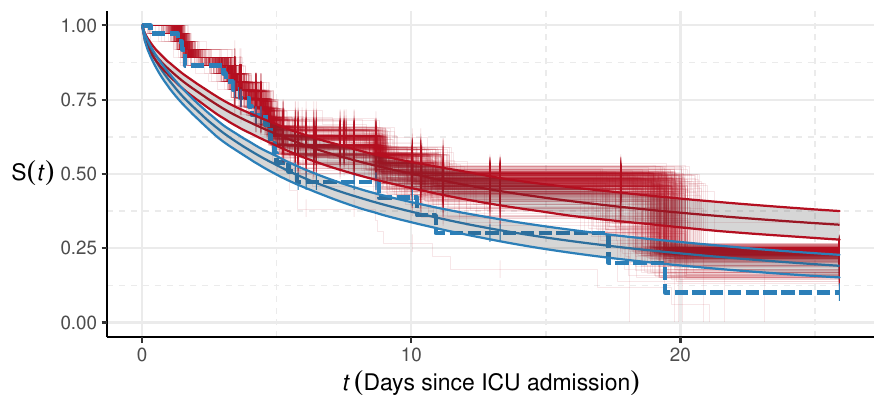} 

}

\caption{Survival curves and mean survival function at time $t$. The red, stepped lines are draws of $\phi_{1 \cap 2}$ from the melded posterior using PoE pooling, converted into survival curves via the Kaplan-Meier estimator. The smooth red line and interval (posterior mean and 95\% credible interval) denote the model-based, mean survival function obtained from the melded posterior (PoE pooling) values of $\psi_{2}$ and $\phi_{2 \cap 3}$. The blue dashed line is the Kaplan-Meier estimate of $\widehat{\phi}_{1 \cap 2}$, and the blue solid line and interval are the corresponding model-based estimate from $\pd(\psi_{2} \mid \widehat{\phi}_{1 \cap 2}, \widehat{\phi}_{2 \cap 3}, Y_{2})$.}\label{fig:kap_meier_pc_plot}
\end{figure}

The marginal changes to the components of \(\psi_{2}\) visible in Figure
\ref{fig:psi_2_comparison_plot} appear small, however the cumulative
effect of such changes becomes apparent when inspecting the posterior of
the survival function. Figure \ref{fig:kap_meier_pc_plot} displays the
model-based, mean survival function under the melded posterior (using
PoE pooling), and corresponding draws of \(\phi_{1 \cap 2}\) converted
into survival curves using the Kaplan-Meier estimator. Also shown are
the Kaplan-Meier estimate of \(\widehat{\phi}_{1 \cap 2}\) and the mean
survival function computed using
\(\pd(\psi_{2} \mid \widehat{\phi}_{1 \cap 2}, \widehat{\phi}_{2 \cap 3}, Y_{2})\).
The posterior survival functions differ markedly, with the 95\%
intervals overlapping only for small values of time. It is also
interesting to see that \(\widehat{\phi}_{1 \cap 2}\), despite being a
reasonable point estimate of \(\pd_{1}(\phi_{1 \cap 2} \mid Y_{1})\), is
not very likely under the melded posterior. Figure
\ref{fig:kap_meier_pc_plot} also suggests that the Weibull hazard is
insufficiently flexible for this example. We discuss the complexities of
other hazards in Section \ref{conclusion}.

\hypertarget{conclusion}{%
\section{Conclusion}\label{conclusion}}

This paper introduces the chained Markov melded model. In doing so we
make explicit the notion of submodels related in a chain-like way,
describe a generic methodology for joining together any number of such
submodels and illustrate its application with our examples. Our examples
also demonstrate the importance of quantifying the uncertainty when
joining submodels; not doing so can produce biased, over-confident
inference. We also present the choices, and their impacts, that users of
chained Markov melding must make which include: the choice of pooling
function, and where required the pooling weights; the choice of
posterior sampler and the design thereof, including the apportionment of
the pooled prior over the stages and stage-specific MCMC techniques.

We have introduced extensions to linear and logarithmic pooling to
marginals of different but overlapping quantities. Linear pooling,
introduced in Section \ref{pooled-prior}, could be extended to induce
dependence between the components of \(\boldsymbol{\phi}\) using
multivariate or vine copulas (Kurowicka and Joe, 2011; Nelsen, 2006), or
other techniques (Lin \emph{et al.}, 2014). Copula methods are
particularly appealing as, depending on the choice of copula, they yield
computationally cheap to evaluate expressions for the density function,
are easy to sample, and induce correlation between an arbitrary number
of marginals.

Our parallel multi-stage sampler currently only considers \(\Nm = 3\)
submodels, rather than the fully generic definition of chained Markov
melding in Equation \eqref{eqn:melded-model-general}. Whilst we
anticipate needing more complex methods in large \(\Nm\) settings, the
value of \(\Nm\) at which the performance of our multi-stage sampler
becomes unacceptable will depend on the specific submodels and data
under consideration. A general method would consider a large and
arbitrary number of submodels in a chain, and initially split the chain
into more/fewer pieces depending on the computational resources
available. Designing such a method is complex, as it would have to:

\begin{itemize}[nosep]
\tightlist
\item avoid requiring the inverse of any component of $\boldsymbol{\phi}$ with a noninvertible definition,
\item estimate the relative cost of sampling each submodel's subposterior, to split the chain of submodels into steps/jobs of approximately the same computational cost,
\item decide the order in which pieces of the chain are combined.
\end{itemize}

\noindent These are substantial challenges. It may be possible to use
combine the ideas in Lindsten \emph{et al.} (2017) and Kuntz \emph{et
al.} (2021), who propose a parallel Sequential Monte Carlo method, with
the aforementioned constraints to obtain a generic methodology. Ideally
we would retain the ability to use existing implementations of the
submodels, however the need to recompute the weights of the particles,
and hence reevaluate previously considered submodels, may preclude this
requirement. Our current sampler is also sensitive to large differences
in location or scale of the target distribution between the stages. The
impact of these differences can be ameliorated using the methodology of
Manderson and Goudie (2022), and, more generally, Sequential Monte Carlo
samplers are likely to perform better in these settings.

Our chained Markov melding methodology is general and permits any form
of uncertainty in the common quantities. In Section
\ref{survival-analysis-with-time-varying-covariates-and-uncertain-event-times-1}
we use our chained melded model to incorporate uncertainty in the event
times and indicators into a survival submodel. Some specific forms of
uncertainty in the event times have been considered in previous work.
These include Wang \emph{et al.} (2020), who consider uncertain event
times arising from record linkage, where the event time is assumed to be
one of a finite number of event times arising from the record linkage;
and Giganti \emph{et al.} (2020), Oh \emph{et al.} (2018), and Oh
\emph{et al.} (2021), who leverage external validation data to account
for measurement error in the event time. However, the general and
Bayesian nature of our methodology readily facilitates any form of
uncertainty in the event times and the event indicators; uncertainty in
the latter is not considered in the cited papers.

The example in Section
\ref{survival-analysis-with-time-varying-covariates-and-uncertain-event-times-1}
has three more interesting aspects to discuss. Firstly, the P/F ratio
data used in the first submodel is obtained by finding all blood gas
measurements from arterial blood samples. Approximately \(20\%\) of the
venous/arterial labels are missing. In these instances a logistic
regression model, fit by the MIMIC team\footnote{The coefficients,
  classification threshold, and the imputation used in the case of
  missing data are supplied in the \texttt{blood-gasses.sql} file in the
  GitHub repository accompanying this paper. No other information is
  available about this model (the data used to produce the coefficients,
  and the performance of the fitted model).}, is used to predict the
missing label based on other covariates. It is theoretically possible to
refit the model in a Bayesian framework and use the chained melded model
to incorporate the uncertainty in the predicted sample label -- adding
another `link' to the chain.

Secondly, the application of our multi-stage sampler to this example is
similar to the two-stage approach used for joint longitudinal and
time-to-event models (see Mauff \emph{et al.}, 2020 for a description of
this approach). In the two-stage approach, the longitudinal model is fit
using standard MCMC methods in stage one, and the samples are reused in
stage two when considering the time-to-event data. This can
significantly reduce the computational effort required to fit the joint
model. However, unlike our multi-stage sampler, the typical two-stage
approach does not target the full posterior distribution, which can lead
to biased estimates (though Mauff \emph{et al.} (2020) extend the
typical two-stage approach to reduce this bias).

Thirdly, we observe a lack of flexibility the baseline hazard, visible
in Figure \ref{fig:kap_meier_pc_plot}. More complex hazards could be
employed, e.g.~modelling the (log-)hazard using a (penalised) B-spline
(Rosenberg, 1995; Royston and Parmar, 2002; Rutherford \emph{et al.},
2015). However, this increased flexibility precludes an analytic form
for the survival function. Whilst numerical integration is possible it
is not trivial, particularly when the hazard is discontinuous, as our
hazard is at the breakpoint. Splines also have more coefficients than
the single parameter of the Weibull hazard. Identifiability issues arise
with a small number of individuals, many of whom are censored, and are
compounded when there are a relatively large number of other parameters
\((\alpha, \boldsymbol{\theta})\). Whilst we do not believe these costs
are worth incurring for our example, for settings with a larger number
of patients and more complicated longitudinal submodels the increased
flexibility may be vital.

\hypertarget{acknowledgements}{%
\subsection*{Acknowledgements}\label{acknowledgements}}
\addcontentsline{toc}{subsection}{Acknowledgements}

We thank Sarah L Cowan for assistance in understanding respiratory
failure, and Anne Presanis and Brian Tom for many helpful discussions
about the methodological aspects of this work. We also thank Luiz Max
Carvalho for comments on an earlier version of this paper. This work was
supported by The Alan Turing Institute under the UK Engineering and
Physical Sciences Research Council (EPSRC) {[}EP/N510129/1{]} and the UK
Medical Research Council {[}programme code MC\_UU\_00002/2{]}.

\renewcommand{\thesubsection}{\Alph{subsection}}
\setcounter{subsection}{0}

\hypertarget{appendices}{%
\section*{Appendices}\label{appendices}}
\addcontentsline{toc}{section}{Appendices}

\hypertarget{log-pooling-gaussian-densities}{%
\subsection{Log pooling Gaussian
densities}\label{log-pooling-gaussian-densities}}

We can exactly compute \(\pd_{\text{pool}}\) when logarithmically
pooling Gaussian densities. Noting that, in the one dimensional case,
\(\text{N}(\phi; \mu, \sigma^2)^{\lambda_{\modelindex}} = \text{N}(\phi; \mu, \frac{\sigma^2}{\lambda_{\modelindex}})\),
we use the results of Bromiley (2003) and write
\begin{gather}
  \Omega_{1} = \left(
    \begin{bmatrix}
      \lambda_{1}^{-1} & 0 \\
      0 & \lambda_{3}^{-1}
    \end{bmatrix}
    \begin{bmatrix}
      \sigma_{1}^{2} & 0 \\
      0 & \sigma_{3}^{2}
    \end{bmatrix}
  \right)^{-1},
  \;\;
  \Omega_{2} = \left(
    \begin{bmatrix}
      \lambda_{2}^{-1} & 0 \\
      0 & \lambda_{2}^{-1}
    \end{bmatrix}
    \begin{bmatrix}
      \sigma_{2}^{2} & \rho\sigma_{2}^{2} \\
      \rho\sigma_{2}^{2} & \sigma_{2}^{2}
    \end{bmatrix}
  \right)^{-1} \label{eqn:log-pooling-gaussian-one} \\
  \Sigma_{\text{log}} = 
  \left(\Omega_{1} + \Omega_{2}\right)^{-1}, \quad
  \mu_{\text{log}} = 
  \Sigma_{\text{log}} 
  \left(
    \Omega_{1}
    \begin{bmatrix}
      \mu_{1} \\
      \mu_{3}
    \end{bmatrix}
    +
    \Omega_{2}
    \begin{bmatrix}
      \mu_{2, 1} \\
      \mu_{2, 1}
    \end{bmatrix}
  \right),
  \label{eqn:log-pooling-gaussian-two}
\end{gather}\noindent
hence
\(\pd_{\text{pool}}(\phi_{1 \cap 2}, \phi_{2 \cap 3}) = \text{N}((\phi_{1 \cap 2}, \phi_{2 \cap 3});\, \mu_{\text{log}}, \, \Sigma_{\text{log}})\).
The choice of \(\lambda_{2}\) is critical; by controlling the
contribution of \(\pd_{2}\) to \(\pd_{\text{pool}}\), \(\lambda_{2}\)
controls the degree of correlation present in the latter. The left hand
column of Figure \ref{fig:pooled_densities_plot} illustrates this
phenomena. When
\(\lambda_{1} = \lambda_{3} = 0 \implies \lambda_{2} = 1\), all
correlation in \(\pd_{2}\) is present in \(\pd_{\text{pool}}\). The
correlation decreases for increasing values of \(\lambda_{1}\) until
\(\lambda_{1} = \lambda_{3} = 0.5 \implies \lambda_{2} = 0\), where no
correlation persists.

\hypertarget{sequential-sampler}{%
\subsection{Sequential sampler}\label{sequential-sampler}}

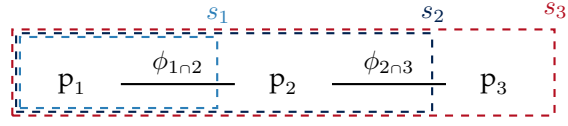
\begin{figure}[htb]
  \centering
  \begin{tikzpicture}[> = stealth, node distance = 1.5cm, thick, state/.style={draw = none, minimum width = 1.25cm}]
  \node [state] (model-1) {$\pd_{1}$};
  \node [state] [right = of model-1](model-2) {$\pd_{2}$};
  \node [state] [right = of model-2](model-3) {$\pd_{3}$};

  \path[-] (model-1) edge node[above] (phi-12) {$\phi_{1 \cap 2}$} (model-2);
  \path[-] (model-2) edge node[above] (phi-23) {$\phi_{2 \cap 3}$} (model-3);

  \node (stage-1) [draw = mymidblue, fit = (model-1) (phi-12), inner sep = 0.05cm, dashed] {};
  \node (stage-1-label) [yshift = 1.75ex, mymidblue] at (stage-1.north east) {$s_{1}$};

  \node (stage-2) [draw = mydarkblue, fit = (model-1) (model-2) (phi-12) (phi-23), inner sep = 0.1cm, dashed] {};
  \node (stage-2-label) [yshift = 1.5ex, mydarkblue] at (stage-2.north east) {$s_{2}$};

  \node (stage-3) [draw = myredhighlight, fit = (model-1) (model-2) (model-3) (phi-12) (phi-23), inner sep = 0.15cm, dashed] {};
  \node (stage-3-label) [yshift = 1.5ex, myredhighlight] at (stage-3.north east) {$s_{3}$};
  \end{tikzpicture}
  \caption{A DAG of the submodels and their common quantities, with the sequential sampling strategy overlaid. The stage one target ($s_{1}$) is encapsulated in the light blue dashed line; stages two and three ($s_{2}, s_{3}$) are in dark blue and red respectively.}
  \label{fig:seq-sampler-dag}
\end{figure}

Figure \ref{fig:seq-sampler-dag} depicts graphically the strategy
employed by the sequential sampler. The sequential sampler assumes that
the pooled prior decomposes such that
\begin{equation}
  \pd_{\text{pool}}(\boldsymbol{\phi}) = 
    \pd_{\text{pool}, 1}(\phi_{1 \cap 2})
    \pd_{\text{pool}, 2}(\phi_{1 \cap 2}, \phi_{2 \cap 3})
    \pd_{\text{pool}, 3}(\phi_{1 \cap 2}, \phi_{2 \cap 3}).
  \label{eqn:sequential-sampler-decomposition}
\end{equation}\noindent
This is necessary to avoid sampling all components of
\(\boldsymbol{\phi}\) in the first stage. All pooled priors trivially
satisfy \eqref{eqn:sequential-sampler-decomposition}, as we can assume
all but \(\pd_{\text{pool}, 3}(\phi_{1 \cap 2}, \phi_{2 \cap 3})\) are
improper, flat distributions. However, including some portion of the
pooled prior in each stage of the sampler can improve performance, and
eliminate computational instabilities when submodel likelihoods contain
little information.

\hypertarget{stage-one-1}{%
\subsubsection{Stage one}\label{stage-one-1}}

Stage one of the sequential sampler targets
\begin{equation}
  \pd_{\text{meld}, 1} (\phi_{1 \cap 2}, \psi_{1} \mid Y_{1}) \propto
  \pd_{\text{pool}, 1}(\phi_{1 \cap 2})
  \frac {
    \pd_{1}(\phi_{1 \cap 2}, \psi_{1}, Y_{1})
  } {
    \pd_{1}(\phi_{1 \cap 2})
  },
  \label{eqn:stage-one-target}
\end{equation}\noindent
using a generic proposal kernel for both \(\phi_{1 \cap 2}\) and
\(\psi_{1}\). The corresponding acceptance probability for a proposed
update from \((\phi_{1 \cap 2}, \psi_{1})\) to
\((\phi_{1 \cap 2}^{*}, \psi_{1}^{*})\) is
\begin{multline}
  \alpha((\phi_{1 \cap 2}^{*}, \psi_{1}^{*}), (\phi_{1 \cap 2}, \psi_{1})) = \\
  \frac {
    \pd_{\text{pool}, 1}(\phi_{1 \cap 2}^{*})
  } {
    \pd_{\text{pool}, 1}(\phi_{1 \cap 2})
  }
  \frac {
    \pd_{1}(\phi_{1 \cap 2}^{*}, \psi_{1}^{*}, Y_{1})
    \pd_{1}(\phi_{1 \cap 2})
  } {
    \pd_{1}(\phi_{1 \cap 2}, \psi_{1}, Y_{1})
    \pd_{1}(\phi_{1 \cap 2}^{*})
  }
  \frac {
    \q(\phi_{1 \cap 2}, \psi_{1} \mid \phi_{1 \cap 2}^{*}, \psi_{1}^{*})
  } {
    \q(\phi_{1 \cap 2}^{*}, \psi_{1}^{*} \mid \phi_{1 \cap 2}, \psi_{1})
  }.
  \label{eqn:stage-one-acceptance-probability}
\end{multline}

\hypertarget{stage-two-1}{%
\subsubsection{Stage two}\label{stage-two-1}}

The stage two target augments the stage one target by including the
second submodel, corresponding prior marginal distribution, and an
additional pooled prior term
\begin{multline}
  \pd_{\text{meld}, 2} (\phi_{1 \cap 2}, \phi_{2 \cap 3}, \psi_{1}, \psi_{2} \mid Y_{1}, Y_{2}) \propto \\
  \pd_{\text{pool}, 1}(\phi_{1 \cap 2})
  \pd_{\text{pool}, 2}(\phi_{1 \cap 2}, \phi_{2 \cap 3})
  \frac {
    \pd_{1}(\phi_{1 \cap 2}, \psi_{1}, Y_{1})
  } {
    \pd_{1}(\phi_{1 \cap 2})
  }
  \frac {
    \pd_{2} (\phi_{1 \cap 2}, \phi_{2 \cap 3}, \psi_{2}, Y_{2})
  } {
    \pd_{2}(\phi_{1 \cap 2}, \phi_{2 \cap 3})
  }.
  \label{eqn:stage-two-target}
\end{multline}\noindent
A Metropolis-within-Gibbs strategy is employed, where the stage one
samples are used as a proposal for \(\phi_{1 \cap 2}\), whilst a generic
proposal kernel is used for \(\psi_{2}\) and \(\phi_{2 \cap 3}\). Thus
the proposal distributions for \(\phi_{1 \cap 2}^{*}\) and
\((\phi_{2 \cap 3}^{*}, \psi_{2}^{*})\) are
\begin{align}
  \phi_{1 \cap 2}^{*}, \psi_{1}^{*} \mid \phi_{2 \cap 3}, \psi_{2} &\sim \pd_{\text{meld}, 1}(\phi_{1 \cap 2}^{*}, \psi_{1}^{*} \mid Y_{1}) \\
  \phi_{2 \cap 3}^{*}, \psi_{2}^{*} \mid \phi_{1 \cap 2}, \psi_{1} &\sim \q(\phi_{2 \cap 3}^{*}, \psi_{2}^{*} \mid \phi_{2 \cap 3}, \psi_{2}).
  \label{eqn:stage-two-gibbs-updates}
\end{align}\noindent
The acceptance probability for this proposal strategy is
\begin{align}
  \alpha & \!\! \begin{multlined}[t]
     \left((\phi_{1 \cap 2}^{*}, \psi_{1}^{*}), (\phi_{1 \cap 2}, \psi_{1})\right) = \\
    \frac {
      \pd_{\text{pool}, 2}(\phi_{1 \cap 2}^{*}, \phi_{2 \cap 3})
    } {
      \pd_{\text{pool}, 2}(\phi_{1 \cap 2}, \phi_{2 \cap 3})
    }
    \frac {
      \pd_{2}(\phi_{1 \cap 2}^{*}, \phi_{2 \cap 3}, \psi_{2}, Y_{2})
      \pd_{2}(\phi_{1 \cap 2}, \phi_{2 \cap 3})
    } {
      \pd_{2}(\phi_{1 \cap 2}, \phi_{2 \cap 3}, \psi_{2}, Y_{2})
      \pd_{2}(\phi_{1 \cap 2}^{*}, \phi_{2 \cap 3})
    }
  \label{eqn:stage-two-acceptance-probabilities-one}
  \end{multlined} \\[1.5ex]
  \alpha & \!\! \begin{multlined}[t]
    \left((\phi_{2 \cap 3}^{*}, \psi_{2}^{*}), (\phi_{2 \cap 3}, \psi_{2})\right) = \\
    \frac {
      \pd_{\text{pool}, 2}(\phi_{1 \cap 2}, \phi_{2 \cap 3}^{*})
    } {
      \pd_{\text{pool}, 2}(\phi_{1 \cap 2}, \phi_{2 \cap 3})
    }
    \frac {
      \pd_{2}(\phi_{1 \cap 2}, \phi_{2 \cap 3}^{*}, \psi_{2}^{*}, Y_{2})
      \pd_{2}(\phi_{1 \cap 2}, \phi_{2 \cap 3})
    } {
      \pd_{2}(\phi_{1 \cap 2}, \phi_{2 \cap 3}, \psi_{2}, Y_{2})
      \pd_{2}(\phi_{1 \cap 2}, \phi_{2 \cap 3}^{*})
    }
    \frac {
      \q(\phi_{2 \cap 3}, \psi_{2} \mid \phi_{2 \cap 3}^{*}, \psi_{2}^{*})
    } {
      \q(\phi_{2 \cap 3}^{*}, \psi_{2}^{*} \mid \phi_{2 \cap 3}, \psi_{2})
    }.
  \label{eqn:stage-two-acceptance-probabilities-two}
  \end{multlined}
\end{align}\noindent
Our judicious choice of proposal distribution has resulted in a
cancellation in Equation
\eqref{eqn:stage-two-acceptance-probabilities-one} which removes all
terms related to \(\pd_{1}\). Similarly, all terms related to
\(\pd_{1}\) are constant -- hence cancel -- in Equation
\eqref{eqn:stage-two-acceptance-probabilities-two}. This eliminates any
need to re-evaluate the first submodel.

\hypertarget{stage-three}{%
\subsubsection{Stage three}\label{stage-three}}

In stage three we target the full melded posterior
\begin{multline}
  \pd_{\text{meld}, 3} (\phi_{1 \cap 2}, \phi_{2 \cap 3}, \psi_{1}, \psi_{2}, \psi_{3} \mid Y_{1}, Y_{2}, Y_{3}) \propto \\ \pd_{\text{meld}} (\phi_{1 \cap 2}, \phi_{2 \cap 3}, \psi_{1}, \psi_{2}, \psi_{3} \mid Y_{1}, Y_{2}, Y_{3}).
  \label{eqn:stage-three-target}
\end{multline}\noindent
The target has now been broadened to include terms from the third
submodel and the entirety of the pooled prior. Again, we employ a
Metropolis-within-Gibbs sampler, with proposals drawn such that
\begin{align}
  \phi_{1 \cap 2}^{*}, \phi_{2 \cap 3}^{*}, \psi_{1}^{*}, \psi_{2}^{*} \mid \psi_{3} 
    &\sim 
    \pd_{\text{meld}, 2}(\phi_{1 \cap 2}^{*}, \phi_{2 \cap 3}^{*}, \psi_{1}^{*}, \psi_{2}^{*} \mid Y_{1}, Y_{2})
  \\
  \psi_{3}^{*} \mid \phi_{1 \cap 2}, \phi_{2 \cap 3}, \psi_{1}, \psi_{2} 
    &\sim
    \q(\psi_{3}^{*} \mid \psi_{3}),
  \label{eqn:stage-three-gibbs-updates}
\end{align}\noindent
which leads to acceptance probabilities of
\begin{align}
  \alpha & \!\! \begin{multlined}[t] \left(
    (\phi_{1 \cap 2}^{*}, \phi_{2 \cap 3}^{*}, \psi_{1}^{*}, \psi_{2}^{*}),
    (\phi_{1 \cap 2}, \phi_{2 \cap 3}, \psi_{1}, \psi_{2})
  \right)
  = \\\frac{
    \pd_{\text{pool}, 3}(\phi_{1 \cap 2}^{*}, \phi_{2 \cap 3}^{*})
  } {
    \pd_{\text{pool}, 3}(\phi_{1 \cap 2}, \phi_{2 \cap 3})
  }
  \frac {
    \pd_{3}(\phi_{2 \cap 3}^{*}, \psi_{3}, Y_{3})
    \pd_{3}(\phi_{2 \cap 3})
  } {
    \pd_{3}(\phi_{2 \cap 3}, \psi_{3}, Y_{3})
    \pd_{3}(\phi_{2 \cap 3}^{*})
  } 
  \end{multlined}
  \\
  \alpha & \left(
    \psi_{3}^{*},
    \psi_{3}
  \right)
  =
  \frac {
    \pd_{3}(\phi_{2 \cap 3}, \psi_{3}^{*}, Y_{3})
    \q(\psi_{3} \mid \psi_{3}^{*})
  } {
    \pd_{3}(\phi_{2 \cap 3}, \psi_{3}, Y_{3})
    \q(\psi_{3}^{*} \mid \psi_{3})
  }.
\end{align}\noindent
The informed choice of proposal distribution for
(\(\phi_{1 \cap 2}, \phi_{2 \cap 3}, \psi_{1}, \psi_{2}\)) has allowed
us to target the full melded posterior without needing to evaluate all
submodels simultaneously.

\hypertarget{normal-approximation-calculations}{%
\subsection{Normal approximation
calculations}\label{normal-approximation-calculations}}

Substituting in the approximations of Section
\ref{normal-approximations-to-submodel-components} to Equation
\eqref{eqn:normal-approx-melded-posterior-target} yields the approximate
melded posterior
\begin{multline}
  \widehat{\pd}_{\text{meld}} (\phi_{1 \cap 2}, \phi_{2 \cap 3}, \psi_{2} \mid Y_{1}, Y_{2}, Y_{3})
  \, \propto \, \\
  \frac {
    \widehat{\pd}_{1}(\phi_{1 \cap 2} \mid \widehat{\mu}_{1}, \widehat{\Sigma}_{1})
  } {
    \widehat{\pd}_{1}(\phi_{1 \cap 2} \mid \widehat{\mu}_{1, 0}, \widehat{\Sigma}_{1, 0})
  }
  \frac {
    \widehat{\pd}_{3}(\phi_{2 \cap 3} \mid \widehat{\mu}_{3}, \widehat{\Sigma}_{3})
  } {
    \widehat{\pd}_{3}(\phi_{2 \cap 3} \mid \widehat{\mu}_{3, 0}, \widehat{\Sigma}_{3, 0})
  }
  \pd_{2}(\phi_{1 \cap 2}, \phi_{2 \cap 3}, \psi_{2} \mid Y_{2}).
  \label{eqn:normal-approximation-approximate-target}
\end{multline}
\noindent
Noting that the product of independent normal densities is an
unnormalised multivariate normal density with independent components, we
rewrite Equation \eqref{eqn:normal-approximation-approximate-target} as
\begin{equation}
\begin{gathered}
  \widehat{\pd}_{\text{meld}} (\phi_{1 \cap 2}, \phi_{2 \cap 3}, \psi_{2} \mid Y_{1}, Y_{2}, Y_{3})
  \propto
  \frac {
    \widehat{\pd}_{\text{nu}}(\phi_{1 \cap 2}, \phi_{2 \cap 3} \mid \widehat{\mu}_{\text{nu}}, \widehat{\Sigma}_{\text{nu}})
  } {
    \widehat{\pd}_{\text{de}}(\phi_{1 \cap 2}, \phi_{2 \cap 3} \mid \widehat{\mu}_{\text{de}}, \widehat{\Sigma}_{\text{de}})
  }
  \pd_{2}(\phi_{1 \cap 2}, \phi_{2 \cap 3}, \psi_{2} \mid Y_{2}), \\[1.5ex]
  \widehat{\mu}_{\text{nu}} = \begin{bmatrix}
    \widehat{\mu}_{1} \\
    \widehat{\mu}_{3}
  \end{bmatrix}\!\!, \quad
  \widehat{\Sigma}_{\text{nu}} = \begin{bmatrix}
    \widehat{\Sigma}_{1} & 0 \\
    0 & \widehat{\Sigma}_{3}
  \end{bmatrix}\!\!, \quad
  \widehat{\mu}_{\text{de}} = \begin{bmatrix}
    \widehat{\mu}_{1, 0} \\
    \widehat{\mu}_{3, 0}
  \end{bmatrix}\!\!, \quad
  \widehat{\Sigma}_{\text{de}} = \begin{bmatrix}
    \widehat{\Sigma}_{1, 0} & 0 \\
    0 & \widehat{\Sigma}_{3, 0}
  \end{bmatrix}\!\!.
\end{gathered}
\label{eqn:normal-approx-nu-de-form}
\end{equation}
  \noindent
The ratio of normal densities is also an unnormalised normal density,
and hence Equation \eqref{eqn:normal-approx-nu-de-form} simplifies to
\begin{equation}
\begin{gathered}
  \widehat{\pd}_{\text{meld}} (\phi_{1 \cap 2}, \phi_{2 \cap 3}, \psi_{2} \mid Y_{1}, Y_{2}, Y_{3})
  \, \propto \,
  \widehat{\pd}(\phi_{1 \cap 2}, \phi_{2 \cap 3} \mid \widehat{\mu}, \, \widehat{\Sigma})
  \pd_{2}(\phi_{1 \cap 2}, \phi_{2 \cap 3}, \psi_{2} \mid Y_{2}), \\[1.5ex]
    \widehat{\Sigma} = \left(
      \widehat{\Sigma}_{\text{nu}}^{-1} - \widehat{\Sigma}_{\text{de}}^{-1}
    \right)^{-1}, \quad
    \widehat{\mu} = \widehat{\Sigma} \left(
      \widehat{\Sigma}_{\text{nu}}^{-1}\widehat{\mu}_{\text{nu}} - \widehat{\Sigma}_{\text{de}}^{-1}\widehat{\mu}_{\text{de}}
    \right).
\end{gathered}
\label{eqn:final-normal-approx-appendix}
\end{equation}

\hypertarget{calculating-the-cumulative-fluid-balance-from-the-raw-fluid-data}{%
\subsection{Calculating the cumulative fluid balance from the raw fluid
data}\label{calculating-the-cumulative-fluid-balance-from-the-raw-fluid-data}}

In the raw fluid data each patient has
\(\tilde{l} = 1, \ldots, \tilde{L}_{i}\) observations. Each observation
\(\tilde{x}_{i, \tilde{l}}\) is typically a small fluid administration
(e.g.~an injection of some medicine in saline solution), or a fluid
discharge (almost always urine excretion). The observations have
corresponding observation times \(\tilde{u}_{i, \tilde{l}}\), with
\(\tilde{\boldsymbol{u}}_{i} = \{\tilde{u}_{i, 1}, \ldots, \tilde{u}_{i, \tilde{L}_{i}}\}\)
and
\(\tilde{\boldsymbol{x}}_{i} = \{\tilde{x}_{i, 1}, \ldots, \tilde{x}_{i, \tilde{L}_{i}}\}\).
We code the fluid administrations/inputs as positive values, and the
excretions/outputs as negative values. Each patient has an enormous
number of raw fluid observations \((L_{i} \ll \tilde{L}_{i})\) and it is
computationally infeasible to consider them all at once. We aggregate
the raw fluid observations into 8-hourly changes in fluid balance. From
these 8-hourly changes we calculate the cumulative fluid balance.

Mathematically, we define an ordered vector of boundary values
\begin{equation}
  \boldsymbol{v}_{i} = (\lfloor \min\{\tilde{\boldsymbol{u}}_{i}\} \rfloor,  \lfloor \min\{\tilde{\boldsymbol{u}}_{i}\} \rfloor + \frac{1}{3}, \ldots, \lceil \max\{\tilde{\boldsymbol{u}}_{i}\} \rceil),
\end{equation} noting that \(\dim(\boldsymbol{v}_{i}) = L_{i} + 1\).
Because the observation times encoded as \emph{days since ICU admission}
and we are interested in the 8-hourly changes, our floor and ceiling
functions round down or up to the appropriate third respectively. The
raw fluid observations are then divided up into \(L_{i}\) subsets of
\(\{\tilde{\boldsymbol{x}}_{i}, \tilde{\boldsymbol{u}}_{i}\}\) based on
which boundary values the observation falls in between: \begin{equation}
  V_{i, l} = \left\{
    \{\tilde{\boldsymbol{x}}_{i}, \tilde{\boldsymbol{u}}_{i}\}
    \mid
    v_{i, l} \leq \tilde{\boldsymbol{u}}_{i} < v_{i, l + 1}
  \right\},
\end{equation} for \(l = 1, \ldots, L_{i}\). Denote
\(N^{V}_{i, l} = \lvert V_{i, l} \rvert \mathop{/} 2\) (dividing by two
as \(V_{i, l}\) contains both the observation and the observation time).
The \(l\)\textsuperscript{th} 8-hourly fluid change \(\Delta_{i, l}\)
and corresponding observation time \(u_{i, l}\) can then be computed as
\begin{equation}
  \Delta_{i, l} = \sum_{s = 1}^{N^{V}_{i, l}} \tilde{x}_{i, s}, \,\, \text{s.t.} \,\, \tilde{x}_{i, s} \in V_{i, l}, \qquad
  u_{i, l} = \frac{1}{N^{V}_{i, l}} \sum_{s = 1}^{N^{V}_{i, l}} \tilde{u}_{i, s}, \,\, \text{s.t.} \,\, \tilde{u}_{i, s} \in V_{i, l}.
\end{equation} Finally, the 8-hourly cumulative fluid balance data are
computed by \(x_{i, l} = \sum_{s = 1}^{l} \Delta_{i, s}\), and we assume
they too are observed at \(u_{i, l}\).

\hypertarget{priors-and-justification-for-the-cumulative-fluid-submodel}{%
\subsection{Priors and justification for the cumulative fluid
submodel}\label{priors-and-justification-for-the-cumulative-fluid-submodel}}

The parameters for the gamma prior for \(\eta^{b}_{1, i}\) and
\(\eta^{a}_{1, i}\) are obtained by assuming that the 2.5-, 50-, and
97.5- percentiles are at 0.5, 5, and 20 (Belgorodski \emph{et al.},
2017). A slope of \(0.5\) (i.e.~the change in cumulative fluid balance
per day) is unlikely but possible due to missing data; a slope of \(20\)
is also unlikely but possible as extremely unwell patients can have very
high respiratory rates and thus require large fluid inputs.

The prior for the breakpoint \(\kappa_{i}\) is derived as follows.
Define \(u_{i, (1)} = \min(\boldsymbol{u}_{i})\) and
\(u_{i, (n)} = \max(\boldsymbol{u}_{i})\), with
\(r_{i} = u_{i, (n)} - u_{i, (1)}\). We reparameterise the breakpoint by
noting that \(\kappa_{i} = \kappa^{\text{raw}}_{i}r_{i} + u_{i, (1)}\),
where \(\kappa^{\text{raw}} \in [0, 1]\). We then set
\(\kappa^{\text{raw}}_{i} \sim \text{Beta}(5, 5)\) to regularise the
breakpoint towards the middle of each individual's stay in ICU. This is
crucial to ensure the submodel is identifiable when there is little
evidence of a breakpoint in the data. Note that this results in the
following analytic expression for \(\pd_{2}(\phi_{2 \cap 3})\)
\begin{equation}
  \pd_{3}(\phi_{2 \cap 3}) = \prod_{i = 1}^{N} \pd(\eta^{b}_{1, i}) \pd(\eta^{a}_{1, i}) \pd(\kappa_{i}), \,\, \text{with} \,\,\,
  \pd(\kappa_{i}) = \pd_{\kappa^{\text{raw}}_{i}}\left(\frac{\kappa_{i} - u_{i, (1)}}{r_{i}}\right) \frac{1}{r_{i}}
\end{equation} by the change of variables formula.

Specifying a prior for \(\eta_{0, i}\), the cumulative fluid balance at
\(\kappa_{i}\), is difficult because it too depends on the length of
stay. Instead, we reparameterise so that \(\eta_{0, i}\) is a function
of the y-intercept \(\eta_{0, i}^{\text{raw}}\). \begin{equation}
  \eta_{0, i} =
    (\eta_{0, i}^{\text{raw}} + \eta^{b}_{1, i} \kappa_{i}) \boldsymbol{1}_{\{0 < \kappa_{i}\}} +
    (\eta_{0, i}^{\text{raw}} + \eta^{a}_{1, i} \kappa_{i}) \boldsymbol{1}_{\{0 \geq \kappa_{i}\}}
\end{equation} We place a \(\text{LogNormal}(1.61, 0.47^2)\) prior on
\(\eta_{0, i}^{\text{raw}}\). These values are obtained assuming that,
\emph{a priori}, the \(2.5\%, 50\%\), and \(99\%\) percentiles of
\(\eta_{0, i}^{\text{raw}}\) are \(0.5, 5\), and \(15\) respectively
(Belgorodski \emph{et al.}, 2017). This is a broad prior that reflects
the numerous possible admission routes into the ICU. We expect those
admitted from the wards to have little pre-admission fluid data. Those
admitted from the operating theatre occasionally have their in-theatre
fluid input recorded after admission into the ICU, with no easy way to
distinguish these records in the data.

\hypertarget{analytic-form-for-the-survival-function}{%
\subsection{Analytic form for the survival
function}\label{analytic-form-for-the-survival-function}}

The hazard at arbitrary time \(t\) is

\begin{gather*}
  h_{i}(t) = \gamma t^{\gamma - 1} \exp\left\{\boldsymbol{w}_{i}^{\top}\boldsymbol{\theta} + \alpha \frac{\partial}{\partial t} m_{i}(t)\right\} \\
  m_{i}(t) = \eta_{0, i} + \eta^{b}_{1, i}(t - \kappa_{i})\boldsymbol{1}_{\{t < \kappa_{i}\}} + \eta^{a}_{1, i}(t - \kappa_{i})\boldsymbol{1}_{\{t \geq \kappa_{i}\}} \\
  \frac{\partial}{\partial t} m_{i}(t) = \eta^{b}_{1, i}\boldsymbol{1}_{\{t < \kappa_{i}\}} + \eta^{a}_{1, i}\boldsymbol{1}_{\{t \geq \kappa_{i}\}}.
\end{gather*} Then, for \(t > \kappa_{i}\), the cumulative hazard is
\begin{align*}
  \int_{0}^{t} h_{i}(u) \text{d}u
  &= \int_{0}^{t}
    \gamma u^{\gamma - 1}
    \exp\left\{
      \boldsymbol{w}_{i}^{\top}\boldsymbol{\theta} +
      \alpha \eta^{b}_{1, i}\boldsymbol{1}_{\{u < \kappa_{i}\}} +
      \alpha \eta^{a}_{1, i}\boldsymbol{1}_{\{u \geq \kappa_{i}\}}
    \right\}
    \text{d}u \\
  &= \gamma \exp\{\boldsymbol{w}_{i}^{\top}\boldsymbol{\theta}\}
    \int_{0}^{t}
      u^{\gamma - 1}
      \exp\left\{
        \alpha \eta^{b}_{1, i}\boldsymbol{1}_{\{u < \kappa_{i}\}} +
        \alpha \eta^{a}_{1, i}\boldsymbol{1}_{\{u \geq \kappa_{i}\}}
      \right\}
    \text{d}u \\
  &= \gamma \exp\{\boldsymbol{w}_{i}^{\top}\boldsymbol{\theta}\}
    \left[
      \int_{0}^{\kappa_{i}}
        u^{\gamma - 1}
        \exp\left\{
          \alpha \eta^{b}_{1, i}
        \right\}
      \text{d}u
      +
      \int_{\kappa_{i}}^{t}
        u^{\gamma - 1}
        \exp\left\{
          \alpha \eta^{a}_{1, i}
        \right\}
      \text{d}u
    \right] \\
  &= \exp\{\boldsymbol{w}_{i}^{\top}\boldsymbol{\theta}\}
    \left[
      \exp\left\{
        \alpha \eta^{b}_{1, i}
      \right\}
      \kappa_{i}^{\gamma}
      +
      \exp\left\{
        \alpha \eta^{a}_{1, i}
      \right\}
      (t^{\gamma} - \kappa_{i}^{\gamma})
    \right]
\end{align*}

and for \(t < \kappa_{i}\)

\begin{align*}
  \int_{0}^{t} h_{i}(u) \text{d}u
  &= \gamma \exp\{\boldsymbol{w}_{i}^{\top}\boldsymbol{\theta}\}
    \left[
      \int_{0}^{t}
        u^{\gamma - 1}
        \exp\left\{
          \alpha \eta^{b}_{1, i}
        \right\}
      \text{d}u
    \right] \\
  &= \exp\{\boldsymbol{w}_{i}^{\top}\boldsymbol{\theta}\}
    \left[
      \exp\left\{
        \alpha \eta^{b}_{1, i}
      \right\}
      t_{i}^{\gamma}
    \right] \\
  &= t_{i}^{\gamma} \exp\{\boldsymbol{w}_{i}^{\top}\boldsymbol{\theta} + \alpha \eta^{b}_{1, i}\}.
\end{align*} The survival functions then have corresponding definitions
for \(t > \kappa_{i}\) and \(t < \kappa_{i}\) as
\(S_{i}(t) = \exp\{-\int_{0}^{t} h_{i}(u) \text{d}u\}\).

\hypertarget{p2-prior-justification}{%
\subsection{Survival submodel prior
justification}\label{p2-prior-justification}}

Our prior for \((\gamma, \alpha, \boldsymbol{\theta})\) must result in a
plausible distribution for \(\pd_{2, i}(T_{i} \mid d_{i} = 1)\), and a
reasonable balance between \(d_{i} = 1\) and \(d_{i} = 0\) events. The
primary concern is unintentionally specifying a prior for which the bulk
of \(\pd_{2, i}(T_{i} \mid d_{i} = 1)\) is very close to zero. In
addition, certain extreme configurations of
\((\gamma, \alpha, \boldsymbol{\theta})\) cause issues for the
methodology of Crowther and Lambert (2013), particularly the numerical
root finding and numerical integration steps. We would like to rule out
such extreme configurations \emph{a priori}. Ideally we would encode
this information a joint prior for
\((\gamma, \alpha, \boldsymbol{\theta})\), but specifying the
appropriate correlation structure for these parameters is prohibitively
challenging. Instead we focus on specifying appropriate marginals for
each of \(\gamma, \alpha\), and \(\boldsymbol{\theta}\), and create
visual prior predictive checks (Gabry \emph{et al.}, 2019; Gelman
\emph{et al.}, 2020) to ensure the induced prior for \((T_{i}, d_{i})\)
is acceptable.

Before justifying our chosen marginal prior, we note that the
\(\exp\{\boldsymbol{x}_{i}\theta + \alpha \frac{\partial}{\partial T_{i}} m_{i}(T_{i})\}\)
term implies that the priors for \(\theta\) and \(\alpha\) are on the
log-scale. Hence the magnitude of these parameters must be small,
otherwise all event times would be very near zero or at infinity. The
asymmetric effect of the transformation from the log scale also implies
that symmetric priors are not obviously sensible. From these
observations we deduce that \(\theta\) and \(\alpha\) must not be too
large in magnitude, however if they are negative then they can be
slightly larger. Hence, we specify the skew-normal priors detailed in
Section \ref{survival-submodel-pd_2}, noting that the skewness parameter
for \(\alpha\) is smaller, because
\(\frac{\partial}{\partial T_{i}} m_{i}(T_{i})\) is strictly positive
and typically between 0.5 and 20, whilst \(\boldsymbol{w}_{i}\) is
standardised to be approximately standard normal. Lastly, if \(\gamma\)
is too far away from \(1\) (in either direction), then the event times
are very small either because the hazard increases rapidly
(\(\gamma \gg 1\)), or because almost all of the cumulative hazard is in
the neighbourhood of 0 (\(\gamma \ll 1\)). We specify a gamma
distribution for \(\gamma\) with the \(1\)\textsuperscript{th}-,
\(50\)\textsuperscript{th}-, and \(99\)\textsuperscript{th}-percentiles
of \(\pd_{2}(\gamma)\) at \(0.2, 1\), and \(2\), allowing for a wide
range of hazard shapes, but removing many of the extremes.

\hypertarget{estimating-submodel-prior-marginal-distributions}{%
\subsection{Estimating submodel prior marginal
distributions}\label{estimating-submodel-prior-marginal-distributions}}

For \(\pd_{1}\), we note that
\(\pd_{1}(\phi_{1 \cap 2}) = \prod_{i = 1}^{N}\pd_{1, i}(T_{i}, d_{i})\),
and that \(\pd_{1, i}(T_{i}, d_{i})\) conditions on each individual's
length of stay (in specifying the location of the knots), as well as the
range, mean, and standard deviation of the P/F data (by standardising
\(\tilde{z}_{i, j}\)). Simple Monte Carlo samples are drawn from
\(\pd_{1}(\phi_{1 \cap 2})\) and used to estimate
\(\widehat{\pd}_{1}(\phi_{1 \cap 2})\). Under the second submodel we
obtain samples from \(\pd_{2}(\phi_{1 \cap 2}, \phi_{2 \cap 3})\) using
the methodology of Crowther and Lambert (2013) as implemented in
\texttt{simsurv} (Brilleman, 2021). These samples are use to estimate
\(\widehat{\pd}_{2}(\phi_{1 \cap 2}, \phi_{2 \cap 3})\).

\hypertarget{pf-submodel}{%
\subsubsection{P/F submodel}\label{pf-submodel}}

\begin{figure}

{\centering \includegraphics{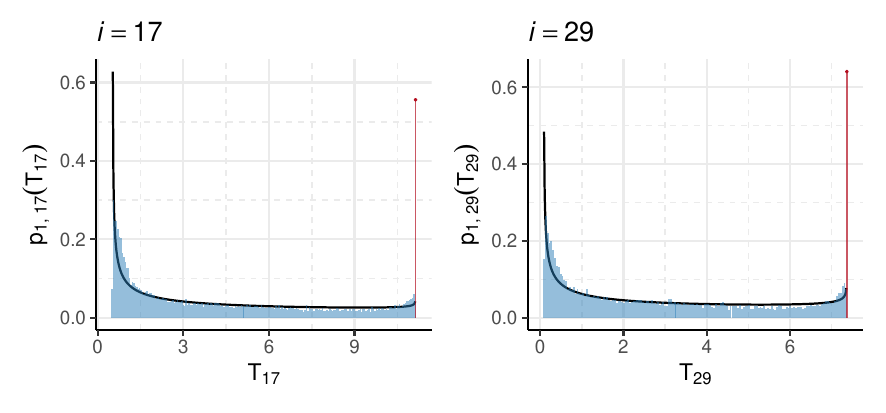} 

}

\caption{Fitted distribution (curve) and Monte Carlo samples drawn from $\pd_{1}(\phi_{1 \cap 2})$ (histogram) for a subset of the individuals in the cohort. The height of the atom at $C_{i}$ (red bar and point) has been set to $1 - \widehat{\pi}_{i}$}\label{fig:pf_prior_fit}
\end{figure}

We approximate \(\pd_{1}(\phi_{1 \cap 2})\) using a mixture of discrete
and continuous distributions, with a discrete spike at \(C_{i}\) for the
censored events and a beta distribution for the (rescaled) event times.
Monte Carlo samples of \(T_{i}\) and \(d_{i}\) are obtained from
\(\pd_{1, i}(T_{i}, d_{i})\) by drawing \(\beta_{0, i}\) and
\(\boldsymbol{\zeta}_{i}\) from their respective prior distributions and
then solving \eqref{eqn:event-time-model-def}. Denoting the estimated
mixture weight \(\widehat{\pi}_{i} \in [0, 1]\), the density estimate is
\begin{equation}
  \widehat{\pd}_{1, i}(T_{i}, d_{i}) =
    \widehat{\pi}_{i} \, \text{Beta}\left(\frac{T_{i}}{C_{i}}; \widehat{a}, \widehat{b}\right) \frac{1}{C_{i}} \boldsymbol{1}_{\{d_{i} = 1\}} +
    (1 - \widehat{\pi}_{i}) \boldsymbol{1}_{\{d_{i} = 0, T_{i} = C_{i}\}}
  \label{eqn:pf-event-time-prior-dist}
\end{equation} where \(\widehat{\pi}_{i}, \widehat{a}_{i}\) and
\(\widehat{b}_{i}\) are maximum likelihood estimates obtained using the
prior samples. Examples of \(\widehat{\pd}_{1, i}(T_{i}, d_{i})\) for a
subset of individuals are displayed in Figure \ref{fig:pf_prior_fit}.

\hypertarget{survival-submodel}{%
\subsubsection{Survival submodel}\label{survival-submodel}}

Our estimate of \(\pd_{2}(\phi_{1 \cap 2}, \phi_{2 \cap 3})\) relies on
the fact that \begin{equation}
  \pd_{2}(\phi_{1 \cap 2}, \phi_{2 \cap 3}) = \prod_{i = 1}^{N}\pd_{2, i}(T_{i}, d_{i}, \kappa_{i}, \eta^{b}_{1, i}, \eta^{a}_{1, i}).
\end{equation} As such we estimate
\(\pd_{2, i}(T_{i}, d_{i}, \kappa_{i}, \eta^{b}_{1, i}, \eta^{a}_{1, i})\)
for each individual and take the product of these estimates. Drawing
samples from
\(\pd_{2, i}(T_{i}, d_{i}, \kappa_{i}, \eta^{b}_{1, i}, \eta^{a}_{1, i})\)
is challenging: we use the approach proposed by Crowther and Lambert
(2013) as implemented in Brilleman (2021). Inspecting the samples
reveals correlation between \((T_{i}, d_{i})\) and
\((\kappa_{i}, \eta^{b}_{1, i}, \eta^{a}_{1, i})\) that we would like to
capture in our estimate. To do so, we fit a mixture of multivariate
normal distributions to transformations of the continuous parameters
with support on \(\mathbb{R}\), \begin{equation}
  \begin{gathered}
    \tilde{T}_{i} = \text{Logit}\left(\frac{T_{i}}{C_{i}}\right), \quad
    \tilde{\kappa}_{i} = \text{Logit}\left(\frac{\kappa_{i} - u_{i, (1)}}{u_{i, (n)} - u_{i, (1)}}\right), \\
    \tilde{\eta}^{b}_{1, i} = \log(\eta^{b}_{1, i}), \quad
    \tilde{\eta}^{a}_{1, i} = \log(\eta^{a}_{1, i}).
  \end{gathered}
\end{equation} The resulting density estimate, with estimated mixture
weight \(\widehat{\theta}_{i} \in [0, 1]\), is \begin{align*}
  \widehat{\pd}_{2}(T_{i}, d_{i}, \kappa_{i}, \eta^{b}_{1, i}, \eta^{a}_{1, i}) =
    \Big[&
      (\widehat{\theta}_{i})
      \text{N}\left(
        \left[\tilde{T}_{i}, \tilde{\kappa}_{i}, \tilde{\eta}^{b}_{1, i}, \tilde{\eta}^{a}_{1, i} \right]^{\mathsf{T}}; \widehat{\mu}_{1, i}, \widehat{\Sigma}_{1, i}
      \right)
      \boldsymbol{1}_{\{d_{i} = 1\}}
      + \\
      & (1 - \widehat{\theta}_{i})
      \text{N}\left(\left[\tilde{\kappa}_{i}, \tilde{\eta}^{b}_{1, i}, \tilde{\eta}^{a}_{1, i} \right]^{\mathsf{T}}; \widehat{\mu}_{0, i}, \widehat{\Sigma}_{0, i} \right)
      \boldsymbol{1}_{\{d_{i} = 0, T_{i} = C_{i}\}}
    \Big]
    J_{i},
\end{align*} where
\(\widehat{\theta}_{i}, \widehat{\mu}_{1, i}, \widehat{\Sigma}_{1, i}, \widehat{\mu}_{0, i}\),
and \(\widehat{\Sigma}_{0, i}\) are maximum likelihood estimates, and
the Jacobian correction \(J_{i}\) is \begin{equation}
  J_{i} = \left[
    \left(
      \frac{1}{C_{i} - T_{i}} +
      \frac{1}{T_{i}}
    \right)^{d_{i}}
    \left(
      \frac{1}{u_{i, (n)} - \kappa_{i}} +
      \frac{1}{\kappa_{i} - u_{i, (1)}}
    \right)
    \left(
      \frac{1}{\eta^{b}_{1, i}}
    \right)
    \left(
      \frac{1}{\eta^{a}_{1, i}}
    \right)
  \right].
\end{equation}

\begin{figure}

{\centering \includegraphics[width=\textwidth]{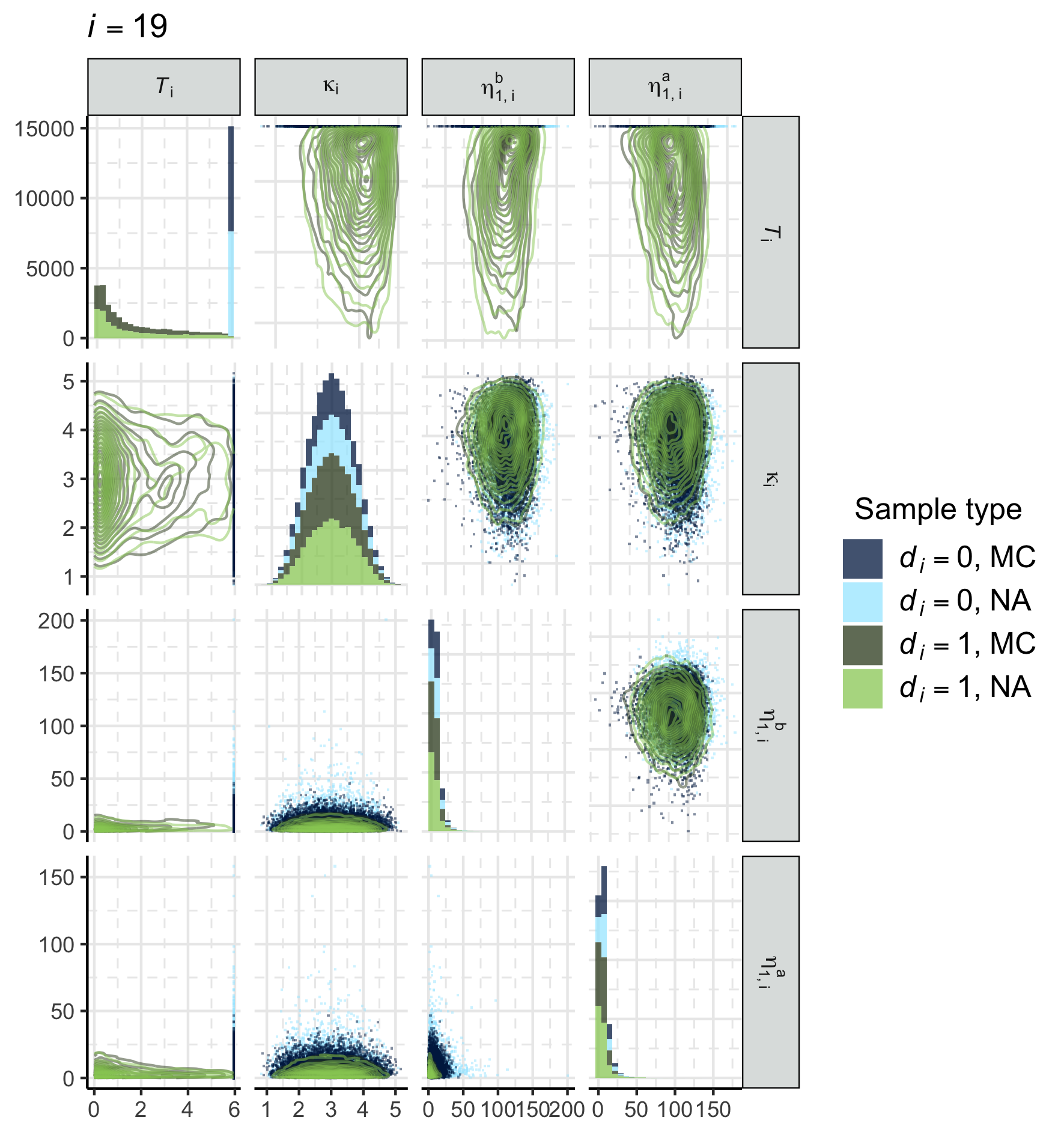} 

}

\caption{Monte Carlo (MC) samples from $\pd_{2, i}(\phi_{1 \cap 2}, \phi_{2 \cap 3})$ obtained using \texttt{simsurv} and samples from the fitted normal approximation (NA) for $i = 19$. The panels on the off diagonal elements contain a 2D kernel density estimate for $d_{i} = 1$ and the samples for $d_{i} = 0$. Diagonal and lower-triangular panels are on their original scales, whilst the upper-triangular panels are on the log scale.}\label{fig:surv_prior_plot_fit}
\end{figure}

We assess the fit of this estimate by drawing samples from
\(\widehat{\pd}_{2, i}(\phi_{1 \cap 2}, \phi_{2 \cap 3})\) and comparing
them to the Monte Carlo samples drawn using \texttt{simsurv}. Our visual
assessment is displayed in Figure \ref{fig:surv_prior_plot_fit} for
individual \(i = 19\). The normal approximation seems to fit the samples
well, with the shape of \(\pd_{2, i}(T_{i} \mid d_{i} = 1)\) closely
matching that of the Monte Carlo samples, and with a similar mix of
\(d_{i} = 0\) and \(d_{i} = 1\) samples.

\begin{figure}

{\centering \includegraphics{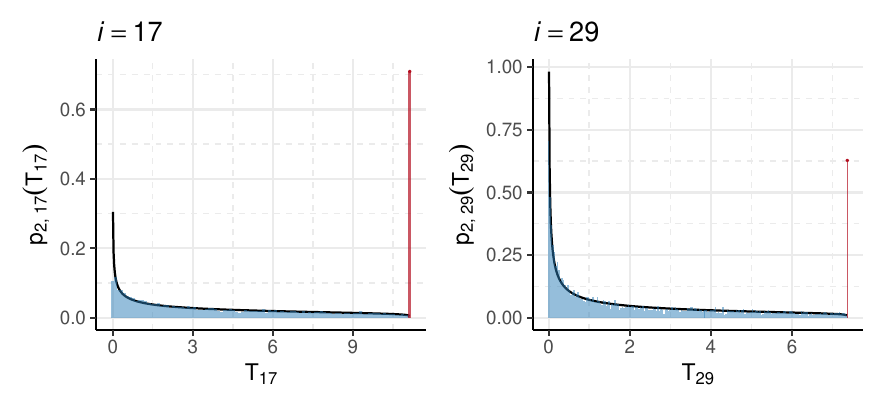} 

}

\caption{Fitted distribution (curve) and Monte Carlo samples drawn from $\pd_{2}(\phi_{1 \cap 2})$ (histogram) for a subset of the individuals in the cohort. The height of the atom at $C_{i}$ (red bar and point) has been set to $1 - \widehat{\pi}_{i}$}\label{fig:surv_prior_phi_12_marginal_plot_fit}
\end{figure}

We also require an estimate of \(\pd_{2}(\phi_{1 \cap 2})\) for
experiments discussed in Section \ref{results}. This is obtained using
the samples generated under the survival submodel prior and the
methodology of Section \ref{pf-submodel}. The raw samples and fit are
displayed in Figure \ref{fig:surv_prior_phi_12_marginal_plot_fit}.

\hypertarget{cohort-selection-criteria}{%
\subsection{Cohort selection criteria}\label{cohort-selection-criteria}}

This appendix details the cohort selection criteria and our rationale
for them. In the text we speak of the \(i\)\textsuperscript{th}
individual. This is because in our final data set (the data that results
from applying the following criteria) we are dealing with unique
individuals, however some individuals in MIMIC have multiple ICU stays.
In this appendix \(i\) represents a single stay in ICU.

\begin{enumerate}
\def\labelenumi{\arabic{enumi}.}
\tightlist
\item
  Each ICU stay must have at least 12 \pfratio~observations
  (\(J_{i} \geq 12\)), with the first 6 being greater than 350
  (\(z_{i, j} > 350\) for \(j = 1, \dots, 1\)).

  \begin{itemize}
  \tightlist
  \item
    This is to ensure we have enough data to fit a B-spline with 7
    internal knots. The restriction on the first 6 observations is to
    avoid selecting those who have already started to experience
    respiratory failure prior to ICU admission.
  \end{itemize}
\item
  The time between any 2 consecutive \pfratio~observations cannot exceed
  2 days.

  \begin{itemize}
  \tightlist
  \item
    This is because we believe this lack of \pfratio~observations is
    likely to be an error in the data: e.g.~what appears as a single ICU
    stay in the data is actually two or more separate stays.
  \end{itemize}
\item
  The fluid observations must be after ICU admission (some observations
  are entered as `Pre-admission intake') and cannot be associated with
  fluid administered in the operating room (OR). Note that this does not
  mean all OR fluid administrations are removed, as some are
  mis-labelled.
\item
  There must be sufficient temporal overlap between the fluid data and
  the \pfratio~data. Specifically, \begin{equation}
   \frac{
     \max\left\{0, \min\left[\max(\boldsymbol{t}_{i}), \max(\tilde{\boldsymbol{u}}_{i})\right] - \max\left[\min(\boldsymbol{t}_{i}), \min(\tilde{\boldsymbol{u}}_{i})\right]\right\}
   } {
     \max\left[\max(\boldsymbol{t}_{i}), \max(\tilde{\boldsymbol{u}}_{i})\right] - \min\left[\min(\boldsymbol{t}_{i}), \min(\tilde{\boldsymbol{u}}_{i})\right]
   }
   > 0.9
   \label{eqn:overlap-def}
    \end{equation}

  \begin{itemize}
  \tightlist
  \item
    The numerator of \eqref{eqn:overlap-def} is strictly positive, and
    the denominator ensures that the quantity is bounded between 0 and
    1.
  \item
    We cannot investigate the relationship between the rate of fluid
    intake and respiratory failure if the latter occurs without
    sufficient fluid data surrounding the event.
  \end{itemize}
\end{enumerate}

\hypertarget{baseline-covariate-information}{%
\subsection{Baseline covariate
information}\label{baseline-covariate-information}}

The baseline covariate vector \(\boldsymbol{w}_{i}\) contains the median
of the measurements taken in the first 24 hours of the ICU stay, which
are then stardardised, of the following covariates: Anion gap,
Bicarbonate, Creatinine, Chloride, Glucose, Hematocrit, Hemoglobin,
Platelet, Partial thromboplastin time, International normalized ratio,
Prothrombin time, Sodium, blood Urea nitrogen, White blood cell count,
Age at ICU admission, and Gender.

\hypertarget{one-at-a-time}{%
\subsection{\texorpdfstring{Updating \(\phi_{1 \cap 2}\) and
\(\phi_{2 \cap 3}\) in stage two
individual-at-a-time}{Updating \textbackslash phi\_\{1 \textbackslash cap 2\} and \textbackslash phi\_\{2 \textbackslash cap 3\} in stage two individual-at-a-time}}\label{one-at-a-time}}

The parallel sampler described in Section \ref{parallel-sampler} is a
MH-within-Gibbs scheme, with each iteration sampling from the
conditionals of the melded model \begin{equation}
\begin{gathered}
  \pd_{\text{meld}}(\phi_{1 \cap 2}, \psi_{1} \mid \boldsymbol{Y}, \psi_{2}, \phi_{2 \cap 3}, \psi_{3}), \\
  \pd_{\text{meld}}(\phi_{2 \cap 3}, \psi_{3} \mid \boldsymbol{Y}, \phi_{1 \cap 2}, \psi_{1}, \psi_{2}), \\
  \pd_{\text{meld}}(\psi_{2} \mid \boldsymbol{Y}, \phi_{1 \cap 2}, \psi_{1}, \phi_{2 \cap 3}, \psi_{3}).
\end{gathered}
\end{equation} We would like to update the first two conditionals
`individual-at-a-time'. For simplicity we will assume the pooled prior
is formed using product-of-experts pooling, and focus on the first
conditional
\(\pd_{\text{meld}}(\phi_{1 \cap 2}, \psi_{1} \mid \boldsymbol{Y}, \psi_{2}, \phi_{2 \cap 3}, \psi_{3})\).
Similar arguments apply to the second conditional and other pooling
types. Recall that \begin{equation}
\begin{aligned}
\phi_{1 \cap 2} &=
  (\phi_{1 \cap 2, 1}, \ldots, \phi_{1 \cap 2, N})  \\
  &= (\phi_{1 \cap 2, 1}(\chi_{1, 1}), \ldots, \phi_{1 \cap 2, N}(\chi_{1, N})) \\
  &= \left((T_{1}, d_{1}), \ldots, (T_{N}, d_{N})\right),
\end{aligned}
\end{equation} and
\(\psi_{1} = (\psi_{1, 1}, \ldots, \psi_{1, N}) = (\omega_{i})_{i = 1}^{N}\).
We would like to update \(\phi_{1 \cap 2}\) and \(\psi_{1}\) by updating
\((T_{i}, d_{i}, \omega_{i})\) for one individual at a time, for a total
of \(N\) `sub-steps'.

\hypertarget{sub-step-1}{%
\subsubsection{Sub-step 1}\label{sub-step-1}}

Suppose we are at step \(t - 1\) of the Markov chain, and we are
proposing values for step \(t\). Without any loss of generality we
assume that we are updating individual \(i = 1\) in sub-step 1 -- in
practice we update the individuals in a random order for each iteration
of the MH-within-Gibbs scheme.

Our target is \begin{equation}
  \pd_{1}(\chi_{1, 1}, \psi_{1, 1} \mid Y_{1}, (\chi_{1, i}, \psi_{1, i})_{i = 2}^{N})
  \pd_{2}(\phi_{1 \cap 2, 1}(\chi_{1, 1}) \mid Y_{2}, (\phi_{1 \cap 2, i}(\chi_{1, i}))_{i = 2}^{N}, \psi_{2}, \phi_{2 \cap 3}).
  \label{eqn:sub-step-target}
\end{equation} The model, detailed in Section
\ref{pf-ratio-submodel-b-spline-pd_1}, uses the conditional independence
between individuals to factorise such that \begin{equation}
  \pd_{1}(\chi_{1}, \psi_{1} \mid Y_{1}) = \prod_{i = 1}^{N} \pd_{1}(\chi_{1, i}, \psi_{1, i} \mid Y_{1}),
\end{equation} which implies\footnote{This property is also true of
  \(\pd_{3}\).}
\((\chi_{1, i}, \psi_{1, i}\indep \chi_{1, i^{'}}, \psi_{1, i^{'}}) \mid Y_{1}\)
for \(i \neq i^{'}\), hence \begin{equation}
  \pd_{1}(\chi_{1, 1}, \psi_{1, 1} \mid Y_{1}, (\chi_{1, i}, \psi_{1, i})_{i = 2}^{N}) =
  \pd_{1}(\chi_{1, 1}, \psi_{1, 1} \mid Y_{1}).
\end{equation} It will be convenient to rewrite
\eqref{eqn:sub-step-target} as \begin{equation}
  \pd_{1}(\chi_{1, 1}, \psi_{1, 1} \mid Y_{1})
  \frac {
    \pd_{2}(\phi_{1 \cap 2, 1}(\chi_{1, 1}), (\phi_{1 \cap 2, i}(\chi_{1, i}))_{i = 2}^{N}, \psi_{2}, \phi_{2 \cap 3} \mid Y_{2})
  } {
    \pd_{2}((\phi_{1 \cap 2, i}(\chi_{1, i}))_{i = 2}^{N}, \psi_{2}, \phi_{2 \cap 3} \mid Y_{2})
  }.
  \label{eqn:sub-step-one-target}
\end{equation}

Suppose there are \(K_{1}\) stage one samples from
\(\pd_{1}(\chi_{1}, \psi_{1} \mid Y_{1})\), and each sample has a
corresponding \(\phi_{1 \cap 2}\). We propose \(\psi_{1, 1}^{*}\) and
\(\chi_{1, 1}^{*}\) (hence \(\phi_{1 \cap 2, 1}^{*})\) by sampling a
random integer \(k_{1}^{*}\) from \(\{1, \ldots, K_{1}\}\), retrieving
the corresponding values of \(\psi_{1}\) and \(\chi_{1}\), and ignoring
\((\chi_{1, 2}, \ldots, \chi_{1, N}, \psi_{1, 2}, \ldots, \psi_{1, N})\).
This is equivalent to proposing from
\(\pd_{1}(\chi_{1, 1}, \psi_{1, 1} \mid Y_{1})\) Under this proposal,
noting that the denominator term in \eqref{eqn:sub-step-one-target} does
not depend on the proposed parameters, the acceptance probability for
this sub-step is \begin{equation}
\begin{aligned}
\alpha&\left((\chi_{1, 1}^{*}, \psi_{1, 1}^{*})_{k_{1}^{*}}, (\chi_{1, 1}, \psi_{1, 1}) \right) \\
&=
\frac {
  \pd_{1}(\chi_{1, 1}^{*}, \psi_{1, 1}^{*} \mid Y_{1})
  \pd_{2}(\phi_{1 \cap 2, 1}(\chi_{1, 1}^{*}), (\phi_{1 \cap 2, i}(\chi_{1, i}))_{i = 2}^{N}, \phi_{2 \cap 3}, \psi_{2} \mid Y_{2})
} {
  \pd_{1}(\chi_{1, 1}, \psi_{1, 1} \mid Y_{1})
  \pd_{2}(\phi_{1 \cap 2, 1}(\chi_{1, 1}), (\phi_{1 \cap 2, i}(\chi_{1, i}))_{i = 2}^{N}, \phi_{2 \cap 3}, \psi_{2} \mid Y_{2})
}
\frac{
  \pd_{1}(\chi_{1, 1}, \psi_{1, 1} \mid Y_{1})
} {
  \pd_{1}(\chi_{1, 1}^{*}, \psi_{1, 1}^{*} \mid Y_{1})
} \\
&= \frac {
  \pd_{2}(\phi_{1 \cap 2, 1}(\chi_{1, 1}^{*}), (\phi_{1 \cap 2, i}(\chi_{1, i}))_{i = 2}^{N}, \phi_{2 \cap 3}, \psi_{2} \mid Y_{2})
} {
  \pd_{2}(\phi_{1 \cap 2, 1}(\chi_{1, 1}), (\phi_{1 \cap 2, i}(\chi_{1, i}))_{i = 2}^{N}, \phi_{2 \cap 3}, \psi_{2} \mid Y_{2})
},
\end{aligned}
\end{equation} which does not depend on \(\pd_{1}\). We store the value
of \(k_{1}^{*}\) associated with the proposal (if it is accepted) in
order to resample the stage one values of \(\psi_{1, 1}\) to reflect the
information in the other submodels.

\hypertarget{sub-step-n}{%
\subsubsection{\texorpdfstring{Sub-step
\(n\)}{Sub-step n}}\label{sub-step-n}}

At sub-step \(n\), for \(1 < n \leq N\) with
\((\chi_{1, i})_{i = N + 1}^{N} = \varnothing\), we have updated
\(((\chi_{1, i}, \psi_{1, i})_{i = 1}^{n - 1})\) to
\(((\chi_{1, i}^{[t]}, \psi_{1, i}^{[t]})_{i = 1}^{n - 1})\). Thus the
target is \begin{align}
  \pd_{1} & \left(\chi_{1, n}, \psi_{1, n} \mid Y_{1}, (\chi_{1, i}^{[t]}, \psi_{1, i}^{[t]})_{i = 1}^{n - 1}, (\chi_{1, i}, \psi_{1, i})_{i = n + 1}^{N}\right) \\
  & \times \pd_{2} \left(\phi_{1 \cap 2, n}(\chi_{1, n}) \mid Y_{2}, \psi_{2}, \phi_{2 \cap 3}, (\phi_{1 \cap 2, i}(\chi_{1, i}^{[t]}))_{i = 1}^{n - 1}, (\phi_{1 \cap 2, i}(\chi_{1, i}))_{i = n + 1}^{N}\right)
  \nonumber \\
  = & \pd_{1} \left(\chi_{1, n}, \psi_{1, n}, \mid Y_{1} \right)
  \frac {
    \pd_{2} \left(\phi_{1 \cap 2, n}(\chi_{1, n}), \psi_{2}, \phi_{2 \cap 3}, (\phi_{1 \cap 2, i}(\chi_{1, i}^{[t]}))_{i = 1}^{n - 1}, (\phi_{1 \cap 2, i}(\chi_{1, i}))_{i = n + 1}^{N} \mid Y_{2}\right)
  } {
    \pd_{2} \left(\psi_{2}, \phi_{2 \cap 3}, (\phi_{1 \cap 2, i}(\chi_{1, i}^{[t]}))_{i = 1}^{n - 1}, (\phi_{1 \cap 2, i}(\chi_{1, i}))_{i = n + 1}^{N} \mid Y_{2}\right)
  }
  \label{eqn:two-to-N-target}
\end{align} We propose \(\chi_{1, n}^{*}, \psi_{1, n}^{*}\) in the same
manner as the previous section. The corresponding acceptance probability
is \begin{equation}
\begin{aligned}
\alpha ((\chi_{1, n}^{*}, \psi_{1, n}^{*}), & (\chi_{1, n}, \psi_{1, n}) )  \\
= &\frac {
  \pd_{1} \left(\chi_{1, n}^{*}, \psi_{1, n}^{*} \mid Y_{1} \right)
} {
  \pd_{1} \left(\chi_{1, n}, \psi_{1, n} \mid Y_{1} \right)
} \\
& \times \frac {
  \pd_{2}(\phi_{1 \cap 2, n}(\chi_{1, n}^{*}), \psi_{2}, \phi_{2 \cap 3}, (\chi_{1, i}^{[t]})_{i = 1}^{n - 1}, (\chi_{1, i})_{i = n + 1}^{N} \mid Y_{2})
} {
  \pd_{2}(\phi_{1 \cap 2, n}(\chi_{1, n}), \psi_{2}, \phi_{2 \cap 3}, (\chi_{1, i}^{[t]})_{i = 1}^{n - 1}, (\chi_{1, i})_{i = n + 1}^{N} \mid Y_{2})
} \\
& \times \frac{
  \pd_{1}(\chi_{1, n}, \psi_{1, n} \mid Y_{1})
} {
  \pd_{1}(\chi_{1, n}^{*}, \psi_{1, n}^{*} \mid Y_{1})
} \\
= &\frac {
  \pd_{2}(\phi_{1 \cap 2, n}(\chi_{1, n}^{*}), \psi_{2}, \phi_{2 \cap 3}, (\chi_{1, i}^{[t]})_{i = 1}^{n - 1}, (\chi_{1, i})_{i = n + 1}^{N} \mid Y_{2})
} {
  \pd_{2}(\phi_{1 \cap 2, n}(\chi_{1, n}), \psi_{2}, \phi_{2 \cap 3}, (\chi_{1, i}^{[t]})_{i = 1}^{n - 1}, (\chi_{1, i})_{i = n + 1}^{N} \mid Y_{2})
}.
\end{aligned}
\end{equation}

\hypertarget{references}{%
\section*{References}\label{references}}
\addcontentsline{toc}{section}{References}

\hypertarget{refs}{}
\begin{CSLReferences}{1}{0}
\leavevmode\vadjust pre{\hypertarget{ref-abadi_estimation_2010}{}}%
Abadi, F., Gimenez, O., Ullrich, B., et al. (2010) Estimation of
immigration rate using integrated population models. \emph{Journal of
Applied Ecology}, \textbf{47}, 393--400. DOI:
\href{https://doi.org/10.1111/j.1365-2664.2010.01789.x}{10.1111/j.1365-2664.2010.01789.x}.

\leavevmode\vadjust pre{\hypertarget{ref-abbas_kullback-leibler_2009}{}}%
Abbas, A. E. (2009) A {Kullback-Leibler} view of linear and log-linear
pools. \emph{Decision Analysis}. {INFORMS}. DOI:
\href{https://doi.org/10.1287/deca.1080.0133}{10.1287/deca.1080.0133}.

\leavevmode\vadjust pre{\hypertarget{ref-ades_multiparameter_2006}{}}%
Ades, A. E. and Sutton, A. J. (2006) Multiparameter evidence synthesis
in epidemiology and medical decision-making: Current approaches.
\emph{Journal of the Royal Statistical Society: Series A (Statistics in
Society)}, \textbf{169}, 5--35. DOI:
\href{https://doi.org/10.1111/j.1467-985X.2005.00377.x}{10.1111/j.1467-985X.2005.00377.x}.

\leavevmode\vadjust pre{\hypertarget{ref-belgorodski_rriskdistributions_2017}{}}%
Belgorodski, N., Greiner, M., Tolksdorf, K., et al. (2017)
{rriskDistributions}: {Fitting} distributions to given data or known
quantiles.

\leavevmode\vadjust pre{\hypertarget{ref-besbeas_integrating_2002}{}}%
Besbeas, P., Freeman, S. N., Morgan, B. J. T., et al. (2002) Integrating
mark\textendash recapture\textendash recovery and census data to
estimate animal abundance and demographic parameters. \emph{Biometrics},
\textbf{58}, 540--547. {{[}Wiley, International Biometric Society{]}}.
DOI:
\href{https://doi.org/10.1111/j.0006-341X.2002.00540.x}{10.1111/j.0006-341X.2002.00540.x}.

\leavevmode\vadjust pre{\hypertarget{ref-brilleman_simsurv_2021}{}}%
Brilleman, S. (2021) Simsurv: {Simulate} survival data.

\leavevmode\vadjust pre{\hypertarget{ref-brilleman_bayesian_2020}{}}%
Brilleman, S. L., Elci, E. M., Novik, J. B., et al. (2020) Bayesian
survival analysis using the rstanarm {R} package. \emph{arXiv:2002.09633
{[}stat{]}}. Available at: \url{https://arxiv.org/abs/2002.09633}.

\leavevmode\vadjust pre{\hypertarget{ref-bromiley_products_2003}{}}%
Bromiley, P. (2003) Products and convolutions of {Gaussian} probability
density functions. \emph{Tina-Vision Memo}, \textbf{3}, 1.

\leavevmode\vadjust pre{\hypertarget{ref-brooks_bayesian_2004}{}}%
Brooks, S. P., King, R. and Morgan, B. J. T. (2004) A {Bayesian}
approach to combining animal abundance and demographic data.
\emph{Animal Biodiversity and Conservation}, \textbf{27}.

\leavevmode\vadjust pre{\hypertarget{ref-burke_meta-analysis_2017}{}}%
Burke, D. L., Ensor, J. and Riley, R. D. (2017) Meta-analysis using
individual participant data: One-stage and two-stage approaches, and why
they may differ. \emph{Statistics in Medicine}, \textbf{36}, 855--875.
DOI: \href{https://doi.org/10.1002/sim.7141}{10.1002/sim.7141}.

\leavevmode\vadjust pre{\hypertarget{ref-carpenter_stan_2017}{}}%
Carpenter, B., Gelman, A., Hoffman, M., et al. (2017) Stan: {A}
probabilistic programming language. \emph{Journal of Statistical
Software}, \textbf{76}, 1--32. DOI:
\href{https://doi.org/10.18637/jss.v076.i01}{10.18637/jss.v076.i01}.

\leavevmode\vadjust pre{\hypertarget{ref-carvalho_bayesian_2022}{}}%
Carvalho, L. M., Villela, D. A. M., Coelho, F. C., et al. (2022)
Bayesian inference for the weights in logarithmic pooling.
\emph{Bayesian Analysis}, 1--29. {International Society for Bayesian
Analysis}. DOI:
\href{https://doi.org/10.1214/22-BA1311}{10.1214/22-BA1311}.

\leavevmode\vadjust pre{\hypertarget{ref-crowther_simulating_2013}{}}%
Crowther, M. J. and Lambert, P. C. (2013) Simulating biologically
plausible complex survival data. \emph{Statistics in Medicine},
\textbf{32}, 4118--4134. DOI:
\href{https://doi.org/10.1002/sim.5823}{10.1002/sim.5823}.

\leavevmode\vadjust pre{\hypertarget{ref-dawid_hyper_1993-1}{}}%
Dawid, A. P. and Lauritzen, S. L. (1993) Hyper {Markov} laws in the
statistical analysis of decomposable graphical models. \emph{The Annals
of Statistics}, \textbf{21}, 1272--1317. {Institute of Mathematical
Statistics}. DOI:
\href{https://doi.org/10.1214/aos/1176349260}{10.1214/aos/1176349260}.

\leavevmode\vadjust pre{\hypertarget{ref-de_valpine_programming_2017}{}}%
de Valpine, P., Turek, D., Paciorek, C. J., et al. (2017) Programming
with models: Writing statistical algorithms for general model structures
with {NIMBLE}. \emph{Journal of Computational and Graphical Statistics},
\textbf{26}, 403--413. DOI:
\href{https://doi.org/10.1080/10618600.2016.1172487}{10.1080/10618600.2016.1172487}.

\leavevmode\vadjust pre{\hypertarget{ref-donnat_bayesian_2020}{}}%
Donnat, C., Miolane, N., Bunbury, F., et al. (2020) A {Bayesian}
hierarchical network for combining heterogeneous data sources in medical
diagnoses. In: \emph{Proceedings of the {Machine Learning} for {Health
NeurIPS Workshop}}, November 2020, pp. 53--84. Proceedings of {Machine
Learning Research}. {PMLR}.

\leavevmode\vadjust pre{\hypertarget{ref-finke_efficient_2019}{}}%
Finke, A., King, R., Beskos, A., et al. (2019) Efficient sequential
{Monte Carlo} algorithms for integrated population models. \emph{Journal
of Agricultural, Biological and Environmental Statistics}, \textbf{24},
204--224. DOI:
\href{https://doi.org/10.1007/s13253-018-00349-9}{10.1007/s13253-018-00349-9}.

\leavevmode\vadjust pre{\hypertarget{ref-gabry_visualization_2019}{}}%
Gabry, J., Simpson, D., Vehtari, A., et al. (2019) Visualization in
{Bayesian} workflow. \emph{Journal of the Royal Statistical Society:
Series A (Statistics in Society)}, \textbf{182}, 389--402. DOI:
\href{https://doi.org/10.1111/rssa.12378}{10.1111/rssa.12378}.

\leavevmode\vadjust pre{\hypertarget{ref-gabry_bayesplot_2021}{}}%
Gabry, J., Mahr, T., Bürkner, P.-C., et al. (2021) Bayesplot: Plotting
for {Bayesian} models.

\leavevmode\vadjust pre{\hypertarget{ref-gelman_bayesian_2020}{}}%
Gelman, A., Vehtari, A., Simpson, D., et al. (2020) Bayesian workflow.
\emph{arXiv:2011.01808 {[}stat{]}}. Available at:
\url{https://arxiv.org/abs/2011.01808}.

\leavevmode\vadjust pre{\hypertarget{ref-genest_characterization_1986}{}}%
Genest, C., McConway, K. J. and Schervish, M. J. (1986) Characterization
of externally {Bayesian} pooling operators. \emph{The Annals of
Statistics}, \textbf{14}, 487--501. {Institute of Mathematical
Statistics}.

\leavevmode\vadjust pre{\hypertarget{ref-giganti_accounting_2020}{}}%
Giganti, M. J., Shaw, P. A., Chen, G., et al. (2020) Accounting for
dependent errors in predictors and time-to-event outcomes using
electronic health records, validation samples, and multiple imputation.
\emph{Annals of Applied Statistics}, \textbf{14}, 1045--1061. DOI:
\href{https://doi.org/10.1214/20-aoas1343}{10.1214/20-aoas1343}.

\leavevmode\vadjust pre{\hypertarget{ref-goudie_joining_2019}{}}%
Goudie, R. J. B., Presanis, A. M., Lunn, D., et al. (2019) Joining and
splitting models with {Markov} melding. \emph{Bayesian Analysis},
\textbf{14}, 81--109. {International Society for Bayesian Analysis}.
DOI: \href{https://doi.org/10.1214/18-BA1104}{10.1214/18-BA1104}.

\leavevmode\vadjust pre{\hypertarget{ref-hastie_generalized_1999}{}}%
Hastie, T. and Tibshirani, R. (1999) \emph{Generalized Additive Models}.
{Boca Raton, Fla}: {Chapman {\(\&\)} Hall/CRC}.

\leavevmode\vadjust pre{\hypertarget{ref-hinton_training_2002}{}}%
Hinton, G. E. (2002) Training products of experts by minimizing
contrastive divergence. \emph{Neural Computation}, \textbf{14},
1771--1800. DOI:
\href{https://doi.org/10.1162/089976602760128018}{10.1162/089976602760128018}.

\leavevmode\vadjust pre{\hypertarget{ref-hooten_making_2019}{}}%
Hooten, M. B., Johnson, D. S. and Brost, B. M. (2019) Making recursive
{Bayesian} inference accessible. \emph{The American Statistician},
1--10. DOI:
\href{https://doi.org/10.1080/00031305.2019.1665584}{10.1080/00031305.2019.1665584}.

\leavevmode\vadjust pre{\hypertarget{ref-jackson_when_2018}{}}%
Jackson, D. and White, I. R. (2018) When should meta-analysis avoid
making hidden normality assumptions? \emph{Biometrical Journal},
\textbf{60}, 1040--1058. DOI:
\href{https://doi.org/10.1002/bimj.201800071}{10.1002/bimj.201800071}.

\leavevmode\vadjust pre{\hypertarget{ref-johnson_mimic-iii_2016}{}}%
Johnson, A. E. W., Pollard, T. J., Shen, L., et al. (2016) {MIMIC-III},
a freely accessible critical care database. \emph{Scientific Data},
\textbf{3}, 160035. {Nature Publishing Group}. DOI:
\href{https://doi.org/10.1038/sdata.2016.35}{10.1038/sdata.2016.35}.

\leavevmode\vadjust pre{\hypertarget{ref-kay_tidybayes_2020}{}}%
Kay, M. (2020) Tidybayes: Tidy data and geoms for {Bayesian} models.
DOI:
\href{https://doi.org/10.5281/zenodo.1308151}{10.5281/zenodo.1308151}.

\leavevmode\vadjust pre{\hypertarget{ref-kedem_statistical_2017}{}}%
Kedem, B., De Oliveira, V. and Sverchkov, M. (2017) \emph{Statistical
Data Fusion}. {World Scientific}. DOI:
\href{https://doi.org/10.1142/10282}{10.1142/10282}.

\leavevmode\vadjust pre{\hypertarget{ref-kharratzadeh_splines_2017}{}}%
Kharratzadeh, M. (2017) Splines in {Stan}. \emph{Stan Case Studies},
\textbf{4}.

\leavevmode\vadjust pre{\hypertarget{ref-kuntz_divide-and-conquer_2021}{}}%
Kuntz, J., Crucinio, F. R. and Johansen, A. M. (2021) The
divide-and-conquer sequential {Monte Carlo} algorithm: Theoretical
properties and limit theorems. \emph{arXiv:2110.15782 {[}math, stat{]}}.
Available at: \url{https://arxiv.org/abs/2110.15782}.

\leavevmode\vadjust pre{\hypertarget{ref-kurowicka_dependence_2011}{}}%
Kurowicka, D. and Joe, H. (eds) (2011) \emph{Dependence Modeling: Vine
Copula Handbook}. {Singapore}: {World Scientific}.

\leavevmode\vadjust pre{\hypertarget{ref-lahat_multimodal_2015}{}}%
Lahat, D., Adali, T. and Jutten, C. (2015) Multimodal data fusion: An
overview of methods, challenges, and prospects. \emph{Proceedings of the
IEEE}, \textbf{103}, 1449--1477. DOI:
\href{https://doi.org/10.1109/JPROC.2015.2460697}{10.1109/JPROC.2015.2460697}.

\leavevmode\vadjust pre{\hypertarget{ref-lauritzen_chain_2002}{}}%
Lauritzen, S. L. and Richardson, T. S. (2002) Chain graph models and
their causal interpretations. \emph{Journal of the Royal Statistical
Society: Series B (Statistical Methodology)}, \textbf{64}, 321--348.
DOI:
\href{https://doi.org/10.1111/1467-9868.00340}{10.1111/1467-9868.00340}.

\leavevmode\vadjust pre{\hypertarget{ref-lebreton_modeling_1992}{}}%
Lebreton, J.-D., Burnham, K. P., Clobert, J., et al. (1992) Modeling
survival and testing biological hypotheses using marked animals: A
unified approach with case studies. \emph{Ecological Monographs},
\textbf{62}, 67--118. DOI:
\href{https://doi.org/10.2307/2937171}{10.2307/2937171}.

\leavevmode\vadjust pre{\hypertarget{ref-lin_recent_2014}{}}%
Lin, G., Dou, X., Kuriki, S., et al. (2014) Recent developments on the
construction of bivariate distributions with fixed marginals.
\emph{Journal of Statistical Distributions and Applications},
\textbf{1}, 14. DOI:
\href{https://doi.org/10.1186/2195-5832-1-14}{10.1186/2195-5832-1-14}.

\leavevmode\vadjust pre{\hypertarget{ref-lindsten_divide-and-conquer_2017}{}}%
Lindsten, F., Johansen, A. M., Naesseth, C. A., et al. (2017)
Divide-and-conquer with sequential {Monte Carlo}. \emph{Journal of
Computational and Graphical Statistics}, \textbf{26}, 445--458. {Taylor
{\(\&\)} Francis}. DOI:
\href{https://doi.org/10.1080/10618600.2016.1237363}{10.1080/10618600.2016.1237363}.

\leavevmode\vadjust pre{\hypertarget{ref-lu_using_1993}{}}%
Lu, C. J. and Meeker, W. Q. (1993) Using degradation measures to
estimate a time-to-failure distribution. \emph{Technometrics},
\textbf{35}, 161--174. {{[}Taylor {\(\&\)} Francis, Ltd., American
Statistical Association, American Society for Quality{]}}. DOI:
\href{https://doi.org/10.2307/1269661}{10.2307/1269661}.

\leavevmode\vadjust pre{\hypertarget{ref-lunn_bugs_2009}{}}%
Lunn, D., Spiegelhalter, D., Thomas, A., et al. (2009) The {BUGS}
project: Evolution, critique and future directions. \emph{Statistics in
Medicine}, \textbf{28}, 3049--3067. {Wiley}. DOI:
\href{https://doi.org/10.1002/sim.3680}{10.1002/sim.3680}.

\leavevmode\vadjust pre{\hypertarget{ref-lunn:etal:13}{}}%
Lunn, D., Barrett, J., Sweeting, M., et al. (2013) Fully {Bayesian}
hierarchical modelling in two stages, with application to meta-analysis.
\emph{Journal of the Royal Statistical Society Series C}, \textbf{62},
551--572. DOI:
\href{https://doi.org/10.1111/rssc.12007}{10.1111/rssc.12007}.

\leavevmode\vadjust pre{\hypertarget{ref-manderson_numerically_2022}{}}%
Manderson, A. A. and Goudie, R. J. B. (2022) A numerically stable
algorithm for integrating {Bayesian} models using {Markov} melding.
\emph{Statistics and Computing}, \textbf{32}, 24. DOI:
\href{https://doi.org/10.1007/s11222-022-10086-2}{10.1007/s11222-022-10086-2}.

\leavevmode\vadjust pre{\hypertarget{ref-massa_combining_2010}{}}%
Massa, M. S. and Lauritzen, S. L. (2010) Combining statistical models.
In \emph{Algebraic Methods in Statistics and Probability {II}}, pp.
239--259. Contemp. {Math}. {Amer. Math. Soc., Providence, RI}. DOI:
\href{https://doi.org/10.1090/conm/516/10179}{10.1090/conm/516/10179}.

\leavevmode\vadjust pre{\hypertarget{ref-mauff_joint_2020}{}}%
Mauff, K., Steyerberg, E., Kardys, I., et al. (2020) Joint models with
multiple longitudinal outcomes and a time-to-event outcome: A corrected
two-stage approach. \emph{Statistics and Computing}, \textbf{30},
999--1014. DOI:
\href{https://doi.org/10.1007/s11222-020-09927-9}{10.1007/s11222-020-09927-9}.

\leavevmode\vadjust pre{\hypertarget{ref-maunder_review_2013}{}}%
Maunder, M. N. and Punt, A. E. (2013) A review of integrated analysis in
fisheries stock assessment. \emph{Fisheries Research}, \textbf{142},
61--74. DOI:
\href{https://doi.org/10.1016/j.fishres.2012.07.025}{10.1016/j.fishres.2012.07.025}.

\leavevmode\vadjust pre{\hypertarget{ref-meng_trio_2014}{}}%
Meng, X.-L. (2014) A trio of inference problems that could win you a
{Nobel Prize} in statistics (if you help fund it). In \emph{Past,
{Present}, and {Future} of {Statistical Science}} (eds X. Lin, C.
Genest, D. L. Banks,et al.), pp. 561--586. {Chapman and Hall/CRC}. DOI:
\href{https://doi.org/10.1201/b16720-52}{10.1201/b16720-52}.

\leavevmode\vadjust pre{\hypertarget{ref-nelsen_introduction_2006}{}}%
Nelsen, R. B. (2006) \emph{An Introduction to Copulas}. Second.
{Springer New York}. DOI:
\href{https://doi.org/10.1007/0-387-28678-0}{10.1007/0-387-28678-0}.

\leavevmode\vadjust pre{\hypertarget{ref-nicholson_interoperability_2021}{}}%
Nicholson, G., Blangiardo, M., Briers, M., et al. (2021)
Interoperability of statistical models in pandemic preparedness:
Principles and reality. \emph{Statistical Science (forthcoming)}.
Available at: \url{https://arxiv.org/abs/2109.13730}.

\leavevmode\vadjust pre{\hypertarget{ref-nimble_development_team_nimble_2019}{}}%
NIMBLE Development Team (2019) {NIMBLE}: {MCMC}, particle filtering, and
programmable hierarchical modeling. DOI:
\href{https://doi.org/10.5281/zenodo.1211190}{10.5281/zenodo.1211190}.

\leavevmode\vadjust pre{\hypertarget{ref-ohagan_uncertain_2006}{}}%
O'Hagan, A., Buck, C. E., Daneshkhah, A., et al. (2006) \emph{Uncertain
Judgements: Eliciting Experts' Probabilities}. Statistics in {Practice}.
{Wiley}. DOI:
\href{https://doi.org/10.1002/0470033312}{10.1002/0470033312}.

\leavevmode\vadjust pre{\hypertarget{ref-oh_considerations_2018}{}}%
Oh, E. J., Shepherd, B. E., Lumley, T., et al. (2018) Considerations for
analysis of time-to-event outcomes measured with error: Bias and
correction with {SIMEX}. \emph{Statistics in medicine}, \textbf{37},
1276--1289. DOI:
\href{https://doi.org/10.1002/sim.7554}{10.1002/sim.7554}.

\leavevmode\vadjust pre{\hypertarget{ref-oh_raking_2021}{}}%
Oh, E. J., Shepherd, B. E., Lumley, T., et al. (2021) Raking and
regression calibration: Methods to address bias from correlated
covariate and time-to-event error. \emph{Statistics in Medicine},
\textbf{40}, 631--649. DOI:
\href{https://doi.org/10.1002/sim.8793}{10.1002/sim.8793}.

\leavevmode\vadjust pre{\hypertarget{ref-parsons_bayesian_2021}{}}%
Parsons, J., Niu, X. and Bao, L. (2021) A {Bayesian} hierarchical
modeling approach to combining multiple data sources: {A} case study in
size estimation. \emph{arXiv:2012.05346 {[}stat{]}}. Available at:
\url{https://arxiv.org/abs/2012.05346}.

\leavevmode\vadjust pre{\hypertarget{ref-plummer_rjags_2019}{}}%
Plummer, M. (2019) Rjags: {Bayesian} graphical models using {MCMC}.

\leavevmode\vadjust pre{\hypertarget{ref-presanis_synthesising_2014}{}}%
Presanis, A. M., Pebody, R. G., Birrell, P. J., et al. (2014)
Synthesising evidence to estimate pandemic (2009) {A}/{H1N1} influenza
severity in 2009\textendash 2011. \emph{Annals of Applied Statistics},
\textbf{8}, 2378--2403. {The Institute of Mathematical Statistics}. DOI:
\href{https://doi.org/10.1214/14-AOAS775}{10.1214/14-AOAS775}.

\leavevmode\vadjust pre{\hypertarget{ref-rizopoulos_joint_2012}{}}%
Rizopoulos, D. (2012) \emph{Joint Models for Longitudinal and
Time-to-Event Data: With Applications in {R}}. {CRC Press}.

\leavevmode\vadjust pre{\hypertarget{ref-rosenberg_hazard_1995}{}}%
Rosenberg, P. S. (1995) Hazard function estimation using {B-splines}.
\emph{Biometrics}, \textbf{51}, 874--887. {{[}Wiley, International
Biometric Society{]}}. DOI:
\href{https://doi.org/10.2307/2532989}{10.2307/2532989}.

\leavevmode\vadjust pre{\hypertarget{ref-royston_flexible_2002}{}}%
Royston, P. and Parmar, M. K. B. (2002) Flexible parametric
proportional-hazards and proportional-odds models for censored survival
data, with application to prognostic modelling and estimation of
treatment effects. \emph{Statistics in Medicine}, \textbf{21},
2175--2197. DOI:
\href{https://doi.org/10.1002/sim.1203}{10.1002/sim.1203}.

\leavevmode\vadjust pre{\hypertarget{ref-rufo_bayesian_2012}{}}%
Rufo, María Jesús, Pérez, C. J. and Martín, J. (2012) A {Bayesian}
approach to aggregate experts' initial information. \emph{Electronic
Journal of Statistics}, \textbf{6}, 2362--2382. {Institute of
Mathematical Statistics and Bernoulli Society}. DOI:
\href{https://doi.org/10.1214/12-EJS752}{10.1214/12-EJS752}.

\leavevmode\vadjust pre{\hypertarget{ref-rufo_log-linear_2012}{}}%
Rufo, M. J., Martín, J. and Pérez, C. J. (2012) Log-linear pool to
combine prior distributions: {A} suggestion for a calibration-based
approach. \emph{Bayesian Analysis}, \textbf{7}, 411--438. {International
Society for Bayesian Analysis}. DOI:
\href{https://doi.org/10.1214/12-BA714}{10.1214/12-BA714}.

\leavevmode\vadjust pre{\hypertarget{ref-rutherford_use_2015}{}}%
Rutherford, M. J., Crowther, M. J. and Lambert, P. C. (2015) The use of
restricted cubic splines to approximate complex hazard functions in the
analysis of time-to-event data: A simulation study. \emph{Journal of
Statistical Computation and Simulation}, \textbf{85}, 777--793. {Taylor
{\(\&\)} Francis}. DOI:
\href{https://doi.org/10.1080/00949655.2013.845890}{10.1080/00949655.2013.845890}.

\leavevmode\vadjust pre{\hypertarget{ref-schaub_integrated_2011}{}}%
Schaub, M. and Abadi, F. (2011) Integrated population models: A novel
analysis framework for deeper insights into population dynamics.
\emph{Journal of Ornithology}, \textbf{152}, 227--237. DOI:
\href{https://doi.org/10.1007/s10336-010-0632-7}{10.1007/s10336-010-0632-7}.

\leavevmode\vadjust pre{\hypertarget{ref-schaub_local_2006}{}}%
Schaub, M., Ullrich, B., Knötzsch, G., et al. (2006) Local population
dynamics and the impact of scale and isolation: A study on different
little owl populations. \emph{Oikos}, \textbf{115}, 389--400. DOI:
\href{https://doi.org/10.1111/j.2006.0030-1299.15374.x}{10.1111/j.2006.0030-1299.15374.x}.

\leavevmode\vadjust pre{\hypertarget{ref-seethala_early_2017}{}}%
Seethala, R. R., Hou, P. C., Aisiku, I. P., et al. (2017) Early risk
factors and the role of fluid administration in developing acute
respiratory distress syndrome in septic patients. \emph{Annals of
Intensive Care}, \textbf{7}, 11. DOI:
\href{https://doi.org/10.1186/s13613-017-0233-1}{10.1186/s13613-017-0233-1}.

\leavevmode\vadjust pre{\hypertarget{ref-soetaert_rootsolve_2020}{}}%
Soetaert, K., Hindmarsh, A. C., Eisenstat, S. C., et al. (2020)
Rootsolve: Nonlinear root finding, equilibrium and steady-state analysis
of ordinary differential equations.

\leavevmode\vadjust pre{\hypertarget{ref-stan_development_team_rstan_2021}{}}%
Stan Development Team (2021) {RStan}: The {R} interface to {Stan}.

\leavevmode\vadjust pre{\hypertarget{ref-the_ards_definition_task_force_acute_2012}{}}%
The ARDS Definition Task Force (2012) Acute respiratory distress
syndrome: The {Berlin} definition. \emph{JAMA}, \textbf{307},
2526--2533. DOI:
\href{https://doi.org/10.1001/jama.2012.5669}{10.1001/jama.2012.5669}.

\leavevmode\vadjust pre{\hypertarget{ref-tom:etal:10}{}}%
Tom, J. A., Sinsheimer, J. S. and Suchard, M. A. (2010) Reuse, recycle,
reweigh: {Combating} influenza through efficient sequential {Bayesian}
computation for massive data. \emph{The Annals of Applied Statistics},
\textbf{4}, 1722--1748. {The Institute of Mathematical Statistics}. DOI:
\href{https://doi.org/10.1214/10-AOAS349}{10.1214/10-AOAS349}.

\leavevmode\vadjust pre{\hypertarget{ref-vehtari_rank-normalization_2020}{}}%
Vehtari, A., Gelman, A., Simpson, D., et al. (2020) Rank-normalization,
folding, and localization: An improved \(\widehat{R}\) for assessing
convergence of {MCMC}. \emph{Bayesian Analysis}. {International Society
for Bayesian Analysis}. DOI:
\href{https://doi.org/10.1214/20-BA1221}{10.1214/20-BA1221}.

\leavevmode\vadjust pre{\hypertarget{ref-wang_shape-restricted_2021}{}}%
Wang, W. and Yan, J. (2021) Shape-restricted regression splines with {R}
package Splines2. \emph{Journal of Data Science}, \textbf{19}, 498--517.
{School of Statistics, Renmin University of China}. DOI:
\href{https://doi.org/10.6339/21-JDS1020}{10.6339/21-JDS1020}.

\leavevmode\vadjust pre{\hypertarget{ref-wang_integrative_2020}{}}%
Wang, W., Aseltine, R., Chen, K., et al. (2020) Integrative survival
analysis with uncertain event times in application to a suicide risk
study. \emph{Annals of Applied Statistics}, \textbf{14}, 51--73.
{Institute of Mathematical Statistics}. DOI:
\href{https://doi.org/10.1214/19-AOAS1287}{10.1214/19-AOAS1287}.

\leavevmode\vadjust pre{\hypertarget{ref-zipkin_synthesizing_2018}{}}%
Zipkin, E. F. and Saunders, S. P. (2018) Synthesizing multiple data
types for biological conservation using integrated population models.
\emph{Biological Conservation}, \textbf{217}, 240--250. DOI:
\href{https://doi.org/10.1016/j.biocon.2017.10.017}{10.1016/j.biocon.2017.10.017}.

\end{CSLReferences}

\end{document}